\documentclass{article}

\usepackage{PRIMEarxiv}

\usepackage[utf8]{inputenc}
\usepackage[T1]{fontenc}
\usepackage{hyperref}
\usepackage{url}
\usepackage{array}
\usepackage{booktabs}
\usepackage{longtable}
\usepackage{amsfonts}
\usepackage{amssymb}
\usepackage{amsmath}
\usepackage{nicefrac}
\usepackage{microtype}
\usepackage{graphicx}
\usepackage{fancyhdr}
\usepackage{enumitem}
\usepackage{multirow}
\usepackage{makecell}
\usepackage{caption}
\usepackage{tikz}
\usetikzlibrary{trees,arrows.meta,positioning,shapes.geometric,fit}

\graphicspath{{media/}{../imgs/figs/}}

\usepackage{xspace}
\usepackage{xcolor}
\usepackage{amsthm}
\newtheorem{example}{Example}

\newcommand{\dbname}{\textsc{SpecDB}\xspace}

\newcommand{\dbgraph}{\textsc{DBModuleGraph}\xspace}

\newcommand{\cooperate}{\;\Longleftrightarrow\;}
\newcommand{\refiningagent}{\texttt{specdb-refining} \xspace}
\newcommand{\dbfeatureextract}{\textsc{DB-Feature-Extract}\xspace}
\newcommand{\dbfeaturemerge}{\textsc{DB-Feature-Merge}\xspace}

\newcommand{\reffig}[1]{Fig.~\ref{fig:#1}}
\newcommand{\refsec}[1]{Sec.~\ref{sec:#1}}
\newcommand{\reftable}[1]{Table~\ref{tab:#1}}

\newcommand{\cm}{\checkmark}

\title{\dbname: LLM-Generated Customized Databases\\via Feature-Oriented Decomposition}

\author{
  Yunkai Lou \\
  Alibaba Group \\
  Hangzhou, China \\
  \texttt{louyunkai.lyk@alibaba-inc.com} \\
  \And
  Longbin Lai \\
  Alibaba Group \\
  Hangzhou, China \\
  \texttt{longbin.lailb@alibaba-inc.com} \\
  \And
  Shunyang Li \\
  Alibaba Group \\
  Hangzhou, China \\
  \texttt{lishunyang.lsy@alibaba-inc.com} \\
  \And
  Zhengping Qian \\
  Alibaba Group \\
  Hangzhou, China \\
  \texttt{zhengping.qzp@alibaba-inc.com} \\
  \And
  Ying Zhang \\
  Zhejiang Gongshang University \\
  Hangzhou, China \\
  \texttt{ying.zhang@zjgsu.edu.cn} \\
}

\begin{document}
\maketitle

\begin{abstract}
Mainstream relational databases ship a uniform feature set across
deployments, although individual workloads exercise only a fraction
of the available subsystems. We investigate whether a database can
instead be generated on demand with a feature set matched to the
target workload.
We present \dbname, a system that uses large language models (LLMs)
to synthesize customized relational databases. We survey 9
production systems and decompose them into 10 functional modules,
each further divided into implementation variants. To capture
dependencies among variants, including cases where implementations
in disjoint subtrees must be co-designed, we adopt the FODA feature
model and extend it with a cooperate edge, yielding a dependency
graph \dbgraph. \dbname operationalizes \dbgraph through a layered
module-construction pipeline in which each module is generated,
validated, and integrated by a dedicated subagent (driven
internally by three inner agents: Main, Tester, Architect), and a
\refiningagent that iteratively repairs and tunes the assembled
database against a user-supplied refining harness with read-only
access to existing database source code. A companion selection
component translates a natural-language workload description into a
set of implementation variants, providing an end-to-end pipeline
from workload description to deployable database.
We evaluate \dbname on TPC-C with BenchmarkSQL. The generated
database (23{,}779 lines of Rust) completes 60-minute TPC-C at 1
and 10 warehouses with zero errors. At 10 warehouses it reaches
tpmC\,=\,130, compared to 128 for PostgreSQL~14.23 and 127
for MySQL~8.0.45, with per-transaction latency comparable to
PostgreSQL and at $\sim$2.9\% of PostgreSQL's and $\sim$2.8\% of
MySQL's server-side code size. Because the agent operates at the
level of module specifications rather than product source, the
pipeline has no product-specific scaffolding to defend and can
in principle combine techniques across system boundaries; the
TPC-C instance here exercises a PG-dominated subset of that
capacity, and a workload-mix demonstration is left to future
work. Paired with falling LLM costs, we see this capacity as a
step toward a future in which generating a purpose-built
database for a target workload becomes straightforward.
\end{abstract}


\section{Introduction}
\label{sec:intro}

Relational database management systems (RDBMSs) are deployed across
workloads with widely varying feature requirements. Representative
examples fall into roughly four categories.
\emph{Embedded and small-scale} applications---a bibliography
manager, a mobile app's local store, a one-off CSV cleanup
script---need simple CRUD over a small table; MVCC, crash
recovery, and a network server are all unnecessary.
\emph{Transactional OLTP} applications---e-commerce checkout, a
banking ledger, multi-player game state---require MVCC,
multi-version isolation, crash recovery, and high-throughput
indexing under many concurrent connections, with tail latency as
the dominant cost.
\emph{Analytical OLAP} applications---BI dashboards, ad-hoc data
exploration, nightly ETL---scan gigabytes-to-terabytes and
aggregate them; columnar storage, vectorized execution, and bulk
ingest matter, whereas concurrent fine-grained updates and full
ACID over single rows do not.
\emph{Latency-bound or specialized} applications---real-time
trading engines, online feature stores---demand aggressive space-for-time optimizations (in-memory
indexes, no-disk-on-read paths, no eviction) that would be
prohibitive for a multi-terabyte dataset.
For any one workload, the required feature set is a strict subset
of the full RDBMS design space.

Mainstream systems such as PostgreSQL~\cite{postgresql} and
MySQL~\cite{mysql} instead adopt a monolithic architecture in which a
single product covers the entire design space. Every installation
ships the same full stack of subsystems (e.g., query optimizer,
storage engine) regardless of which of these the workload
actually exercises. Workload-specific tuning may then be
applied on top of such monolithic systems to improve performance
on the given tasks~\cite{autoadmin,ottertune,selfdriving}. This produces three
costs: unused subsystems consume code and runtime resources,
system complexity raises the barrier to use, extension, and
correct configuration, and the system's functional boundary is
fixed at release time---new scenarios that require a different
feature subset cannot be accommodated without forking the
codebase or accepting the overhead of features the scenario
never exercises.

These costs share a common cause: the system is designed once and then
adapted to each workload through configuration. An alternative is to
fix the feature set at generation time, designing the system per
workload rather than at run time~\cite{endofarchera}. Recent advances in code-generation
LLMs make this approach feasible: SpecFS~\cite{specfs} demonstrated
that a complete file system can be synthesized on demand from modular
specifications. The analogous move for databases is to generate, from
a workload description, a database whose feature set matches the
workload, exposing only the configuration surface that the chosen
feature set requires.

Realizing this in practice, however, is harder than it appears.
Asking an LLM to implement each database subsystem purely from a
workload description and the model's own priors faces two
compounding problems. The first is algorithmic: left to its
priors, an LLM tends to select algorithms with worse time complexity when
more efficient alternatives
exist~\cite{abbassi2025inefficiencies,qiu2025enamel}; in the
database context it is particularly costly because hot paths
execute millions of times per second. A heap allocator that scans
pages linearly for free space instead of maintaining a free-space
map, an MVCC layer that omits a vacuum horizon and leaks dead
versions, a B$^+$-tree that takes an exclusive latch on every
internal node instead of using a shared-exclusive intent latch for
descent---each of these is a natural ``obvious'' implementation
that is nonetheless far from production quality.

The second problem is the repair cascade that follows. An LLM can
produce a compilable, unit-test-passing implementation in minutes,
but the first realistic workload run exposes failures. Fixing one
defect shifts the bottleneck to the next; the model patches
locally without restructuring, and each patch introduces new
regressions. The loop rarely converges: after enough point fixes
the subsystem's internal invariants are too entangled to repair
incrementally, and the only path forward is a full rewrite of the
component. 

\begin{figure}[t]
    \centering
    \includegraphics[width=\columnwidth]{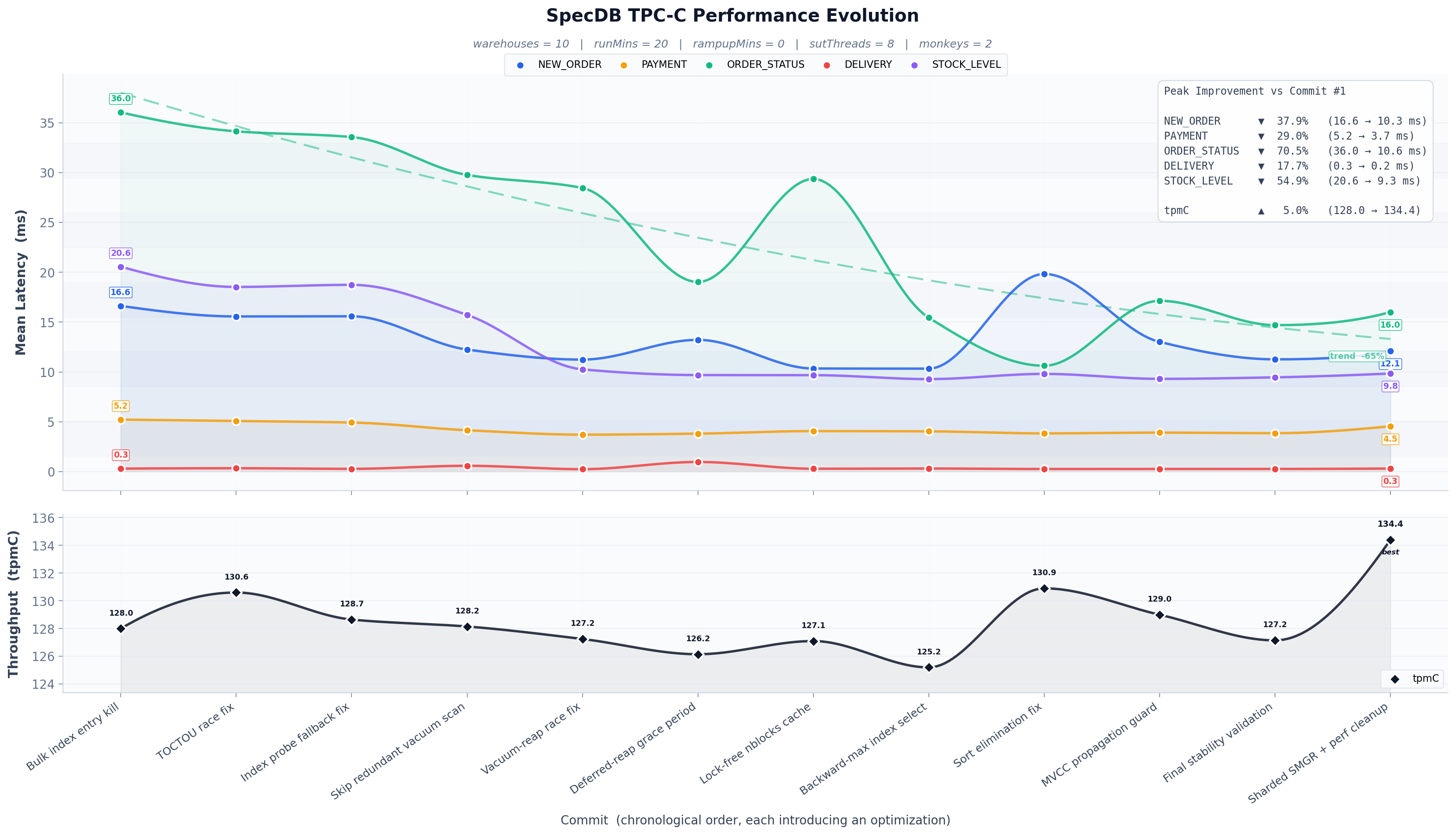}
    \caption{Refining progression: per-transaction mean latency (top)
    and tpmC throughput (bottom) across 12 representative commits
    during the refining phase (wh=10, \texttt{runMins=20}). Latency
    trends downward and throughput trends upward as the agent
    iteratively fixes performance defects.}
    \label{fig:refining-curve}
\end{figure}

We address the problems by changing what the LLM is asked to produce.
Rather than synthesize each subsystem from a workload description and
the model's priors, we ask it to borrow: for every architectural slot,
to select a design that an existing production system has shipped, and
to re-implement it in the generated database. Production RDBMSs have
explored which storage models, concurrency-control strategies,
recovery protocols, and access paths perform well on real hardware;
borrowing from these designs reduces the search space the LLM must
cover. Drawing on the DB-Engines
ranking\footnote{\url{https://db-engines.com/en/ranking}}, we select
9 open-source relational databases (PostgreSQL, MySQL, MariaDB,
SQLite, DuckDB, TiDB, ClickHouse, YugabyteDB, CockroachDB).
These 9 systems are chosen so that the seven classification
dimensions of the relational design space (workload, deployment,
scale, transactional guarantees, data model, SQL surface,
distribution topology) are jointly covered (\reftable{db-coverage}): each principal
alternative along each dimension is realized by at least one of
the surveyed systems. Taking the data model dimension as an example, the 9 systems
collectively span the major data-model alternatives a relational
engine can expose: pure relational tables of scalar columns
(MySQL, MariaDB, SQLite, TiDB), object-relational extensions
with composite, array, and range types (PostgreSQL, CockroachDB,
YugabyteDB, DuckDB), multi-model engines that ship JSON, spatial,
and vector types as first-class (PostgreSQL, DuckDB, ClickHouse),
and schema-flexible JSON-column storage (supported by eight of the
nine). The other six dimensions are covered with the same logic. \refsec{dbtree} explains the
dimensions and the selection in detail, and reports the full
per-system coverage in \reftable{db-coverage}. We then analyze the
module structure and implementation techniques of each surveyed
system.
The result is consolidated into a module tree that records, per
module (storage, indexing, concurrency, etc.), the design each
surveyed database adopts. PostgreSQL's append-only \texttt{heap} and
MySQL InnoDB's clustered B$^+$-tree, for example, both appear under
storage as alternative designs. Different modules can be filled from
different databases: an analytical workload, for instance, may pair
ClickHouse-style columnar storage with PostgreSQL-style query
optimization.

We operationalize the borrowing recipe by handing the module tree to
a coding agent and asking it to assemble the chosen designs into a
working database. This raises two challenges.

\paragraph{Challenge 1: Cross-module dependencies.}
Borrowing a known-good design for each module does not by itself
ensure that the resulting subsystems fit together. The implementation
of one module often depends on choices made in another module in a
disjoint part of the tree. The buffer pool (under storage) must not
flush a dirty page until the corresponding WAL record (under
recovery) has reached stable storage; the optimizer's access-path
enumeration must match the indexing module's leaf inventory. If each
module is generated in isolation, integration requires substantial
rewrites.

\paragraph{Challenge 2: Harness engineering.}
A coding agent producing one-shot code rarely yields a database that
compiles, passes its tests, and meets the workload's performance
target in a single pass. Under the borrowing recipe, each module's
implementation must also be reconciled with the borrowed behaviour
of its cooperating modules. A single-agent loop provides no
separation between code generation, testing, integration, and
post-deployment performance tuning, and no mechanism to consult the
production codebases the designs were borrowed from when problems
arise.

We address Challenge 1 by adopting the FODA feature
model~\cite{kang1990feature} on top of the module tree and extending
it with cross-module edges. Two are inherited from classical FODA
(requires, mutex-with) and two are new: reference, for soft
directional consistency, and cooperate, for the symmetric case where
two modules share an invariant that neither owns alone. The result
is the dependency graph \dbgraph. For Challenge 2, we extend
SpecFS's generation approach into a layered module-construction
pipeline in which each selected module is handled by a dedicated
subagent, and the subagents in the same dependency layer run in
parallel. Each subagent drives three specialized inner agents in
sequence: a Main Agent that generates Rust code for the module, a
Tester Agent that writes and runs unit and integration tests and
reports structured results, and an Architect Agent that performs
cross-module integration and fixes correctness or performance
regressions. A subsequent
\refiningagent takes a user-supplied refining harness and target metrics
(e.g., mean latency $\le$50\,ms) and iteratively repairs and tunes
the database until the targets are met. The \refiningagent has
read-only access to the source repositories of the 9 surveyed
databases, allowing it to consult how PostgreSQL, MySQL, or other
systems handle a given situation and adapt their solution. This
extends the borrowing loop into the post-generation debugging phase.
\reffig{refining-curve} illustrates the process of using
\texttt{specdb-refining} to refine the generated database against
the TPC-C benchmark: per-transaction latency decreases and
throughput increases across successive commits as the agent
iteratively diagnoses and resolves performance defects.

This paper makes the following contributions:
\begin{enumerate}[nosep]
    \item We introduce per-workload customized databases generated
    via a borrowing recipe. Rather than ship a single monolithic
    system, we generate per workload a database containing only the
    features the workload needs, and we restrict each module's
    implementation to designs already shipped by production RDBMSs.
    This avoids the performance inefficiencies that LLM-generated
    code exhibits when written purely from a workload description.
    \item We survey 9 production RDBMSs and consolidate their
    module structures into a unified module tree whose leaves are
    implementation variants drawn from the surveyed systems. To
    capture cross-module dependencies among these variants, we
    adopt the FODA feature model and extend it with two edge types,
    reference (soft directional) and cooperate (symmetric joint
    contract), yielding the dependency graph \dbgraph.
    \item We implement \dbname as a pipeline that operationalizes
    \dbgraph. The pipeline includes a workload-driven selection
    component that maps a natural-language description to nodes in
    \dbgraph, a layered module-construction stage in which each
    module is generated, validated, and integrated by a dedicated
    subagent (driven internally by Main, Tester, and Architect
    inner agents), and a \refiningagent that closes the loop
    against a user-supplied refining harness with read-only access to
    the surveyed databases' source code.
    \item We evaluate \dbname on TPC-C with BenchmarkSQL. The
    generated database, 23{,}779 lines of Rust, completes 60-minute
    TPC-C at 1 and 10 warehouses with zero errors. At 10
    warehouses it reaches tpmC\,=\,130, compared to 128 for
    PostgreSQL~14.23 and 127 for MySQL~8.0.45, with
    per-transaction latency comparable to PostgreSQL and at
    $\sim$2.9\% of PostgreSQL's and $\sim$2.8\% of MySQL's
    server-side LOC.
\end{enumerate}

The remainder of this paper is organized as follows.
\refsec{related} reviews related work on LLM-based system synthesis.
\refsec{dbtree} describes how \dbgraph is constructed from 9
production database codebases and how we extend FODA with the
cooperate edge. \refsec{design} presents the \dbname pipeline.
\refsec{experiment} reports TPC-C results. \refsec{limitations}
discusses limitations and future work. \refsec{conclude} concludes.

\section{Related Work}
\label{sec:related}

\subsection{Workload-Driven DBMS Tuning}
\label{sec:related-tuning}

A substantial body of prior work adapts a DBMS to a target
workload at deployment or runtime through configuration choices,
physical-design advisors, or autonomous in-engine planners,
without modifying the engine's source code.
\emph{AutoAdmin}~\cite{autoadmin}, initiated at Microsoft
Research in 1996 and incorporated into SQL Server as the Index
Tuning Wizard and the Database Engine Tuning Advisor, automates
the selection of indexes, materialized views, and partitioning
through cost-based search over the optimizer's ``what-if''
interface. \emph{OtterTune}~\cite{ottertune} applies factor
analysis, Lasso regression, and Gaussian-process recommendation
to tune the configuration knobs exposed by modern DBMSs; it
operates as an external service that connects to the target
system via JDBC, recommends knob settings, and iterates over an
observation loop. \emph{Peloton}~\cite{selfdriving} integrates
the tuning loop within the engine: its self-driving architecture
monitors workload behaviour, forecasts demand, and autonomously
applies actions drawn from a fixed action set (e.g., adding
indexes, adjusting knobs).

All three approaches share a common assumption: the engine code
is implemented in advance, and the automated component, whether
a cost model, a machine-learning recommender, or a forecasting
planner, only \emph{selects} among pre-implemented options. The
achievable optimization space is therefore bounded by the
mechanisms the engine already supports, so both the optimization
capability and the scalability to new mechanisms are limited. \dbname pursues the same objective of
producing a workload-specific DBMS, but moves the synthesis
target from \emph{configuration} to \emph{code}: rather than
select among pre-built knobs, indexes, or actions, the pipeline
synthesizes the corresponding subsystems from a cross-database
specification library. The optimization space is therefore
bounded by what the specification library and the LLM can
jointly express, not by what an existing engine already provides.

\subsection{LLM-Driven System Synthesis}
\label{sec:related-synthesis}

Beyond editor-level completion provided by tools such as GitHub
Copilot~\cite{copilot} and Claude Code~\cite{claude}, two recent
lines of work demonstrate that an LLM agent can produce an
end-to-end system. SpecFS~\cite{specfs} synthesizes a complete
FUSE-based concurrent file system by replacing free-form
natural-language prompts with a structured specification written
in Hoare-style pre/post-conditions and invariants, from which an
LLM toolchain emits the corresponding C code.
Bespoke~OLAP~\cite{bespoke} synthesizes a ``one-size-fits-one''
OLAP engine from a workload contract consisting of parameterised
SQL query templates plus a Parquet dataset, and reports
$11.78\times$/$9.76\times$ TPC-H/CEB total-runtime speedups over
DuckDB by hard-coding workload-specific columnar layouts and
per-template C++ execution kernels. This ``one workload, one
engine'' thesis is consistent with the earlier empirical argument
of Stonebraker et al.~\cite{endofarchera} that purpose-built
engines can outperform general-purpose RDBMSs by orders of
magnitude when freed from features the workload never exercises.
Stonebraker's argument, however, operates at the granularity of
major \emph{markets} (e.g., OLTP, OLAP, stream processing)
and treats per-market specialization as worthwhile only when the
market is, in their phrasing, ``of significant enough size to
warrant the investment'' in a hand-built engine. With LLM-driven
generation, that investment threshold collapses to the cost of a
generation run, so the unit of specialization can shrink
further---per workload, per team, per benchmark, or per class of
queries.

\dbname is related to SpecFS and Bespoke-OLAP. From SpecFS we adopt the specification
style: each leaf in \dbgraph (\refsec{dbgraph-extract}) carries
Hoare-style pre/post-conditions and invariants describing an
implementation technique (e.g., PostgreSQL's append-only heap or
MySQL/InnoDB's clustered B$^+$-tree). The generated artefacts,
however, differ in scope.
SpecFS generates a file system, whose external surface is the POSIX
vfs interface and whose internal modules form a layered stack.
Bespoke generates a workload-specific fast path that does not parse
SQL, omits a conventional query optimizer, and excludes indexes,
statistics, materialized views, concurrency control, crash recovery,
replication, and a wire protocol; out-of-contract queries are routed
to a fallback general-purpose SQL processor. \dbname
generates a standalone relational DBMS that exposes SQL, provides
ACID under concurrent access, and spans parser, optimizer, storage
model, indexing, concurrency control, recovery, and replication. The
resulting module set carries cross-module invariants (e.g.,
WAL-before-page ordering, snapshot-visibility consistency) that no
module owns alone.

\subsection{LLM Agent Harnesses for Code Generation}
\label{sec:related-harness}

Beyond single-prompt code completion, a body of work studies how to
\emph{orchestrate} an LLM into a reliable code-producing agent. We
group prior work into three threads. \emph{Multi-agent role
decomposition} assigns distinct roles---designer, coder,
tester---to separate LLM instances that communicate via structured
messages: ChatDev~\cite{chatdev} casts software development as a
chat between specialized agents, MetaGPT~\cite{metagpt} encodes
standard-operating-procedure templates so that each role's output
matches a fixed schema, and AutoGen~\cite{autogen} provides a
general framework for conversable agents that can also call tools
and humans. \emph{Iterative self-repair} closes a feedback loop
between generated code and execution signals: Reflexion~\cite{reflexion}
turns runtime failures into verbal critiques that are appended to
the next prompt, and Self-Debug~\cite{selfdebug} teaches the model
to inspect intermediate state and rewrite faulty code without
external supervision. \emph{Test-driven flow engineering and
agent-computer interfaces} structure the entire generation
pipeline around verifiable artefacts: AlphaCodium~\cite{alphacodium}
splits competition-style code generation into a spec-then-test-then-code
flow that markedly improves correctness over single-prompt
prompting, while SWE-agent~\cite{sweagent} engineers the interface
between an LLM and a real software-engineering environment (file
editing, command execution, search) to make autonomous bug fixing
feasible.

\dbname's harness specializes these threads for full-system database
synthesis rather than function- or repository-level edits. Role
decomposition is instantiated per module: each selected module is
handled by a dedicated subagent driven by three specialized inner
agents (Main, Tester, Architect) that mirror the
generate-validate-integrate loop required by \dbgraph's cross-module
edges. The Architect Agent addresses cross-module integration over
symmetric joint contracts, which prior chat-style frameworks do not
surface as a first-class construct.
Iterative repair is extended beyond unit-test failures to performance
regressions: the \refiningagent runs a user-supplied workload harness
with explicit target metrics and turns latency or throughput
shortfalls into repair tasks. Where AlphaCodium and SWE-agent rely on
the model's priors when a fix is non-obvious, the \refiningagent has
read-only access to the source repositories of the 9 surveyed
databases, so that borrowing also drives post-generation debugging
and tuning.

\section{Constructing \dbgraph from Production Codebases}
\label{sec:dbtree}

We extract modules and implementation techniques from 9 production
relational systems (SQLite, DuckDB, TiDB, PostgreSQL, MariaDB,
ClickHouse, MySQL, YugabyteDB, CockroachDB) into a unified module
tree, and turn this tree into the dependency graph \dbgraph by adding
cross-module edges.

\subsection{Background: the FODA Feature Model}
\label{sec:foda}

FODA (Feature-Oriented Domain Analysis)~\cite{kang1990feature} models
a software family as a \emph{feature tree}. Each non-root node is
annotated with its relationship to its parent: \emph{mandatory}
($\bullet$, present whenever the parent is present), \emph{optional}
($\circ$, may or may not be present), \emph{alternative}/XOR (a
sibling group of which exactly one is selected), or \emph{or} (a
sibling group of which one or more are selected). The tree
therefore captures both the common backbone of the family and the
variability points at which family members differ. FODA also
admits \emph{cross-tree constraints} between non-sibling features
to restrict valid selections; we defer the discussion of edges
to \refsec{dbgraph-edges}.

\paragraph{Why FODA fits the database design space.}
Three properties make FODA a fit for this setting. First, production
relational databases form a product family: they share a common
backbone (parser, query processor, storage engine, recovery, etc.)
but differ in which implementation each module ships
(B$^+$-tree-clustered storage vs.\ append-only heap, MVCC vs.\ 2PL,
WAL vs.\ rollback journal). FODA's mandatory markers capture the
backbone, and alternative/or groups capture the variability. Second,
customising a database for a workload is an instance of FODA's
configuration problem: choose, for every variability point, which
sub-feature to include. This makes the user-facing operation in
\dbname (picking an implementation per module) a projection of FODA
semantics onto the database design space. Third, FODA admits
cross-tree constraints between non-sibling features, which database
modules require (storage talks to recovery, the optimizer talks to
indexing), and provides a vocabulary that we extend
in~\refsec{dbgraph-edges}.

\subsection{Bottom-Up Extraction from 9 Codebases}
\label{sec:dbgraph-extract}

\paragraph{Selection of the 9 systems.}
Before describing the extraction procedure, we explain why these
particular 9 databases were surveyed. The selection targets
\emph{coverage of the relational design space} rather than market
size. We characterize that design space along seven user-facing
dimensions, each split into its principal alternatives, and select
systems so that every alternative is realized by at least one
production implementation. The seven dimensions are workload
positioning (OLTP / OLAP / HTAP), deployment shape (embedded /
single-server / distributed cluster), scale envelope (vertical /
read-replica / sharded / native horizontal / geo-distributed),
transactional guarantees (single-node ACID / distributed ACID /
strict serializable / tunable or BASE), data model (pure relational
/ object-relational / multi-model / schema-flexible), SQL surface
(broad ANSI / stored procedures / advanced features such as CTE and
window functions / limited or specialized dialect), and distribution
topology (single instance / single-leader / multi-leader /
per-shard consensus / geo-replicated). The taxonomy intentionally
covers the major mainstream categories; several less mainstream
ones, such as workload-specialized engines, cloud-native architectures with separated compute
and storage, and federated query engines, are left outside the
scope of this paper to keep the design space tractable.
\reftable{db-coverage} reports the resulting coverage: every one
of the 28 alternatives is realized by at least one of the 9
systems. This is the precondition for the borrowing recipe: if any
alternative had no production exemplar, the corresponding XOR slot
in the merged tree would have to be filled by LLM priors rather
than by re-implementing a proven design.

\begin{table*}[t]
\centering
\caption{Coverage of seven classification dimensions across the 9
surveyed databases. \cm\ indicates that the system covers the
alternative. The 28 alternatives are split into three panels (a)--(c)
for readability; every alternative is realized by at least one
production system.}
\label{tab:db-coverage}
\footnotesize
\setlength{\tabcolsep}{4pt}
\renewcommand{\arraystretch}{1.0}

\textit{(a) Workload positioning, deployment shape, and scale envelope.}\\[0.2em]
\resizebox{\textwidth}{!}{%
\begin{tabular}{@{}l|ccc|ccc|ccccc@{}}
\toprule
& \multicolumn{3}{c|}{\textbf{Workload}}
& \multicolumn{3}{c|}{\textbf{Deployment}}
& \multicolumn{5}{c}{\textbf{Scale}} \\
\cmidrule(lr){2-4}\cmidrule(lr){5-7}\cmidrule(lr){8-12}
\textbf{DB}
& OLTP & OLAP & HTAP
& Embedded & \makecell{Single-\\server} & Distributed
& Vertical & \makecell{Read-\\replica} & Sharded & Horizontal & \makecell{Geo-\\distributed} \\
\midrule
PostgreSQL  & \cm &     &     &     & \cm &     & \cm & \cm &     &     &     \\
MySQL       & \cm &     &     &     & \cm &     & \cm & \cm &     &     &     \\
MariaDB     & \cm &     & \cm &     & \cm &     & \cm & \cm & \cm &     &     \\
SQLite      & \cm &     &     & \cm &     &     & \cm &     &     &     &     \\
DuckDB      &     & \cm &     & \cm &     &     & \cm &     &     &     &     \\
TiDB        &     &     & \cm &     &     & \cm &     &     &     & \cm & \cm \\
ClickHouse  &     & \cm &     &     & \cm & \cm & \cm &     & \cm & \cm & \cm \\
YugabyteDB  & \cm &     &     &     &     & \cm &     &     &     & \cm & \cm \\
CockroachDB & \cm &     &     &     &     & \cm &     &     &     & \cm & \cm \\
\bottomrule
\end{tabular}%
}

\vspace{0.5em}

\textit{(b) Transactional guarantees and data model.}\\[0.2em]
\begin{tabular}{@{}l|cccc|cccc@{}}
\toprule
& \multicolumn{4}{c|}{\textbf{Guarantees}}
& \multicolumn{4}{c}{\textbf{Data model}} \\
\cmidrule(lr){2-5}\cmidrule(lr){6-9}
\textbf{DB}
& \makecell{Single-\\node ACID} & \makecell{Distributed\\ACID} & \makecell{Strict\\serializable} & Tunable
& \makecell{Pure\\relational} & \makecell{Object-\\relational} & \makecell{Multi-\\model} & \makecell{Schema-\\flexible} \\
\midrule
PostgreSQL  & \cm &     &     &     &     & \cm & \cm & \cm \\
MySQL       & \cm &     &     &     & \cm &     &     & \cm \\
MariaDB     & \cm &     &     &     & \cm &     &     & \cm \\
SQLite      & \cm &     &     &     & \cm &     &     & \cm \\
DuckDB      & \cm &     &     &     &     & \cm & \cm & \cm \\
TiDB        &     & \cm &     &     & \cm &     &     & \cm \\
ClickHouse  &     &     &     & \cm &     &     & \cm & \cm \\
YugabyteDB  &     & \cm & \cm &     &     & \cm &     & \cm \\
CockroachDB &     & \cm & \cm &     &     & \cm &     & \cm \\
\bottomrule
\end{tabular}

\vspace{0.5em}

\textit{(c) SQL surface and distribution topology.}\\[0.2em]
\begin{tabular}{@{}l|cccc|ccccc@{}}
\toprule
& \multicolumn{4}{c|}{\textbf{SQL surface}}
& \multicolumn{5}{c}{\textbf{Topology}} \\
\cmidrule(lr){2-5}\cmidrule(lr){6-10}
\textbf{DB}
& \makecell{Broad\\ANSI} & \makecell{Stored\\procedures} & \makecell{Advanced\\features} & \makecell{Limited\\dialect}
& \makecell{Single\\instance} & \makecell{Single-\\leader} & \makecell{Multi-\\leader} & \makecell{Per-shard\\Raft} & \makecell{Geo-\\replicated} \\
\midrule
PostgreSQL  & \cm & \cm & \cm &     &     & \cm &     &     &     \\
MySQL       & \cm & \cm & \cm &     &     & \cm & \cm &     &     \\
MariaDB     & \cm & \cm & \cm &     &     & \cm & \cm &     &     \\
SQLite      &     &     & \cm & \cm & \cm &     &     &     &     \\
DuckDB      & \cm &     & \cm &     & \cm &     &     &     &     \\
TiDB        & \cm & \cm & \cm &     &     &     &     & \cm & \cm \\
ClickHouse  &     &     & \cm & \cm &     & \cm &     &     & \cm \\
YugabyteDB  & \cm & \cm & \cm &     &     &     &     & \cm & \cm \\
CockroachDB & \cm & \cm & \cm &     &     &     &     & \cm & \cm \\
\bottomrule
\end{tabular}
\end{table*}

\begin{figure*}[t]
\centering
\includegraphics[width=\textwidth]{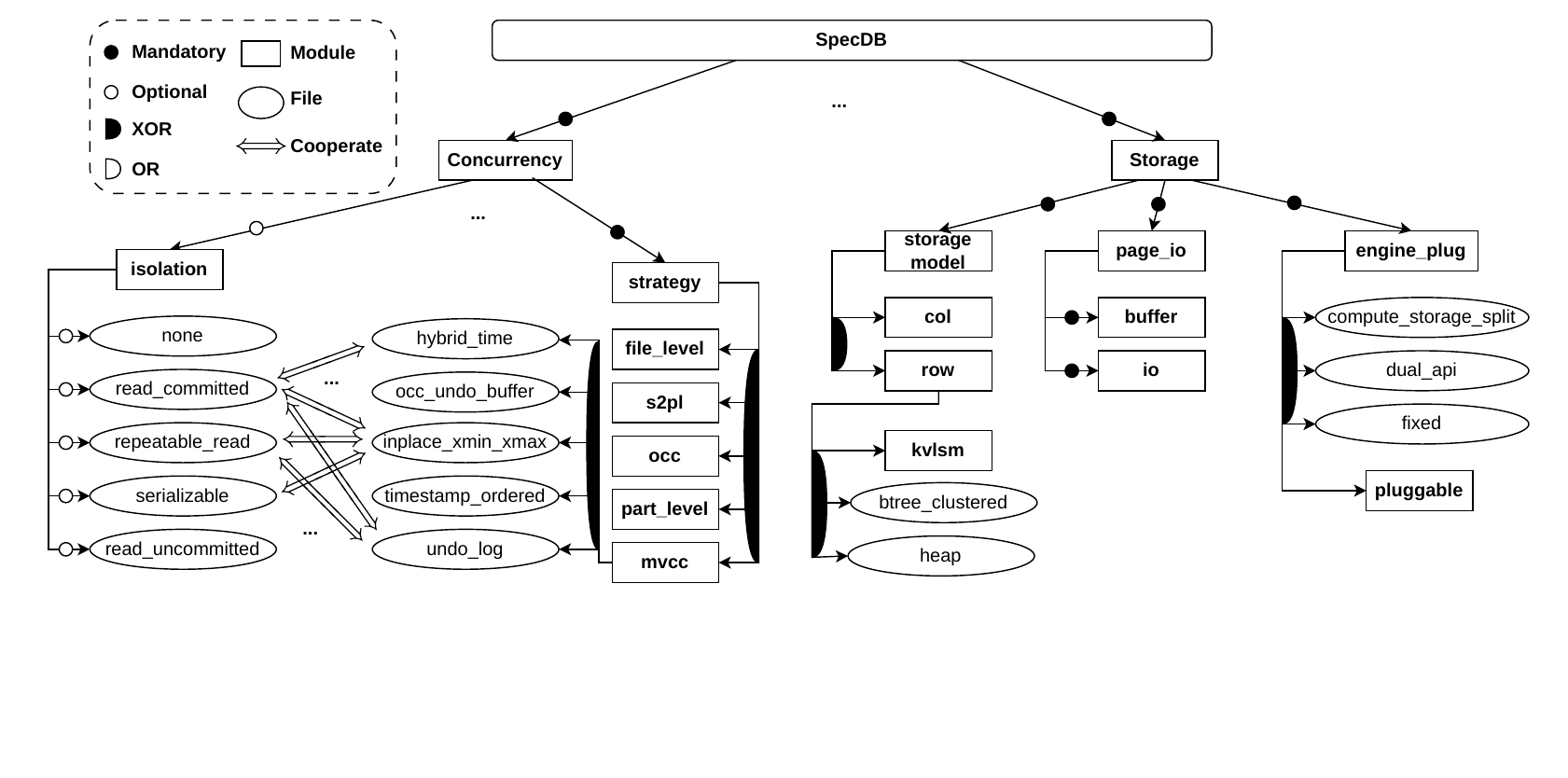}
\caption{Two representative branches of the \dbname module
tree---\texttt{concurrency} and \texttt{storage}---illustrating the
notation used throughout. Node shapes distinguish \emph{modules}
(rectangles, internal subtrees) from \emph{files} (ovals, leaves). On parent--child
edges, a filled circle ($\bullet$) marks the child as
\emph{mandatory} and an empty circle ($\circ$) marks it
\emph{optional}; a filled arc spanning a sibling group marks an
\emph{alternative} (XOR, exactly one selected), and an open arc
marks \emph{or} (at least one selected). Double-headed arrows
($\Leftrightarrow$) are \emph{cooperate} edges. For example, both
\texttt{concurrency} and \texttt{storage} are mandatory under
\dbname. Within \texttt{concurrency}, \texttt{strategy} is
mandatory and its children form an XOR group.
The remaining top-level modules and deeper levels of the
two shown branches are omitted for space.}
\label{fig:foda-tree}
\end{figure*}

\paragraph{Extraction overview.}
With the 9 systems, the remaining task is to extract their
module structures into a unified tree. We do this bottom-up from
the source trees rather than top-down, since a top-down authoring
would require us to anticipate the design space ahead of time. Two
automation skills, run over each of the 9 repositories in turn,
fold the heterogeneous codebases into a single unified feature
tree; \reffig{foda-tree} shows two representative branches.

\paragraph{Per-database extraction (\dbfeatureextract).}
For each of the 9 databases, the \dbfeatureextract skill runs three
phases over their source codebases.

\textit{(i) Module extraction.} The agent reads the top-level
directory structure, build files (\texttt{CMakeLists.txt},
\texttt{Makefile}, \texttt{go.mod}), and any
\texttt{ARCHITECTURE.md}/\texttt{README} files, and identifies the
architectural modules the codebase contains rather than projecting a
fixed template onto it. PostgreSQL's source tree, for instance,
surfaces \texttt{src/backend/storage}, \texttt{src/backend/access},
\texttt{src/backend/optimizer}, and \texttt{src/backend/replication},
which align onto the storage/indexing/query-processing/replication
axes; MySQL's \texttt{storage/innobase} reveals InnoDB as a pluggable
engine, recorded as a separate dimension under storage.

\textit{(ii) Hierarchical decomposition.} Each module is decomposed
by following the source code into its implementation choices: storage
model (row/column/KV), data organisation (heap, sorted file, LSM,
B$^+$-tree-clustered), buffer-pool replacement (clock-sweep, LRU,
ARC), and so on, until each leaf is a concrete implementation. The
skill applies a split-vs-don't-split rule set: configuration values
(page size, buffer-pool size) stay as leaf annotations, while
behaviourally distinct choices (WAL vs.\ rollback journal, hash join
vs.\ merge join, MVCC vs.\ 2PL vs.\ OCC) become subtree splits. This
keeps parameter knobs and architectural decisions on separate axes.

\textit{(iii) FODA relationship annotation.} Each parent-child edge
is labelled mandatory, optional, alternative (XOR), or or based on
whether the source code unconditionally initialises the child, gates
it behind a compile flag or plugin selector, or selects one of
several mutually exclusive variants.

Running the three phases on each of the 9 codebases gives, per
database, an annotated \emph{module tree}: the architectural
modules of that database, hierarchically decomposed into their
implementation choices. To present this output in a form that supports
cross-database comparison, we render each module tree as a table
whose rows are $\langle$database, module, module-tree-path,
technical description$\rangle$ tuples. The complete per-database tables are
collected in Appendix~\ref{sec:appendix:per-db-tables}.
\reftable{storage-cross-db} shows one such example: the
storage-module rows for PostgreSQL and MySQL, with each row giving
one storage-axis path and a brief description of how the database
realises it.

\paragraph{Cross-database merge (\dbfeaturemerge).}
The 9 module trees describe 9 parallel design spaces in different
vocabularies. To produce a single design space, the
\dbfeaturemerge skill folds them into the unified module tree.

The merge is necessary because the per-database trees, while
faithful to their source codebases, are not directly comparable. The
same design choice appears under different names (e.g., PostgreSQL's
``shared buffer pool'' and MySQL's ``InnoDB buffer pool''), at
different levels of the tree (PostgreSQL files its row-storage
variant under \texttt{storage\_model/row/heap}, while MySQL files
the same slot under \texttt{innodb/storage\_model/row\_format}), and
as different mandatory children of the same parent (PostgreSQL's
heap is mandatory under row storage; MySQL's clustered B$^+$-tree is
mandatory under the same slot). The merge identifies these
relationships: which mechanisms are different names for the same
concept, which co-exist as independent capabilities, and which are
alternative solutions to the same design problem and therefore
become selection knobs in the unified tree.

The merge realises this in two phases that run together over
each pair of trees. The first phase performs \emph{module
alignment}: it matches concepts that play the same role across the
two trees, regardless of name, granularity, or parent, and records
the matches in an alignment table that the second phase consults.
Alignment is not constrained to nodes at the same depth: when one
source documents a single leaf where the other expands the same
concept into a multi-level subtree of internal sub-mechanisms,
alignment matches the leaf to the subtree's root and the merge
keeps the deeper source's subtree in place beneath the aligned
node, treating the shallower source's leaf as one contribution to
that subtree. Refinements therefore accumulate across passes
rather than collapsing down to the coarsest common denominator.
The second phase performs \emph{relationship reconciliation}: when
the two sources mandate different children of an aligned parent,
the merge does not pick one over the other, nor does it silently
drop either; it demotes both to \emph{alternative} siblings of the
merged parent, turning the disagreement into an explicit XOR
variability point. Identical mandatories remain mandatory, and
optional children carry over unchanged.
{\footnotesize
\renewcommand{\arraystretch}{1.15}
\begin{longtable}{@{}>{\raggedright\arraybackslash}m{0.09\textwidth} >{\raggedright\arraybackslash}m{0.13\textwidth} >{\raggedright\arraybackslash}p{0.30\textwidth} >{\raggedright\arraybackslash}p{0.42\textwidth}@{}}
\caption{Storage-module rows extracted from the per-database extraction tables (Appendix~\ref{sec:appendix:per-db-tables}); for space we show only PostgreSQL and MySQL. Each row is one $\langle$database, module, module-tree-path, technical description$\rangle$ tuple. Paths are reproduced verbatim from each database's source module tree.}\\
\label{tab:storage-cross-db}\\
\toprule
\textbf{Database} & \textbf{Module} & \textbf{Module-Tree Path} & \textbf{Technical Description} \\
\midrule
\endfirsthead
\multicolumn{4}{c}{\footnotesize\itshape continued from previous page}\\
\toprule
\textbf{Database} & \textbf{Module} & \textbf{Module-Tree Path} & \textbf{Technical Description} \\
\midrule
\endhead
\midrule
\multicolumn{4}{r}{\footnotesize\itshape continued on next page}\\
\endfoot
\bottomrule
\endlastfoot
PostgreSQL & \multirow{5}{*}{Storage Engine} & Storage Model / Row Store / Row Organization / Heap / Heap Page Management & Each heap relation is paired with a Free Space Map that tracks remaining room per page in a tree structure, letting inserts pick a non-full page in near O(1), and a Visibility Map that uses 2 bits per page to mark all-visible and all-frozen pages so MVCC visibiliyty checks and vacuum work can be skipped. \\
 &  & Storage Model / Row Store / Row Organization / Heap / TOAST & Oversized attributes are sliced into chunks stored in a per-relation TOAST table and compressed inline; the default codec is the historical pglz, while LZ4 is offered from PG 14+ as a faster optional alternative selectable per column. \\
 &  & Table Access Method & PostgreSQL ships exactly one built-in Table AM, the heap, which all base relations use unless a third-party Table AM is registered through the pluggable API. \\
 &  & Storage Manager & Block-level I/O goes through a thin storage manager indirection layer (smgr) that dispatches to one built-in backend, the magnetic-disk backend, which simply maps relation forks onto regular operating-system files. \\
 &  & Async I/O & Asynchronous I/O can be performed by the legacy synchronous read path (method\_sync), by a pool of dedicated I/O worker processes (method\_worker), or, on Linux from PG 18+, by submitting requests through io\_uring (method\_io\_uring) for kernel-side batching. \\
\midrule
MySQL & \multirow{4}{*}{Storage Engine} & Storage Engine Selection / InnoDB / Storage Model / Row Formats & InnoDB supports four on-disk row formats: REDUNDANT preserves the original 5.0 layout, COMPACT shrinks header overhead, DYNAMIC stores variable-length columns off-page when they overflow, and COMPRESSED applies page-level zlib compression. \\
 &  & Storage Engine Selection / InnoDB / InnoDB Page Management & The doublewrite buffer absorbs partial-write torn pages by staging full page images before writing them in place. The change buffer (formerly insert buffer) defers and merges secondary-index modifications when target pages are not in the buffer pool. \\
 &  & Storage Engine Selection / InnoDB / InnoDB Tablespace & InnoDB stores data across the shared system tablespace ibdata1, file-per-table .ibd files, user-defined general tablespaces shared by multiple tables, dedicated undo tablespaces for rollback segments, and a temporary tablespace for intrinsic and user temporary tables. \\
 &  & Storage Engine Selection / MyISAM & MyISAM is a classic non-transactional row store that keeps data in MYD files and indexes in MYI files, optimised for read-mostly workloads without MVCC or crash recovery. \\
\end{longtable}
}

\begin{example}[Merging PostgreSQL and MySQL storage models]
\label{ex:pg-mysql-storage-merge}
We illustrate the two phases on the \texttt{storage\_model}
rows of \reftable{storage-cross-db}.

\emph{Alignment.} PostgreSQL files its row variant under
\texttt{storage\_model/row/heap}, while MySQL files the
same design choice under \texttt{storage\_engine/innodb/storage\_model/row\_formats} as a clustered B$^+$-tree.
The two paths agree on the role they play---they both decide how
a row is laid out on disk---but disagree on packaging: MySQL
wraps the storage model inside an InnoDB-specific subtree because
historically each MySQL engine ships its own row layout, whereas
PostgreSQL exposes the storage model directly because there is
only one engine to wrap. Module alignment recognizes the
intermediate \texttt{storage\_engine/innodb} level as
a packaging artefact rather than a storage-model concept and
matches both sources onto a single merged node
\texttt{storage/storage\_model/row/}.

\emph{Reconciliation.} On the merged node, PostgreSQL contributes
\texttt{heap} and MySQL contributes \texttt{btree\_clustered},
and each source had marked its contribution as the mandatory
child of the row-storage parent. Relationship reconciliation
refuses to pick a winner: both children are demoted to
alternative siblings, yielding the variability point
\texttt{storage/storage\_model/row/}~$\in$~\{\texttt{heap},
\texttt{btree\_clustered}\}. The two formerly-mandatory choices
become an explicit XOR knob in the merged tree.

Folding in the remaining seven databases extends this knob with a
\texttt{kvlsm/} sibling (CockroachDB's Pebble, TiDB's RocksDB)
and adds a \texttt{col/} sibling under \texttt{storage\_model/}
itself (ClickHouse's MergeTree variants, DuckDB's row-group
columnar storage), without changing the row/$\{$\texttt{heap},
\texttt{btree\_clustered}$\}$ alternation derived from the
PostgreSQL/MySQL pair. The cumulative \texttt{storage\_model}
subtree is the upper half of the storage branch shown in
\reffig{foda-tree}.
\end{example}

\paragraph{Result.}
After 8 merge passes (one per additional database), the unified
module tree contains 10 top-level modules (\texttt{architecture},
\texttt{concurrency}, \texttt{core}, \texttt{indexing},
\texttt{parser}, \texttt{query\_processing}, \texttt{recovery},
\texttt{replication}, \texttt{storage}, \texttt{value\_types}) and
several hundred leaves. \reffig{foda-tree} shows two representative
branches (\texttt{concurrency} and \texttt{storage}) to illustrate
the hierarchical shape and the FODA annotations; the remaining
eight modules follow the same pattern. \refsec{dbgraph-edges}
augments this tree with cross-module edges to produce \dbgraph.

\subsection{From the Module Tree to \dbgraph: Cross-Module Edges}
\label{sec:dbgraph-edges}

The module tree captures, for each module independently, what the
module is built from. Database modules do not stand alone: storage
talks to recovery, the optimizer talks to indexing, and the parser
talks to the lock manager. The mandatory/optional/XOR/or annotations
on parent-child edges do not express these cross-module
relationships. To turn the module tree into a dependency graph that
\dbname's code generator can walk, we use four cross-module edge
types. Two are inherited from classical FODA's cross-tree
constraints (requires, mutex-with), and two are new: reference, for
soft directional consistency, and cooperate, for symmetric co-design.

\paragraph{Requires and mutex-with.}
Classical FODA's two cross-tree constraints carry over directly.
Requires ($A \to B$) is a hard dependency: $A$ cannot be implemented
without $B$, so the generation pipeline must finish $B$ first. For
example, \texttt{value\_types/compound/array} $\to$
\texttt{parser/syntax/expression/subscript}: the array value type
is unreachable from SQL unless the parser accepts the subscript
grammar \texttt{arr[i]} that reads array elements, so the array
type cannot be exercised end-to-end until the subscript productions
are in place. Mutex-with
($A \otimes B$) prohibits coexistence in the same configuration
(e.g., \texttt{storage\_model/row} $\otimes$
\texttt{storage\_model/col}). These two cover the must-have and the
cannot-coexist cases.

\paragraph{Reference: soft directional dependency.}
Between hard requires and outright mutex-with, we observe a third
pattern in production codebases: $A$ does not need $B$ to exist, but
if $B$ is selected, $A$ must stay consistent with $B$'s interface.
Encoding this as a requires edge overstates the dependency and
forces unnecessary co-selection; dropping it loses an integration
constraint that production code enforces. We therefore add reference
($A \dashrightarrow B$), a soft directional consistency edge.
Examples include:

\begin{itemize}[nosep]
    \item join-order search $\dashrightarrow$
    statistics module: join-order search can
    fall back to default selectivities and still produce a plan,
    but if a statistics module is selected, the search must consume
    the statistics' distribution summaries and cardinality estimates. The
    dependency is soft (search runs without statistics) but
    enforced when both are present.
    \item parser productions for DDL \texttt{CREATE DATABASE} $\dashrightarrow$ catalog
    implementation, (with three
    XOR variants: heap-relation, versioned-descriptor,
    engine-dictionary): the namespace depth that
    \texttt{CREATE DATABASE} parses (hierarchical
    \texttt{db.schema.table} for the first two variants vs flat
    \texttt{schema.table} for the engine-dictionary variant)
    depends on the catalog variant in use. The parser productions
    can be written without a catalog, but once a catalog variant
    is selected, its hierarchy constrains which name forms the
    parser must accept.
\end{itemize}

Together with requires and mutex-with, the three directional edges
express ``$A$ must / may / must-not depend on $B$'', a caller-callee
relationship in which one side defines the contract and the other
consumes it.

\paragraph{Cooperate: a symmetric joint contract.}
A directional edge cannot express the case where two leaves share an
invariant that neither owns alone, so that neither side is caller
nor callee. Forcing such a case into a directional edge picks an
arbitrary side, hides the joint contract, and creates a generation
order that the other side later has to revise. We add cooperate
($A \cooperate B$), a symmetric joint contract on a shared
invariant (a binary layout, an ordering rule, or a conflict policy)
that neither side can specify in isolation. The two leaves must be
generated as a single unit; otherwise they are written independently
and only meet at integration time, producing the integration costs
that motivate this section. Our design contains 61 cooperate edges;
representative cases include:

\begin{itemize}[nosep]
    \item MVCC strategy $\cooperate$ recovery strategy: the on-disk
    version-record format (xmin/xmax fields, undo-pointer encoding,
    redo/undo split) is split between MVCC's visibility logic and
    the recovery log's redo path; the layout is owned by neither
    side and any change requires both to agree.
    \item checkpoint $\cooperate$ buffer pool: redo correctness
    rests on two halves of one invariant---the buffer pool flushes
    a dirty page only after its WAL record is durable, and the
    checkpoint records exactly the WAL prefix consistent with what
    is on disk---neither module enforces both halves alone.
    \item optimizer $\cooperate$ indexing: the access-path
    contract---which predicates each index answers, what bounds and
    statistics it exposes---is split between the optimizer's
    enumeration and the indexing module's implementation; the
    optimizer cannot enumerate paths the indexing module does not
    implement, and the indexing module cannot expose capabilities
    the optimizer does not consume.
    \item MVCC inplace $\cooperate$ isolation level: the conflict
    policy (abort vs.\ wait, visibility cut-off, in-place version
    overwrite rule) lives in neither module on its own.
\end{itemize}

\paragraph{Edges from production-database evidence.}
Like the tree, the four edge types are grounded in the 9 surveyed
codebases rather than designed top-down. For each edge, at least
one production system documents or implements the corresponding
constraint, which we identify by reading source code and
architecture documentation with LLM assistance.

These cross-module edges admit more rigorous, programmatic
recovery, which we leave as future work. In principle, requires
edges can be recovered from cross-module function calls and
imports in the source tree; mutex-with edges from compile-time or
build-time exclusivity (e.g., a build configuration that admits
exactly one storage model); reference edges from
interface-coupling points where one module's API is consumed by
another only when both are selected (e.g., cost-model dependence
on statistics histograms, catalog-to-wire type identity); and
cooperate edges from architectural invariants documented in
production-DB READMEs that the corresponding source files
transitively enforce, with READMEs serving as the entry points
and source as the verification target. Compared with the LLM-assisted reading we currently use, such a
pipeline would yield edges more rigorously and verifiably grounded
in the source itself, and would let \dbgraph track upstream changes
in the surveyed codebases automatically.

The module tree augmented with these cross-module edges is the
dependency graph \dbgraph. Each leaf carries a Hoare-style
specification following SpecFS~\cite{specfs}. The downstream
pipeline (\refsec{design}) consumes \dbgraph for workload-driven
node selection, cooperate-cycle decomposition, and code generation.

\section{System Design}
\label{sec:design}

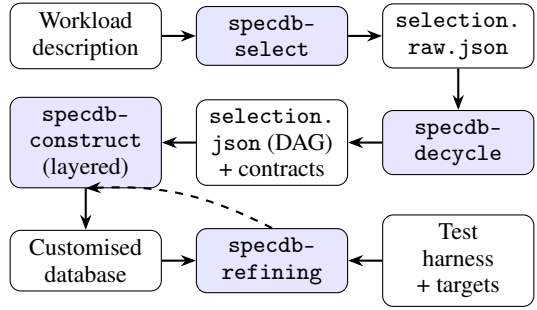
\begin{figure}[t]
\centering
\begin{tikzpicture}[
    node distance=0.35cm and 0.45cm,
    box/.style={draw, rounded corners, minimum height=0.65cm, minimum width=2.0cm, font=\small, align=center},
    skill/.style={draw, rounded corners, minimum height=0.65cm, minimum width=2.0cm, font=\small, fill=blue!10, align=center},
    arr/.style={-{Stealth[length=2mm]}, thick},
    every node/.style={font=\small}
]
\node[box] (user) {Workload\\description};
\node[skill, right=of user] (sel) {\texttt{specdb-}\\\texttt{select}};
\node[box, right=of sel] (raw) {\texttt{selection.}\\\texttt{raw.json}};
\node[skill, below=0.55cm of raw] (dec) {\texttt{specdb-}\\\texttt{decycle}};
\node[box, left=of dec] (sel2) {\texttt{selection.}\\\texttt{json} (DAG)\\+ contracts};
\node[skill, left=of sel2] (sdb) {\texttt{specdb-}\\\texttt{construct}\\(layered)};
\node[box, below=0.55cm of sdb] (db) {Customised\\database};
\node[skill, right=of db] (bug) {\texttt{specdb-}\\\texttt{refining}};
\node[box, right=of bug] (harness) {Test\\harness\\+ targets};

\draw[arr] (user) -- (sel);
\draw[arr] (sel) -- (raw);
\draw[arr] (raw) -- (dec);
\draw[arr] (dec) -- (sel2);
\draw[arr] (sel2) -- (sdb);
\draw[arr] (sdb) -- (db);
\draw[arr] (db) -- (bug);
\draw[arr] (harness) -- (bug);
\draw[arr, dashed] (bug.north) to[bend right=15] (sdb.south);
\end{tikzpicture}
\caption{End-to-end \dbname workflow. Skills and agents (shaded)
operate over \dbgraph: \texttt{specdb-select} maps a workload to a
node selection, \texttt{specdb-decycle} merges cooperate cycles into
contract nodes, \texttt{specdb-construct} drives the layered module construction, and the
\refiningagent closes the loop against a user harness. The dashed
arrow indicates that the \refiningagent may re-invoke
\texttt{specdb-construct} when a missing module is the root cause.}
\label{fig:overview}
\end{figure}

\dbname operationalizes \dbgraph (\refsec{dbtree}) as a four-stage
pipeline that maps a natural-language workload description to a
deployed, performance-tuned database. \reffig{overview} shows the
stages: workload-driven node selection (\refsec{selection}),
cooperate-cycle decomposition (\refsec{decycle}), layered module
construction (\refsec{pipeline}), and a refining loop against a
user-supplied refining harness (\refsec{refining}). Each stage is
implemented as a skill: \texttt{specdb-select},
\texttt{specdb-decycle}, \texttt{specdb-construct}, and
\refiningagent respectively.
Then, a coding-agent is invoked to implement the customized database with these skills.

\subsection{Workload-Driven Selection}
\label{sec:selection}

A user typically arrives with workload requirements rather than
database expertise: they know what data they have, what queries
they intend to run, and what latency or throughput they need, but
not which storage model, concurrency-control strategy, or recovery
protocol matches those needs. Mainstream databases expose a
configuration surface large enough that even users with some
database background struggle to map requirements onto settings,
and \dbgraph (\refsec{dbtree}), with several hundred leaves and
four edge types, is correspondingly hard to navigate by hand. We
therefore design the \texttt{specdb-select} skill to translate
user requirements directly into the corresponding modules and
implementation variants in \dbgraph, without exposing the FODA
structure to the user.

The skill takes four classes of input from the user: dataset
characteristics (e.g., table count, row counts, column types, key
distributions); query patterns (point lookup vs.\ range, join
frequency, read/write ratio); performance goals (p50/p99 latency,
throughput, memory budget); and workload type, which takes one of
three forms: a free-text description of the intended use case, the
path to a representative dataset (SQL dump, CSV/Parquet directory,
or an existing database), or the name of a standard benchmark
(e.g., TPC-C, TPC-H, YCSB).

A coding agent then invokes \texttt{specdb-select} with these
inputs and walks \dbgraph one variability point at a time, picking
the appropriate variant for each XOR or Or group (e.g., choosing the storage model and value types supported by the database). The picks cannot be made in isolation, because
the cross-module edges of \refsec{dbgraph-edges} constrain
neighboring choices.
For instance, when a node is selected, every node it
requires must also be selected, and every node it conflicts with via a
mutex-with edge must be excluded. The
skill therefore closes the selection under these edges before
returning.

The output of \texttt{specdb-select} is a JSON file containing the
picked nodes and the requires, reference, cooperate, and mutex-with
edges among them. This file is the input to the cycle-resolution
stage of \refsec{decycle}.

\subsection{Resolving Cooperate Cycles}
\label{sec:decycle}

Once \texttt{specdb-select} returns the chosen nodes and the edges
among them, the next stage prepares the graph for code generation.
Code generation requires a topological order over the selected
nodes: when $A$ requires $B$, the pipeline must finish $B$ first so
that $A$'s implementation can build on $B$'s interface. Requires,
reference, and mutex-with edges are all directional and, by
construction, do not form cycles, so they admit such an order via
topological sort. Cooperate edges, however, are
symmetric (\refsec{dbgraph-edges}). We call a maximal connected
subgraph of nodes linked by cooperate edges a
\emph{cooperate component}; any cooperate component forms a cycle
that the scheduler cannot break, because implementing any member
requires referencing the others' implementation. Breaking the cycle by picking a side would discard the
co-design constraint that cooperate encodes. We design the
\texttt{specdb-decycle} skill to resolve this: each cooperate
component is merged into a single contract node that the generation
pipeline treats as one unit.

\begin{figure}[t]
\centering
\begin{tikzpicture}[
    node distance=0.35cm and 0.65cm,
    leaf/.style={draw, rounded corners, fill=green!12, font=\scriptsize, align=center, minimum width=2.4cm, minimum height=0.6cm},
    contract/.style={draw, rounded corners, thick, fill=orange!18, font=\small, align=center, minimum width=3.0cm, minimum height=0.9cm},
    coop/.style={<->, thick, dashed},
    req/.style={-{Stealth[length=2mm]}, thick},
    every node/.style={font=\scriptsize}
]
\node[leaf] (mvcc) {\texttt{mvcc/}\\\texttt{inplace\_xmin\_xmax}};
\node[leaf, below=0.7cm of mvcc] (rc) {\texttt{isolation/}\\\texttt{repeatable\_read}};

\draw[coop] (mvcc) -- (rc) node[midway, right, font=\tiny, inner sep=1pt] {coop};

\node[font=\small, below=0.4cm of rc] (lab1) {(a) raw cooperate component};

\node[leaf, right=2.6cm of mvcc] (vac) {\texttt{gc/}\\\texttt{heap\_lazy\_vacuum}};
\node[contract, below=0.55cm of vac] (c1) {\texttt{contract/}\\\texttt{inplace\_rr\_coord}};
\node[leaf, below=0.55cm of c1] (buf) {\texttt{page\_io/}\\\texttt{buffer/shared\_pool}};

\draw[req] (vac) -- node[midway, right, font=\tiny, inner sep=1pt] {req} (c1);
\draw[req] (c1) -- node[midway, right, font=\tiny, inner sep=1pt] {req} (buf);

\node[font=\small, below=0.4cm of buf] (lab2) {(b) merged contract node};
\end{tikzpicture}
\caption{\texttt{specdb-decycle}: a 2-leaf cooperate component
binding the inplace MVCC variant (\texttt{xmin/xmax} tuple header)
and the Repeatable Read isolation level (left) collapses into the
single contract node \texttt{contract/inplace\_rr\_coord} (right).
The contract carries the visibility predicate (a row is visible
iff its xmin committed before the transaction's snapshot and its
xmax did not) that neither member specifies alone. Requires edges
crossing the component (e.g., \texttt{heap\_lazy\_vacuum}
requiring inplace MVCC, and inplace MVCC requiring the buffer pool
to read/write tuple bytes) survive the rewrite with their
endpoints redirected to the contract node, while the cooperate
edge disappears from the output graph.}
\label{fig:decycle}
\end{figure}
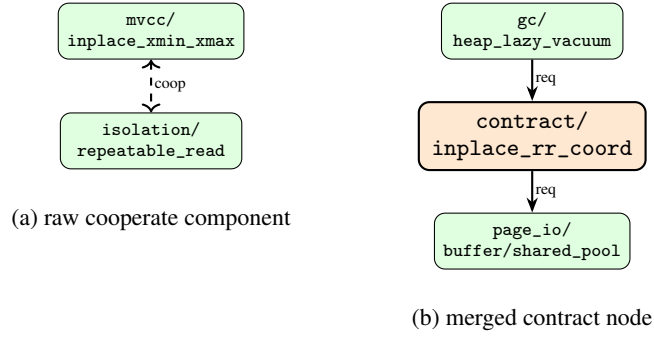

\reffig{decycle} illustrates how \texttt{specdb-decycle}
coordinates the cooperate between two selected techniques: the MVCC
strategy \texttt{concurrency/strategy/mvcc/inplace\_xmin\_xmax} and
the isolation level \texttt{concurrency/isolation/repeatable\_read},
joined into the contract \texttt{inplace\_rr\_coord}. Both are
implementation variants picked from XOR groups in \dbgraph:
\texttt{inplace\_xmin\_xmax} is one of several MVCC strategies
(as shown in \reffig{foda-tree}), and \texttt{repeatable\_read}
is one of several isolation levels.
The former decides the on-disk tuple header (the xmin and xmax
fields and their hint bits), and the latter decides what each
transaction is allowed to see (a single snapshot fixed at
transaction begin and reused by every statement in the transaction).
Neither module specifies the visibility predicate alone---MVCC alone
produces tuples but cannot say which versions a query sees, and
\texttt{repeatable\_read} alone names a guarantee but writes no
bytes; the joint contract ``a row is visible iff its xmin committed
before the transaction's snapshot and its xmax did not'' lives in
neither module.

We run Union-Find over all cooperate edges in the raw selection
(plus any $k$-way groups the user declared as a single entry rather
than $\binom{k}{2}$ pairs), identify these two leaves as one
connected component, and merge them into a single contract node
\texttt{contract/inplace\_rr\_coord} (right). The contract inherits
its members' spec paths, and its own spec is rendered from
a template that fixes the joint implementation order: define the
shared protocol first, read every member spec, implement each
member's primitives, write the coordination glue, and finally run
the cross-module test points. We then rewrite remaining edges: any
requires, reference, or mutex-with endpoint that lay inside the
component is redirected to the new contract node (in the figure,
the \texttt{heap\_lazy\_vacuum} module that requires
\texttt{inplace\_xmin\_xmax}, and the buffer pool that
\texttt{inplace\_xmin\_xmax} reads/writes pages through, both have
their endpoints rewired to \texttt{inplace\_rr\_coord});
self-loops and duplicate pairs are dropped, all cooperate edges
are removed, and the original member nodes are removed. The output
is a DAG with only requires, reference, and mutex-with edges, ready
for the topological scheduler in \refsec{pipeline}.

\subsection{Layered Module Construction}
\label{sec:pipeline}

With the input now a DAG, code generation can in principle proceed
in topological order: each node is generated only after all its
requires/reference predecessors are done. To parallelise independent
work---unrelated modules whose implementations do not depend on
each other---we further group the topologically sorted nodes into
\emph{layers}. A layer is a maximal set of nodes whose every
predecessor in the DAG lies in an earlier layer, so nodes within
the same layer are pairwise independent and can be generated
concurrently. Running this layering on the TPC-C selection yields
13 generation layers. The selection picks 69 leaves and 21 cooperate
edges; \texttt{specdb-decycle}'s Union-Find merges the cooperate
edges into 6 cooperate components, each collapsed into a contract
node, so the resulting DAG has 49 nodes (43 leaves + 6 contracts).
The layered schedule is:

\begin{small}
\begin{verbatim}
Layer 1:  core/{wal, checkpoint}
Layer 2:  value_types/{binary, boolean/native_bool,
                       numeric/{exact_decimal/bounded_fixed_point,
                                overflow_behavior/error_on_overflow,
                                signedness/signed_only},
                       string/{collation_scope/server_wide,
                               length_semantics/char_length},
                       temporal/{timestamp_encoding/epoch_micros,
                                 timezone_model/naive_local_only}}
Layer 3:  storage/{storage_model/row/heap, engine_plug/fixed,
                   page_io/{buffer/shared_pool, io/smgr}}
Layer 4:  indexing/{equality/btree,
                    modifiers/{multicolumn, covering}}
Layer 5:  concurrency/{strategy/mvcc/inplace_xmin_xmax,
                       gc/heap_lazy_vacuum,
                       locking/row/record,
                       locking/infrastructure/{lwlock, mutex}}
Layer 6:  recovery/{strategy/wal_only, auxiliary/checkpoint}
Layer 7:  concurrency/{transaction/model/single_node_acid,
                       isolation/repeatable_read}
Layer 8:  parser/implementation/flex_bison,
          parser/syntax/{constraints/{check, foreign_keys},
                         ddl/{create_index, create_table},
                         dml/{prepared_statements,
                              select_clauses/{for_share,
                                              for_update},
                              siud},
                         expression/{arithmetic_operators,
                                     comparison_operators},
                         functions/aggregate,
                         query_features/subqueries},
          contract/{varchar_len_coord,
                    collation_cmp_coord,
                    overflow_arith_coord}
Layer 9:  query_processing/optimizer/{
            framework/single_cbo,
            strategies/logical/{column_pruning,
                                predicate_pushdown},
            strategies/physical/{cost_model, search_greedy,
              statistics/{analyze, histograms}}}
Layer 10: query_processing/{execution/model/volcano,
                            aggregation/hash,
                            joins/{index, lookup, nested_loop},
                            mutation/constraint_check,
                            operators/{filter, limit, projection},
                            scans/{index, index_only, index_range,
                                   point_get, sequential},
                            sort/{external_merge, in_memory_quick,
                                  top_n}},
          contract/optimizer_index_coord
Layer 11: contract/oltp_durability_coord (13 members:
          mvcc/inplace_xmin_xmax + isolation/repeatable_read
           + recovery/{wal_only, checkpoint}
           + transaction/single_node_acid
           + page_io/buffer/shared_pool
           + locking/row/record + execution/volcano
           + scans/{index, index_range, sequential}
           + mutation/constraint_check
           + parser/dml/select_clauses/for_update)
Layer 12: architecture/{catalog/heap_relations,
                        model/multi_thread/thread_pool,
                        networking/{transports/tcp,
                                    wire_protocols/pg_fe_be}},
          contract/pg_wire_type_identity_coord
Layer 13: architecture/engine_core/session_object
          + main.rs (terminal integration)
\end{verbatim}
\end{small}

Names like \texttt{isolation/repeatable\_read} and
\texttt{recovery/wal\_only} appear twice in this schedule: once as a
leaf in their own dependency-driven layer, and once as a member
inside a contract's parenthesised list at a later layer. The
repetition reflects how large contracts are actually generated.
When a contract has many members (e.g.,
\texttt{oltp\_durability\_coord} has 13), handling the entire contract in a single subagent yields
limited benefit: the larger the joint scope, the more
specifications, source files, and test artefacts share one
context, and the resulting output quality degrades relative to a
smaller-scoped task. Therefore, large contracts are split into two phases. Some members are implemented at their own dependency-driven
leaf layer (e.g., \texttt{isolation/repeatable\_read} at Layer 7,
\texttt{recovery/wal\_only} at Layer 6), and the contract's own
layer is used to wire the already-built members together behind a
unified API and to implement the joint protocol (e.g.,
lock-then-fetch for \texttt{oltp\_durability\_coord}) that no
member owns alone. Small contracts whose members fit
in a single subagent's budget (e.g., \texttt{varchar\_len\_coord} at Layer 8 with 2 members) are not
split: members and glue are written together in the contract's
own layer.

Given this schedule, the pipeline processes layers in order, and
within each layer every module---whether a leaf or a contract---is
handled by its own subagent, so the subagents in a layer run in
parallel. Each subagent drives its assigned module end-to-end
through three specialized inner agents that take turns: the Main
Agent first reads the sub-skill spec for the node, the specifications of
upstream ancestors, and any contract-node members the node depends
on, and writes Rust source under the target directory; when the node is
a contract, the Main Agent additionally writes the wiring code
that ties its members together behind a unified API and any
joint-protocol code that no single member owns.
The Tester Agent then writes unit and integration tests, runs them
via \texttt{cargo test} and \texttt{cargo bench}, and reports
pass/fail counts, latency histograms, and error messages; for
contract nodes the tests include the cross-module test points. The Architect Agent finally reviews
code quality and wires the module into the engine's main
query-processing flow, so that incoming SQL actually reaches the
new code; without this wiring the module compiles in isolation but
is dead code at run time. If correctness or performance fails, the
Architect Agent patches locally or sends the module back to the
Main Agent for regeneration.

All three inner agents run for every module: dropping the Tester
Agent leaves races and visibility bugs undetected at module
boundaries, and dropping the Architect Agent produces modules that
compile in isolation but are unreachable when the engine accepts a
query.
Please note that to accelerate generation, the Tester and Architect agents may be merged into a single Reviewer subagent, trading off strict verification rigor for reduced latency.
Each module cycles through a test-integrate-retest loop (up to three
rounds), and if it still fails, the Main Agent regenerates from
scratch (up to three regeneration attempts) before the
\refiningagent (\refsec{refining}) takes over.

\subsection{Test-Harness-Driven Refinement}
\label{sec:refining}

The construction pipeline produces an end-to-end database, but the
resulting artefact is rarely ready to deploy, for a variety of
reasons: cross-module integration bugs may hide on interaction
surfaces that unit tests fail to cover, the measured performance
may miss the user's target, \texttt{specdb-select} may have omitted
a capability the workload turns out to need (e.g., group commit,
plan cache, HOT updates), and so on.

To close this loop, we design the \refiningagent. The skill takes
three inputs: the problem to fix, the optimization target, and a
test script. The problem can be either a correctness bug or a
performance shortfall---we treat the latter as a first-class
problem on equal footing with crashes and wrong answers. The
optimization target makes the success criterion measurable: a
specific bug fixed, a latency below a threshold, zero error lines
over a 60-minute run, and so on; each repair attempt is checked
against this criterion. The test script is what the skill runs
after every fix to decide whether the criterion now holds. The
problem and the optimization target can either be carried over
from the context of earlier stages (for instance, a regression
flagged by the Tester Agent) or be supplied directly by the user;
the test script can likewise be synthesized from the workload
description and earlier test artefacts, or be supplied by the
user.

For each diagnosed problem, the skill commits to one of two repair
routes, decided by cross-checking the sub-skill catalog before any
code change:

\begin{itemize}[nosep]
    \item Route A (existing-code issue). The relevant module
    already exists in our codebase, and the issue---a correctness
    bug or a performance shortfall---lives inside it. The skill
    first attempts a direct edit. If a direct edit does not yield
    a reproducible fix (which is common for concurrency races,
    MVCC visibility glitches, or any bug whose trigger is
    intermittent on the production code), the skill constructs a
    minimal standalone Rust test that exercises only the
    misbehaving code path, locates the root cause and a candidate
    fix on this minimal reproducer, and then ports the fix back to
    the production code and data. Throughout, the skill consults
    the source of the 9 surveyed relational databases, examines how they avoid or solve the same issue,
    and adapts the fix accordingly. This is a second form of
    borrowing, complementary to the design borrowing of
    \refsec{intro}---there we borrow module \emph{designs} at
    generation time, here we consult their source code directly
    to guide bug fixes and performance tuning at repair time.
    \item Route B (missing-module issue). The diagnosis is that a
    module the workload turns out to need was not selected by
    \texttt{specdb-select} and was therefore never generated. The skill locates the corresponding
    module in \dbgraph, computes the closure of its requires and
    reference dependencies, and routes the resulting subgraph
    through the layered module construction stage
    (\refsec{pipeline}) so that the missing module and its
    dependencies are generated and integrated through the same
    quality gate as the original modules. 
\end{itemize}

Each cycle reproduces the problem, identifies the subsystem,
selects a route, applies the route's repair procedure, rebuilds, and
re-runs the test script. Once the criterion holds, the skill
performs a performance audit: every measured path that still
misses the stated target becomes a new task on the queue and is
routed in the same way. The loop has no upper iteration bound, and
new error classes can be pushed onto the queue and processed in
the same session. The only graceful exit other than success is a
stuck rule: if the same error class appears at the same magnitude
for three consecutive iterations after non-trivial code edits, the
skill surfaces the diagnosis to the user, since this typically
signals that a deeper architectural or specification change is
required (only edits that touch the diagnosed subsystem's code
path reset the counter).

\section{Evaluation}
\label{sec:experiment}

We evaluate \dbname end-to-end with TPC-C, with PostgreSQL~14.23
and MySQL~8.0.45 as baselines. We address three questions: (1) Can \dbname be
generated for the target workload? (2) Does the generated database
match the baseline systems on that workload? (3) How does the
code-size cost compare?

\subsection{Experimental Setup}
\label{sec:setup}

\paragraph{Hardware and host.}
All experiments run on a dual-socket Intel Xeon Silver 4314
server (2$\times$16 cores, 64 logical threads, 503\,GiB DDR4
DRAM) running
Ubuntu~22.04.5 LTS. The database server,
BenchmarkSQL~\cite{benchmarksql} client, and metrics collection
share the host. No virtualisation or container is used; the
databases are run sequentially on bare metal.

\paragraph{Baselines.}
We compare against two production relational databases. PostgreSQL
is a widely used open-source object-relational DBMS that serves as
a single-node OLTP reference across academia and industry; we use
the 14.23 release. MySQL is an equally widely used open-source
RDBMS, deployed at scale in web and application backends, with
InnoDB as its default storage engine; we use the official 8.0.45
release. Both baselines run with their out-of-the-box default
configuration; the only adjustment is
\texttt{max\_connections=200} on PostgreSQL to accommodate the
client thread count.

\paragraph{Workload.}
TPC-C~\cite{tpcc} is the canonical OLTP benchmark: it simulates a
wholesale supplier processing customer orders across $N$ identical
warehouses, with five transaction types
(\texttt{NEW\_ORDER}, \texttt{PAYMENT}, \texttt{ORDER\_STATUS},
\texttt{DELIVERY}, \texttt{STOCK\_LEVEL}) issued in fixed
proportions; the primary throughput metric, \emph{tpmC}, counts
\texttt{NEW\_ORDER} transactions per minute. Each warehouse holds
roughly 100\,MB of data. We drive the workload with
BenchmarkSQL~\cite{benchmarksql}, which connects to each system via
JDBC and pins all client-side parameters identically across runs
(\reftable{client-config}). We evaluate two scale factors,
1~warehouse (wh=1) and 10~warehouses (wh=10). Each run is 60~minutes with no rampup.

\begin{table}[t]
\centering
\caption{BenchmarkSQL client/workload parameters (identical across all DB runs).}
\label{tab:client-config}
\small
\setlength{\tabcolsep}{6pt}
\begin{tabular}{@{}lr|lr@{}}
\toprule
\textbf{Parameter} & \textbf{Value} & \textbf{Parameter} & \textbf{Value} \\
\midrule
\texttt{runMins}                & 60   & \texttt{paymentWeight}     & 43.2 \\
\texttt{rampupMins}             & 0    & \texttt{orderStatusWeight} & 4.2  \\
\texttt{sutThreads}             & 8    & \texttt{deliveryWeight}    & 4.2  \\
\texttt{monkeys}                & 2    & \texttt{stockLevelWeight}  & 4.2  \\
\texttt{loadWorkers}            & 4    & \texttt{rollbackPercent}   & 1.01 \\
\texttt{terminalMultiplier}     & 1    & warehouses                 & 1 or 10 \\
\texttt{keyingTimeMultiplier}   & 1.0  &                            &      \\
\texttt{thinkTimeMultiplier}    & 1.0  &                            &      \\
\bottomrule
\end{tabular}
\end{table}

\subsection{\dbname Generation}
\label{sec:gen}

The TPC-C-ready instance of \dbname is produced by chaining the
construction skills introduced in \refsec{design}, invoked from a
single Claude Code session driven by Claude Opus~4.7. 

\paragraph{Prompts.}
Four user prompts drive the pipeline end-to-end:

\begin{itemize}[nosep]
    \item \texttt{/specdb-select}: ``I want to generate a database
    where TPC-C queries execute without errors, minimizing
    transaction latency and maximizing throughput. Please use
    \texttt{/specdb-select} to choose the modules to implement.''
    \item \texttt{/specdb-decycle}: ``decycle with
    \texttt{/specdb-decycle}.''
    \item \texttt{/specdb-construct}: ``Begin generating the database.''
    \item \texttt{/specdb-refining}: ``Use
    \texttt{/specdb-refining} to fix the bugs, eliminate all
    errors, and speed up the database; you can borrow ideas from
    the existing databases per \texttt{/specdb-refining}.''
\end{itemize}

\begin{table*}[t]
    \centering
    \caption{Refining-phase borrowings: source design, originating
    database, purpose, observed effectiveness, and whether adopted
    in the shipped artefact.}
    \label{tab:borrowings}
    \footnotesize
    \setlength{\tabcolsep}{3pt}
    \renewcommand{\arraystretch}{1.15}
    \begin{tabular}{@{}p{0.25\linewidth} p{0.09\linewidth} p{0.27\linewidth} p{0.22\linewidth} p{0.09\linewidth}@{}}
    \toprule
    \textbf{Design} & \textbf{Source} & \textbf{Purpose} & \textbf{Effective?} & \textbf{Adopted?} \\
    \midrule
    \texttt{heap\_hot\_search\_buffer} (chain walk under one held buffer latch) & PG & Fix sporadic ``not found'' on hot-updated rows & Yes & Yes \\
    CLOG \texttt{committed\_floor()} for visibility shortcut & PG & Fix latent MVCC visibility bug under 1\% TPC-C rollback & Yes & Yes \\
    \texttt{take\_snapshot\_advancing} with TLS fast-path & PG & Advance snapshot xmin per statement so vacuum reclaim is not naptime-bound & Yes (as part of the shipped stack; flat in isolation) & Yes \\
    Horizon-triggered vacuum (\texttt{horizon\_trigger\_xids}) & PG & Trigger eager vacuum by XID count, not just naptime & Yes & Yes \\
    Btree re-descent on probe gap & PG & Cut STOCK\_LEVEL mean latency 51.4$\to$31.9\,ms ($-38\%$) & Yes & Yes \\
    \texttt{FIND\_LAST\_VERSION} / EvalPlanQual & PG & Make \texttt{SELECT \ldots FOR UPDATE} read latest committed; remove 23\% NEW\_ORDER FU 0-row class & Yes & Yes \\
    \texttt{ComputeXidHorizons} oldest-active-xid floor & PG & Floor vacuum horizon by oldest in-flight transaction's xmin & Yes & Yes \\
    \texttt{buf\_page\_make\_young} (sticky-young) for btree-leaf pin sites & InnoDB & Btree-leaf retention in buffer pool & No (STOCK\_LEVEL mean latency $+18\%$: 33$\to$39\,ms) & No \\
    Rosenkrantz wait-die in record lock manager & InnoDB / CRDB / TiDB & Record-lock conflict policy & No ($\sim$50 \texttt{DeadlockAborted} per 30\,min on hot rows) & No \\
    Bgwriter + freelist offload of CLOCK sweep & PG & Speed eviction path under load & No (5 STOCK errors $+$ 0 perf win) & No \\
    Cross-page \texttt{LP\_REDIRECT} splice for index-resolve after prune & PG & Keep index-pointed root resolvable after opportunistic prune & No (11--14$\times$ regression when enabled) & No \\
    \bottomrule
    \end{tabular}
\end{table*}

\paragraph{Construction \& Refining.}
\texttt{specdb-select} resolves the workload to a selection over
\dbgraph; \texttt{specdb-decycle} folds 21 cooperate edges into 6
contract nodes (\refsec{pipeline}); and \texttt{specdb-construct}
generates the 13-entry schedule of \refsec{pipeline},
producing an initial Rust database with unit and integration
tests. Construction completes with all tests passing but with
realistic-workload failures still present.
The \refiningagent (\refsec{refining}) then iterates on the
artefact under the user's instruction to ``fix the bugs,
eliminate all the errors, and speedup,'' using the TPC-C
benchmark to validate the effect of each fix. Refining takes the
bulk of the session and proceeds in two phases: a correctness
phase that closes realistic-workload bugs invisible to unit
tests, and a performance phase that reduces per-transaction
latency. 

\paragraph{Borrowings.}
Throughout construction and refining, code is generated by
referencing the surveyed databases. In the
\emph{construction} phase, \texttt{/specdb-select} identifies
TPC-C as matching the ``Server OLTP (PostgreSQL-like)''
archetype defined in the skill---single-node, read-heavy,
point+range+secondary scans, strict ACID---and selects the
PG-style variants prescribed by that archetype rather than those
of alternative archetypes (e.g., ``MySQL/InnoDB-style OLTP'',
which targets server deployments dominated by write-heavy point
updates on hot rows). Under the chosen archetype, the selected variants---storage,
MVCC, vacuum, recovery, isolation, parser, and wire
protocol---all trace to PostgreSQL. In the
\emph{refining} phase, the agent consults the same codebases for
targeted bug fixes and performance tuning, with each borrowing
subject to a Yes/No effectiveness check on the TPC-C harness
before adoption: some hit (e.g., PG's
\texttt{heap\_hot\_search\_buffer} fixes a sporadic ``not
found'' on hot-updated rows; \texttt{take\_snapshot\_advancing}
with a TLS fast-path speeds vacuum reclamation), others are
tested and reverted (e.g.,
InnoDB's \texttt{buf\_page\_make\_young} pushes STOCK\_LEVEL
mean latency up by $\sim$18\% (33$\to$39\,ms); Rosenkrantz wait-die from
InnoDB/CRDB/TiDB triggers $\sim$50 \texttt{DeadlockAborted} per
30-minute run on hot rows). 
\reftable{borrowings} lists the refining-phase borrowings.

\paragraph{Refining progression.}
To illustrate how the refining loop improves performance over
time, we select 12 representative commits from the refining
history and benchmark each under the same configuration as
\reftable{client-config} with \texttt{runMins=20}.
\reffig{refining-curve} plots per-transaction mean latency (top
panel) and tpmC throughput (bottom panel) across these commits.
The horizontal axis lists commits in chronological order, each
labelled by the optimization it introduces (e.g., ``Bulk index
entry kill'' removes redundant index entries during vacuum so
that subsequent index scans touch fewer dead tuples;
``Deferred-reap grace period'' delays page compaction by one
vacuum cycle to avoid racing with concurrent readers). As
refining proceeds, overall latency decreases and throughput
increases monotonically in trend, confirming that the coding
agent is able to identify performance defects in the generated
code and progressively optimize them.

\subsection{TPC-C Throughput and Latency}
\label{sec:tpcc-results}

With the artefact of \refsec{gen} in place under the setup of
\refsec{setup}, we now measure its end-to-end TPC-C performance
against the PostgreSQL and MySQL baselines and address the
questions from the start of \refsec{experiment}: does the
generated \dbname function correctly, and does it match the
baselines on throughput and latency?

\paragraph{Correctness.}
All tested systems complete the 60-minute TPC-C workload at both
wh=1 and wh=10 with zero errors across all five transaction
types, indicating that \dbname functions correctly under concurrent
load.

\paragraph{Throughput and latency.}
\reftable{tpcc-tpmc} reports throughput,
\reftable{tpcc-latency-wh1} and \reftable{tpcc-latency} report
per-transaction latency at wh=1 and wh=10 respectively, for
\dbname, PostgreSQL, and MySQL.

\begin{table}[t]
\centering
\caption{TPC-C 60-minute throughput. tpmC = NEW\_ORDER per minute
(TPC-C primary metric). totalTPM = all transactions per minute. NO\#
= NEW\_ORDER count over the 60-minute window.}
\label{tab:tpcc-tpmc}
\small
\begin{tabular}{@{}lrrr|rrr@{}}
\toprule
 & \multicolumn{3}{c|}{\textbf{wh=1}} & \multicolumn{3}{c}{\textbf{wh=10}} \\
\textbf{System} & \textbf{tpmC} & \textbf{totTPM} & \textbf{NO\#}
                & \textbf{tpmC} & \textbf{totTPM} & \textbf{NO\#} \\
\midrule
\dbname            &  13 & 29 &  756 & 130 & 288 & 7{,}771 \\
PostgreSQL~14.23   &  13 & 29 &  766 & 128 & 288 & 7{,}697 \\
MySQL~8.0.45       &  13 & 29 &  792 & 127 & 287 & 7{,}646 \\
\bottomrule
\end{tabular}
\end{table}

At wh=1, the three systems produce identical throughput (13\,tpmC,
29\,totTPM); throughput is bounded by keying and think time rather
than engine efficiency. At wh=10, \dbname (130), PostgreSQL (128),
and MySQL (127) are all within 2.4\% of each other. 

The latency picture (\reftable{tpcc-latency-wh1} and
\reftable{tpcc-latency}) reveals per-transaction differences that
tpmC aggregation hides. At both scale factors, \dbname and
PostgreSQL are within $\sim$1\,ms on NEW\_ORDER and PAYMENT.
MySQL's default \texttt{REPEATABLE READ} isolation with next-key
locks inflates NEW\_ORDER mean latency to $\sim$2$\times$ that of
\dbname/PostgreSQL (29\,ms vs.\ 12--13\,ms at wh=10; 33\,ms vs.\
14--15\,ms at wh=1) and STOCK\_LEVEL to $\sim$3--6$\times$
(34\,ms vs.\ 5.5--11\,ms at wh=10; 32\,ms vs.\ 6--8\,ms at
wh=1), because range locks on hot rows block concurrent writers
longer than row-level locks.

\begin{table*}[t]
\centering
\caption{TPC-C wh=1 per-transaction latency (ms), 60-minute run,
8 SUT threads. Bold values mark per-row best.}
\label{tab:tpcc-latency-wh1}
\small
\begin{tabular}{@{}ll|rrrrrr@{}}
\toprule
\textbf{Transaction} & \textbf{System} & \textbf{count} & \textbf{mean} & \textbf{p50} & \textbf{p95} & \textbf{p99} & \textbf{max} \\
\midrule
\multirow{3}{*}{NEW\_ORDER}
& \dbname            &  756 & \textbf{13.7} & \textbf{13.2} & \textbf{18.4} & \textbf{26.9} & 79.0 \\
& PostgreSQL         &  766 & 14.8 & 14.1 & 21.1 & 32.1 & 84.0 \\
& MySQL              &  792 & 33.2 & 32.2 & 49.3 & 66.2 & 90.0 \\
\midrule
\multirow{3}{*}{PAYMENT}
& \dbname            &  770 & \textbf{5.6} & \textbf{5.1} & \textbf{7.1} & \textbf{11.1} & \textbf{17.0} \\
& PostgreSQL         &  765 & 6.6 & 6.1 & 10.1 & 14.8 & 33.0 \\
& MySQL              &  684 & 11.5 & 11.1 & 17.2 & 25.4 & 81.0 \\
\midrule
\multirow{3}{*}{ORDER\_STATUS}
& \dbname            &   75 & 7.1 & 7.1 & 10.1 & 19.0 & 38.0 \\
& PostgreSQL         &   84 & 7.3 & 7.1 & 11.0 & 12.5 & \textbf{14.0} \\
& MySQL              &   86 & \textbf{6.0} & \textbf{4.0} & \textbf{6.1} & 74.9 & 76.0 \\
\midrule
\multirow{3}{*}{DELIVERY}
& \dbname            &   63 & 0.4 & 1.0 & 1.0 & 1.0 & \textbf{1.0} \\
& PostgreSQL         &   80 & 0.4 & 1.0 & 1.0 & 1.2 & 2.0 \\
& MySQL              &   74 & \textbf{0.3} & 1.0 & 1.0 & 1.0 & 1.0 \\
\midrule
\multirow{3}{*}{STOCK\_LEVEL}
& \dbname            &   79 & 7.9 & 8.1 & \textbf{9.1} & \textbf{10.3} & 11.0 \\
& PostgreSQL         &   58 & \textbf{6.3} & \textbf{6.1} & 10.1 & 11.1 & \textbf{11.0} \\
& MySQL              &   82 & 32.3 & 32.2 & 35.4 & 36.5 & 37.0 \\
\bottomrule
\end{tabular}
\end{table*}

\begin{table*}[t]
\centering
\caption{TPC-C wh=10 per-transaction latency (ms), 60-minute run,
8 SUT threads. Bold values mark per-row best.}
\label{tab:tpcc-latency}
\small
\begin{tabular}{@{}ll|rrrrrr@{}}
\toprule
\textbf{Transaction} & \textbf{System} & \textbf{count} & \textbf{mean} & \textbf{p50} & \textbf{p95} & \textbf{p99} & \textbf{max} \\
\midrule
\multirow{3}{*}{NEW\_ORDER}
& \dbname            & 7{,}771 & \textbf{12.3} & \textbf{11.1} & 19.2 & 66.1 & \textbf{93.0} \\
& PostgreSQL         & 7{,}697 & 13.1 & 13.2 & \textbf{18.2} & \textbf{19.2} & 213.0 \\
& MySQL              & 7{,}646 & 28.6 & 28.2 & 42.6 & 50.0 & 94.0 \\
\midrule
\multirow{3}{*}{PAYMENT}
& \dbname            & 7{,}343 & 5.5 & \textbf{5.1} & 9.1 & 11.1 & \textbf{79.0} \\
& PostgreSQL         & 7{,}297 & \textbf{5.3} & \textbf{5.1} & \textbf{7.1} & \textbf{8.1} & 125.0 \\
& MySQL              & 7{,}437 & 9.6 & 9.1 & 12.2 & 24.4 & 206.0 \\
\midrule
\multirow{3}{*}{ORDER\_STATUS}
& \dbname            &  745 & 16.3 & 17.2 & 30.5 & 34.0 & \textbf{40.0} \\
& PostgreSQL         &  727 & 10.8 & 11.1 & 12.2 & 18.7 & 55.0 \\
& MySQL              &  710 & \textbf{4.5} & \textbf{4.0} & \textbf{5.1} & \textbf{6.1} & 178.0 \\
\midrule
\multirow{3}{*}{DELIVERY}
& \dbname            &  708 & 0.3 & 1.0 & 1.0 & 1.0 & 2.0 \\
& PostgreSQL         &  787 & 0.3 & 1.0 & 1.0 & 1.0 & 2.0 \\
& MySQL              &  717 & 0.3 & 1.0 & 1.0 & 1.0 & \textbf{1.0} \\
\midrule
\multirow{3}{*}{STOCK\_LEVEL}
& \dbname            &  712 & 11.2 & 11.1 & 15.3 & 19.1 & 23.0 \\
& PostgreSQL         &  784 & \textbf{5.5} & \textbf{5.1} & \textbf{7.1} & \textbf{9.1} & \textbf{11.0} \\
& MySQL              &  720 & 33.9 & 34.4 & 37.3 & 52.6 & 101.0 \\
\bottomrule
\end{tabular}
\end{table*}

\paragraph{Takeaways.}
PostgreSQL excels on read-only scan latency (STOCK\_LEVEL
5.5\,ms, ORDER\_STATUS 10.8\,ms), while MySQL achieves the
lowest ORDER\_STATUS latency (4.5\,ms). \dbname reaches
comparable or better performance overall: it slightly exceeds
both on tpmC (130 vs.\ PG 128, MySQL 127) and leads on
NEW\_ORDER mean latency (12.3\,ms vs.\ PG 13.1, MySQL 28.6),
demonstrating that an LLM-generated database can match
hand-engineered production systems on TPC-C throughput and
surpass them on individual transaction types.

\subsection{Code Size}
\label{sec:loc}

\reftable{loc} compares server-side LOC across \dbname and 9
production databases. We use \texttt{cloc 2.08}, exclude blank lines,
comments, build files, tests, client libraries, documentation, and
vendored third-party code. C/C++ headers are included so that
template/inline-function bodies are not double-discounted. For
PostgreSQL and MySQL the LOC count is taken from the exact release
benchmarked above (14.23 and 8.0.45); the other 8 are from full git
clones at the recorded commits.

\begin{table*}[t]
\centering
\begin{minipage}[t]{0.52\linewidth}
\centering
\caption{Server-side LOC across \dbname and 9 production databases.}
\label{tab:loc}
\small
\begin{tabular}{@{}lrr@{}}
\toprule
\textbf{System} & \textbf{Lang.} & \textbf{LOC} \\
\midrule
\dbname             & Rust    &     23{,}779 \\
SQLite 3.54.0                          & C       &    144{,}325 \\
DuckDB v1.5.2                          & C++     &    355{,}541 \\
PostgreSQL 14.23                       & C       &    813{,}802 \\
MySQL 8.0.45                           & C/C++   &    838{,}843 \\
TiDB                                   & Go      &  1{,}106{,}256 \\
MariaDB 13.0.1-gamma                   & C/C++   &  1{,}285{,}857 \\
ClickHouse 26.4                        & C++     &  1{,}402{,}620 \\
YugabyteDB                             & C/C++   &  2{,}232{,}704 \\
CockroachDB                            & Go      &  2{,}660{,}760 \\
\bottomrule
\end{tabular}
\end{minipage}%
\hfill
\begin{minipage}[t]{0.44\linewidth}
\centering
\caption{Server-side LOC per tpmC at wh=10.}
\label{tab:loc-per-tpmc}
\small
\begin{tabular}{@{}lrrr@{}}
\toprule
\textbf{System} & \textbf{tpmC} & \textbf{LOC} & \textbf{LOC/tpmC} \\
\midrule
\dbname            & 130 &     23{,}779 &  \textbf{183} \\
PostgreSQL 14.23   & 128 &    813{,}802 &       6{,}358 \\
MySQL 8.0.45       & 127 &    838{,}843 &       6{,}605 \\
\bottomrule
\end{tabular}
\end{minipage}
\end{table*}

\dbname comprises 23{,}779 lines of Rust, roughly 34$\times$
smaller than the two benchmarked baselines (PostgreSQL 813{,}802,
MySQL 838{,}843) and 6$\times$ smaller than even the most compact
production system in the table (SQLite, 144{,}325). Despite this
size difference, \reftable{loc-per-tpmc} shows that \dbname
slightly exceeds PostgreSQL's TPC-C throughput with $\sim$2.9\%
of PostgreSQL's server-side LOC and $\sim$2.8\% of MySQL's. This metric should be read narrowly: it measures TPC-C OLTP
capability per server LOC. Production databases ship many features
(e.g., replication, partitioning, full SQL surface, extensions, security,
distributed transactions) that do not exercise on a single-node
TPC-C run. The metric isolates the cost of those features for a
workload that does not exercise them: well over 95\% of the production
code does not contribute to this run. This supports the case for
generating a per-workload feature set rather than shipping the union.

\subsection{Discussion}
\label{sec:discussion}

\dbname's most significant practical limitation is that each new
target workload triggers a fresh generation run---module selection,
cycle resolution, layered construction, and refinement---which
takes hours of LLM compute and requires a user-supplied refining harness
before the refining loop can converge. This one-time cost is
non-negligible, yet it should be weighed against what the
alternative actually costs. Stonebraker et al.~\cite{endofarchera}
demonstrated that general-purpose relational systems impose the
full overhead of a design tuned for the broadest possible workload
on every application, including the vast majority that use only a
narrow, fixed slice of the available features, and that stripping
the portions a given workload does not need---as
H-Store~\cite{hstore} did for OLTP---recovers more than an order
of magnitude in throughput. Our own
code-size analysis (\refsec{loc}) reaches the same conclusion from
a different angle: well over 95\% of PostgreSQL's and MySQL's server
code does not contribute to a single-node TPC-C run. For the
overwhelming majority of deployed services---which drive a database
server with a fixed set of parameterised transaction
templates---there is no reason to install, operate, or license a
general-purpose engine (let alone an enterprise product such as
Oracle) when the application will never exercise most of what it
ships. Against that baseline, \dbname's generation step is expensive
precisely once; the resulting artefact is a small (23{,}779-line),
self-contained binary whose maintenance and operational complexity
remain proportional to what the workload actually needs, with no
ongoing feature tax. 

Looking forward, as LLM capability grows and inference costs fall,
the one-time generation overhead will shrink to the point where
producing a bespoke database engine becomes straightforward. At
that point the natural model
becomes \emph{generate-once-deploy-forever}: describe the workload,
let the pipeline converge, and ship a purpose-built artefact rather
than configuring a general-purpose system from the outside. The
Bespoke-OLAP work~\cite{bespoke} pursues a similar philosophy for
analytical engines; \dbname extends the idea to full OLTP stacks
generated from a cross-database feature model
(\refsec{related-synthesis}). We believe this mode of database
provisioning will become the default as LLM-driven development
matures.

This is not a claim that prior database systems research has become
redundant. Quite the opposite: during generation we found that
asking an LLM to implement a subsystem from a blank slate reliably
produces code that is correct in isolation but fails under concurrent
load---heap pages without a free-space map fill up silently, MVCC
without a proper vacuum horizon leaks dead versions, B-tree descent
without an intent latch serialises concurrent readers. Grounding
each module in at least one production implementation consistently
closes those gaps. Decades of engineering decisions encoded in
production codebases are not an obstacle to \dbname but its primary
raw material: a high-quality specification library the LLM
recombines rather than reinvents. What the approach changes is the way that knowledge is reused.
The deeper consequence is principled: a coding agent operating
at the level of module specifications has no product-specific
scaffolding---wire protocol handlers, catalogs, planner state,
crash-recovery bookkeeping---to defend, so techniques can in
principle be combined across system boundaries. The TPC-C
instance presented here lands predominantly on PG-derived
variants (\refsec{gen}) and does not by itself exhibit
cross-system recombination; demonstrating that capacity on a
workload mix that draws from multiple archetypes is left to
future work.

\section{Limitations and Future Work}
\label{sec:limitations}

\dbname demonstrates that an LLM-driven pipeline can generate a
TPC-C-competitive relational database from modular specifications.
Several limitations remain.

\paragraph{Refining-harness supervision.}
The \refiningagent (\refsec{refining}) requires
the user to provide a refining harness and target metrics.
Supervision gives the loop its strength: the harness defines what
``good'' means, and convergence is unambiguous. Without such a
harness, the skill cannot guide tuning, and automatically
synthesizing refining harnesses from a workload description or
trace data is an open direction. Even when the harness is in place,
we observed that bug fixing under the loop frequently requires many
iterations: with one of the strongest available coding models in our
session (Claude Opus~4.7), the agent can still lock onto an
incorrect hypothesis, fail to identify the true root cause of the
problem, and not converge within reasonable wall time (the stuck
rule
(\refsec{refining}) fired on STOCK\_LEVEL after three such
no-progress iterations in our TPC-C run). Designing harness-level
mechanisms that detect such reasoning lock-in and steer the agent
toward alternative hypotheses is a complementary future
direction.

\paragraph{Correctness scoped to the target workload.}
Because the generated database is customized for a specific workload,
many features and code paths in mainstream systems are omitted or
simplified. Correctness for queries outside the refining harness is not
guaranteed. For example, \dbname's TPC-C-targeted instance does not
implement TEXT/VARCHAR faithfully, omits some constraint forms, and
supports only the SQL features TPC-C exercises. Extending coverage
incrementally, driven by additional harness queries, is
straightforward; offering general SQL guarantees would require
richer specifications and broader testing.

\paragraph{Synthesis of existing techniques rather than novel design.}
The modules and their variants in \dbgraph are distilled from 9
production databases (\refsec{dbgraph-extract}). Code generation
proceeds by recombining and tailoring existing techniques rather than
inventing new ones, which fits the customisation goal but limits
\dbname to integrating known techniques rather than producing novel
database research. Closing that gap (e.g., generating new operator
algorithms or storage formats) would require the LLM to drive both
specification evolution and implementation.

\section{Conclusion}
\label{sec:conclude}

We presented \dbname, a system that uses LLMs to generate customized
relational databases from a feature graph (\dbgraph) distilled from
9 production systems. \dbgraph extends the FODA
model~\cite{kang1990feature} with a cooperate edge that captures
cross-subtree implementation dependencies. A cycle-resolution stage
collapses cooperate components into contract nodes, a layered
module-construction stage generates Rust code one topological
layer at a time using per-module subagents (each driven by Main,
Tester, and Architect inner agents), and the \refiningagent closes
the loop against a user-supplied refining harness with read-only access
to the surveyed databases' source code.

The evaluation validates the borrowing recipe. On a 60-minute TPC-C
workload at wh=10, the generated database (23{,}779 lines of Rust)
reaches tpmC=130, slightly ahead of PostgreSQL~14.23 (128) and
MySQL~8.0.45 (127), at $\sim$2.9\% of
PostgreSQL's and $\sim$2.8\% of MySQL's server-side LOC. All
three systems achieve comparable throughput. These results
confirm that an LLM-generated database can match hand-engineered
production systems on a target workload while shipping only the
code that workload actually requires.

Two observations from the generation process point toward a broader
implication. First, grounding modules in production references
proved critical in our experience: modules written from a blank
slate were often correct in isolation but underperformed or broke
under concurrent load, while borrowing from production designs
(predominantly PostgreSQL in this TPC-C case) consistently closed
the gap. Decades of database engineering are
not made obsolete by LLM-driven generation---they become its primary
raw material. Second, because the agent operates at the level of
module specifications rather than product source, the pipeline
has no product-specific scaffolding to defend and can in
principle combine techniques across system boundaries. The
TPC-C case presented here exercises a PG-dominated subset of
that capacity; demonstrating recombination across multiple
production systems on a workload mix is left to future work.
Paired with falling LLM inference costs and growing model
capability, this capacity points toward a future in which
generating a purpose-built database for a target workload
becomes straightforward.

\bibliographystyle{unsrt}
\bibliography{references}

\appendix
\section{Per-database Extraction Tables}
\label{sec:appendix:per-db-tables}

This appendix lists the complete per-database extraction tables produced by the \texttt{db-feature-model} skill on each of the 9 production codebases (\refsec{dbgraph-extract}). Each table contains exactly one row per implemented (\checkmark) leaf the skill discovered in the source tree, in the form $\langle$module, sub-axis, method, description$\rangle$. The cross-database aggregation of the storage rows used as a worked example in \refsec{dbgraph-extract} is shown in \reftable{storage-cross-db}.

\subsection{PostgreSQL}
\label{sec:appendix:per-db-tables:postgresql}
{\footnotesize
\renewcommand{\arraystretch}{1.1}
\begin{longtable}{@{}p{0.16\textwidth} p{0.32\textwidth} p{0.46\textwidth}@{}}
\caption{PostgreSQL per-database extraction (rows merged by (module, path); only implemented leaves shown). \textbf{Module-Tree Path} is the chain of bolded ancestors in PostgreSQL's module tree; \textbf{Technical Description} summarises how PostgreSQL realises that path.}\\
\label{tab:appendix:postgresql}\\
\toprule
\textbf{Module} & \textbf{Module-Tree Path} & \textbf{Technical Description} \\
\midrule
\endfirsthead
\multicolumn{3}{c}{\footnotesize\itshape continued from previous page}\\
\toprule
\textbf{Module} & \textbf{Module-Tree Path} & \textbf{Technical Description} \\
\midrule
\endhead
\midrule
\multicolumn{3}{r}{\footnotesize\itshape continued on next page}\\
\endfoot
\bottomrule
\endlastfoot
Storage Engine & Storage Model / Row Store / Row Organization / Heap / Heap Page Management & Each heap relation is paired with a Free Space Map that tracks remaining room per page in a tree structure, letting inserts pick a non-full page in near O(1), and a Visibility Map that uses 2 bits per page to mark all-visible and all-frozen pages so MVCC visibility checks and vacuum work can be skipped. \\
Storage Engine & Storage Model / Row Store / Row Organization / Heap / TOAST & Oversized attributes are sliced into chunks stored in a per-relation TOAST table and compressed inline; the default codec is the historical pglz, while LZ4 is offered from PG 14+ as a faster optional alternative selectable per column. \\
Storage Engine & Table Access Method & PostgreSQL ships exactly one built-in Table AM, the heap, which all base relations use unless a third-party Table AM is registered through the pluggable API. \\
Storage Engine & Storage Manager & Block-level I/O goes through a thin storage manager indirection layer (smgr) that dispatches to one built-in backend, the magnetic-disk backend, which simply maps relation forks onto regular operating-system files. \\
Storage Engine & Async I/O & Asynchronous I/O can be performed by the legacy synchronous read path (method\_sync), by a pool of dedicated I/O worker processes (method\_worker), or, on Linux from PG 18+, by submitting requests through io\_uring (method\_io\_uring) for kernel-side batching. \\
Indexing & Index Access Methods / BRIN (Block Range Index) & Block Range Indexes summarise ranges of heap pages; the supplied operator classes are minmax for monotonic columns, minmax\_multi for multi-modal data, inclusion for containment of geometric or range types, and bloom for low-cardinality equality filtering. \\
Indexing & Index Features & On top of every access method PostgreSQL supports partial indexes filtered by a WHERE predicate, expression indexes keyed on arbitrary functions, covering indexes that attach non-key INCLUDE columns (PG 11+), and composite multicolumn indexes. \\
Parser & DDL Syntax & The grammar accepts the full DDL surface: CREATE/DROP TABLE for relations, CREATE VIEW that the rewriter later expands, CREATE INDEX for any registered AM, CREATE TRIGGER for row and statement triggers, ALTER TABLE for online schema change, CREATE FUNCTION/PROCEDURE for user-defined routines, and CREATE EXTENSION for packaged modules. \\
Parser & DML Syntax & All standard mutations are parsed: SELECT, INSERT, UPDATE, DELETE, and from PG 15+ the SQL-standard MERGE statement that combines insert/update/delete in a single source-driven pass. \\
Parser & Query Syntax Features & The query language covers scalar and table Subqueries, recursive and non-recursive CTEs, Window Functions, Compound SELECT via UNION/INTERSECT/EXCEPT, UPSERT through INSERT ... ON CONFLICT, Generated Columns, native JSON Path operators on the built-in jsonb type, EXPLAIN/EXPLAIN ANALYZE for plan inspection, LATERAL joins for correlated derived tables, and GROUPING SETS/CUBE/ROLLUP for multi-grouping aggregation. \\
Parser & Constraint Syntax & Declared constraints include Foreign Keys with referential actions, CHECK predicates, UNIQUE and PRIMARY KEY enforced via B-Tree indexes, and EXCLUDE constraints that use a GiST index to forbid overlapping values for arbitrary operators. \\
Parser & Built-in Functions & A large standard library of Date/Time, Math, and String functions is built in, alongside the canonical Aggregate Functions SUM, AVG, COUNT, MIN, MAX, and friends. \\
Query Processing & -- & The raw parse tree is run through an Analyzer/Rewriter that does name resolution, type checking, rule application, view expansion, and row-level security to produce the canonical Query Tree; an optional LLVM-based JIT Compilation can compile expressions and tuple deforming at runtime when costs exceed thresholds; an extended Statistics subsystem captures multi-column dependencies and MCV lists gathered by the ANALYZE command. \\
Query Processing & Optimizer & The Standard Planner enumerates access paths, performs preprocessing, and generates plans using cost-based dynamic programming over join orders; when the number of relations exceeds the geqo\_threshold (default 12), GEQO replaces exhaustive search with a semi-random evolutionary search over join orderings. \\
Query Processing & Executor / Join Operators & Three physical join operators are available: Nested Loop (with optional inner index), Hash Join with a build-and-probe in-memory or partitioned hash table, and Merge Join over presorted inputs. \\
Query Processing & Executor / Scan Operators & The executor offers Sequential Scan over heap pages, Index Scan that pipelines tuples through the heap, Index Only Scan that uses the Visibility Map to skip the heap entirely, Bitmap Scan that ORs index results into a TID bitmap before a heap fetch, and TID Scan for direct ctid lookups. \\
Query Processing & Executor / Other Operators & The remaining stream operators include external-merge Sort, hash and grouped Aggregate, a Hash node that builds the table feeding Hash Join, Materialize for caching a subplan, Memoize for parameterised result caching, WindowAgg for window functions, and Limit. \\
Concurrency Control & Isolation Mechanism & MVCC is realized by storing row versions in-place inside the heap with xmin/xmax fields in each tuple header; readers pick visible versions via snapshot comparison, so old versions linger in the table itself (rather than in an undo log as in Oracle/MySQL) until VACUUM reclaims dead tuples. \\
Concurrency Control & Transaction Management & Per-transaction state is tracked across several SLRU stores: the CLOG (Commit Log) records committed/aborted/in-progress status keyed by xid, a Subtransactions directory maps savepoints back to their parents, Two-Phase Commit persists PREPARE TRANSACTION state for distributed coordinators, and an optional Commit Timestamp store records the wall-clock commit time. \\
Concurrency Control & Lock Management & Four lock layers cooperate: Spinlocks for ultra-short critical sections backed by hardware atomics, LWLocks (lightweight read/write locks) protecting shared-memory data structures, the Regular heavyweight lock manager exposing eight lock modes with deadlock detection, and Predicate Locks used exclusively by Serializable Snapshot Isolation. \\
Concurrency Control & Deadlock Handling & Deadlock Detection runs periodically against a wait-for graph built from blocked processes; when a cycle is found one transaction is chosen as the victim and aborted with a deadlock error. \\
Concurrency Control & Isolation Levels & Three isolation levels are supported: Read Committed, the default, takes a fresh snapshot per command; Repeatable Read uses a single transaction-long snapshot for true snapshot isolation; Serializable layers SSI on top with predicate locks and read-write conflict tracking. \\
Recovery and Logging & -- & Durability rests on a WAL (Write-Ahead Log) subsystem that inserts, reads, and replays records during crash and archive recovery; a Checkpointer periodically flushes dirty pages to bound recovery work; a Background Writer pre-cleans dirty buffers ahead of replacement; WAL Archiving hands completed segments off for long-term retention; PITR (Point-In-Time Recovery) replays archived WAL on top of a base backup to any chosen time; from PG 17+ a WAL Summarizer maintains incremental-backup metadata. \\
Replication & Physical Replication & Streaming Replication has the WAL sender on the primary stream freshly written WAL bytes to a WAL receiver on each standby, giving block-level physical replication; Synchronous Replication can optionally hold COMMIT until designated standbys acknowledge, with the mode configurable per transaction or globally. \\
Replication & Logical Replication & Logical Decoding parses WAL into a stream of row-level changes that downstream output plugins consume; built-in Publication/Subscription DDL ships changes selectively; pgoutput is the default plugin used by native subscribers; from PG 17+ a Conflict Resolution layer detects and reports apply-time conflicts on the subscriber. \\
Replication & -- & Replication Slots are persistent registrations on the primary that pin the oldest required WAL position (and, for logical slots, the catalog xmin) so that needed segments are not recycled before downstream consumers have read them. \\
Memory Management & Buffer Pool / Buffer Replacement & The shared buffer pool uses a Clock Sweep replacement policy that decrements a usage counter on each pass over the buffer ring and evicts the first buffer reaching zero, approximating LRU without per-access bookkeeping. \\
Memory Management & Memory Context / Allocator Implementations & Four memory-context allocator strategies plug into the hierarchical palloc API: AllocSet, the default general-purpose allocator using power-of-two size classes; Generation, optimised for cohort allocations freed together; Slab, for fixed-size objects; and Bump, an append-only arena that does not support per-chunk free. \\
Memory Management & Shared Memory & In addition to the static main shared-memory segment, a DSM (Dynamic Shared Memory) subsystem lets backends create and attach to named segments at runtime, which is what parallel query and extensions use to share state. \\
Catalog and Metadata & -- & All metadata lives in regular heap relations such as pg\_class, pg\_attribute, pg\_type, and pg\_proc; because the System Catalogs are themselves tables, any metadata can be inspected with ordinary SQL. \\
Catalog and Metadata & Cache Layer & Four cooperating caches keep catalog access cheap: a per-row Catalog Cache for syscache lookups, a Relation Cache holding parsed table and index descriptors, a Plan Cache memoising prepared-statement plans, and a Cache Invalidation bus that broadcasts changes across backends so stale entries are evicted. \\
Process Architecture & Process Model & PostgreSQL is Process-per-connection: a postmaster supervisor accepts incoming connections and forks a dedicated backend process per client, isolating memory and crashes at OS-process granularity. \\
Process Architecture & Background Processes & A fixed cast of background processes runs alongside backends: Autovacuum launcher and workers reclaim dead tuples, the Checkpointer flushes dirty buffers, the Background Writer pre-cleans them, the WAL Writer flushes WAL, the Stats Collector / Activity subsystem tracks cumulative counters, and the Syslogger captures stderr into rotated log files. \\
Process Architecture & -- & A general Background Workers framework lets extensions register custom long-running processes, and the Parallel Query infrastructure spawns leader-and-worker groups that share state through DSM to parallelise scans, joins, and aggregates. \\
Networking and Protocol & -- & Clients speak the proprietary PostgreSQL Frontend/Backend Protocol (not HTTP) implemented over a binary message-formatting layer; connections may be wrapped in SSL/TLS via OpenSSL; before authentication, the Host-Based Access Control (HBA) engine matches the client against pg\_hba.conf rules to decide which auth method to require. \\
Networking and Protocol & Authentication & Many authentication methods are bundled: Trust (no challenge), Password in md5 or plaintext, SCRAM-SHA-256 (the recommended default since PG 10+), GSSAPI/Kerberos for enterprise SSO, Certificate (SSL) via TLS client certs, LDAP, RADIUS, PAM, and from PG 18+ OAuth bearer tokens. \\
Procedural Languages & -- & SQL Functions are built in and need no installation; PL/Perl, PL/Python (PL/Python3u), and PL/Tcl ship as in-tree modules that embed the Perl, Python, and Tcl interpreters respectively as procedural-language handlers. \\
Procedural Languages & PL/pgSQL & PL/pgSQL has its own pipeline independent of the SQL engine: a dedicated Parser tokenises and parses block-structured source, a Compiler lowers it to an internal execution tree, an Executor interprets statements including loops, conditionals, and exception handling, and a Function Manager dispatches function and trigger invocations into that interpreter. \\
Extensibility & -- & Packaged functionality is delivered through the Extension Framework managed by CREATE/ALTER/DROP EXTENSION, with the pg\_extension catalog tracking installed versions and a control file plus SQL script defining each extension; orthogonally, function-pointer Hooks at strategic points (planner\_hook, ExecutorStart\_hook, ExecutorRun\_hook, ProcessUtility\_hook, emit\_log\_hook) let loaded modules intercept or wrap core behaviour. \\
Extensibility & Custom Access Methods & Two pluggable AM APIs are exposed: the Index AM API (used by contrib modules such as bloom) for new index types, and from PG 12+ the Table AM API for alternative table storage formats. \\
Extensibility & Foreign Data Wrapper & Two reference Foreign Data Wrappers are bundled: postgres\_fdw connects to remote PostgreSQL servers with predicate and join pushdown, and file\_fdw exposes local CSV or text files as foreign tables. \\
Extensibility & Official Extensions -- contrib / & The contrib directory ships dozens of officially maintained extensions: monitoring and introspection (pg\_stat\_statements, pg\_buffercache, pg\_freespacemap, pg\_visibility, pg\_walinspect, pgstattuple, pg\_prewarm, pageinspect, auto\_explain, amcheck, pg\_surgery, vacuumlo, sslinfo); index and operator-class support (bloom, btree\_gin, btree\_gist, pg\_trgm); data types (hstore, ltree, cube, seg, citext, isn, uuid-ossp); text-search dictionaries (dict\_int, dict\_xsyn, unaccent, fuzzystrmatch); security and crypto (pgcrypto, passwordcheck, sepgsql, auth\_delay); cross-database and tabular tools (dblink, tablefunc, earthdistance, xml2, tcn); and the test\_decoding logical-decoding output plugin. \\
\end{longtable}
}

\subsection{MySQL}
\label{sec:appendix:per-db-tables:mysql}
{\footnotesize
\renewcommand{\arraystretch}{1.1}
\begin{longtable}{@{}p{0.16\textwidth} p{0.32\textwidth} p{0.46\textwidth}@{}}
\caption{MySQL per-database extraction (rows merged by (module, path); only implemented leaves shown). \textbf{Module-Tree Path} is the chain of bolded ancestors in MySQL's module tree; \textbf{Technical Description} summarises how MySQL realises that path.}\\
\label{tab:appendix:mysql}\\
\toprule
\textbf{Module} & \textbf{Module-Tree Path} & \textbf{Technical Description} \\
\midrule
\endfirsthead
\multicolumn{3}{c}{\footnotesize\itshape continued from previous page}\\
\toprule
\textbf{Module} & \textbf{Module-Tree Path} & \textbf{Technical Description} \\
\midrule
\endhead
\midrule
\multicolumn{3}{r}{\footnotesize\itshape continued on next page}\\
\endfoot
\bottomrule
\endlastfoot
Parser & DDL Syntax & The parser accepts CREATE/DROP TABLE, CREATE VIEW, CREATE INDEX, CREATE TRIGGER, CREATE FUNCTION/PROCEDURE, and CREATE EVENT for the timed event scheduler. ALTER TABLE drives InnoDB Online DDL so most schema changes proceed without blocking concurrent DML. \\
Parser & DML Syntax & SELECT, INSERT, UPDATE, and DELETE form the core DML surface. INSERT ... ON DUPLICATE KEY UPDATE provides upsert semantics, REPLACE offers delete-then-insert behaviour, and 8.0+ approximates MERGE through the same ON DUPLICATE KEY mechanism rather than a dedicated MERGE statement. \\
Parser & Query Syntax Features & The 8.0 parser adds recursive CTEs, window functions over OVER clauses, LATERAL derived tables (8.0.14+), JSON path expressions, and optimizer hint comments alongside legacy subqueries, UNION/INTERSECT/EXCEPT compound selects, virtual and stored generated columns, and EXPLAIN/EXPLAIN ANALYZE plan inspection. \\
Parser & Constraint Syntax & UNIQUE and PRIMARY KEY constraints are universally enforced, foreign keys are honoured only by InnoDB (MyISAM ignores them), and CHECK constraints became enforced expressions starting in 8.0.16 rather than parser-only annotations. \\
Parser & Built-in Functions & The function library covers date/time, math, string, regular expression, and aggregate operators including GROUP\_CONCAT, plus 30+ JSON functions over the native JSON type, spatial GIS predicates such as ST\_Contains and ST\_Distance built on Boost.Geometry, and MATCH ... AGAINST full-text search syntax. \\
Parser & Partitioning Syntax & Tables can be horizontally partitioned by RANGE, LIST, HASH, or KEY, with LINEAR variants of HASH and KEY that apply a power-of-two distribution function for cheaper partition addition. \\
Storage Engine & Storage Engine Selection / InnoDB / Storage Model / Row Formats & InnoDB supports four on-disk row formats: REDUNDANT preserves the original 5.0 layout, COMPACT shrinks header overhead, DYNAMIC stores variable-length columns off-page when they overflow, and COMPRESSED applies page-level zlib compression. \\
Storage Engine & Storage Engine Selection / InnoDB / InnoDB Page Management & The doublewrite buffer absorbs partial-write torn pages by staging full page images before writing them in place. The change buffer (formerly insert buffer) defers and merges secondary-index modifications when target pages are not in the buffer pool. \\
Storage Engine & Storage Engine Selection / InnoDB / InnoDB Tablespace & InnoDB stores data across the shared system tablespace ibdata1, file-per-table .ibd files, user-defined general tablespaces shared by multiple tables, dedicated undo tablespaces for rollback segments, and a temporary tablespace for intrinsic and user temporary tables. \\
Storage Engine & Storage Engine Selection / MyISAM & MyISAM is a classic non-transactional row store that keeps data in MYD files and indexes in MYI files, optimised for read-mostly workloads without MVCC or crash recovery. \\
Indexing & Index Types & The default access method is a B+Tree used by both InnoDB and MyISAM. Hash indexes back the Memory engine and InnoDB's adaptive hash, full-text indexes provide MATCH...AGAINST search in InnoDB and MyISAM, and R-Tree indexes serve spatial GIS queries in both engines. \\
Indexing & Index Organization (InnoDB) & The clustered primary-key index physically stores the row, secondary indexes carry the primary-key value as their tuple identifier and require a back-lookup, and the adaptive hash index is automatically promoted by InnoDB over hot B+Tree pages to short-circuit equality probes. \\
Indexing & Index Features & Beyond covering indexes, MySQL offers prefix indexes that index only the first N characters of a column, descending indexes (8.0+) for true reverse-order traversal, invisible indexes (8.0+) hidden from the optimizer for safe drop trials, and functional indexes (8.0.13+) over expressions. \\
Query Processing & -- & Semantic analysis performs name resolution and type checking after parsing. Through the handlerton interface, the secondary-engine offload mechanism can redirect eligible queries to accelerators such as HeatWave while keeping the primary engine for transactional state. \\
Query Processing & Optimizer & The 8.0+ Hypergraph optimizer represents join relationships as a hypergraph and unifies plans under an Access Path abstraction, complementing the legacy heuristic-plus-cost optimizer. The range optimizer drives index range scans, Index Merge, Skip Scan and Group Skip Scan, and partition pruning, while equi-height histograms, index cardinality, and ANALYZE TABLE feed cost estimation. \\
Query Processing & Execution Engine / Join Operators & Joins are executed by nested loop as the universal fallback, Hash Join for equi-joins since 8.0.18, and Batch Key Access which buffers join keys and feeds them to the storage engine in batches to amortise random I/O. \\
Query Processing & Execution Engine / Scan Operators & The executor exposes full table scan, full index scan, index range scan, ref lookups for unique key equality, and Index Merge which intersects or unions row-id sets from several single-column indexes when no composite index covers the predicate. \\
Concurrency Control & Isolation Mechanism & InnoDB implements MVCC by chaining old row versions through the undo log rather than retaining old tuples in the heap as PostgreSQL does. MyISAM falls back to coarse table-level locking, and Memory and Archive run effectively without locking for their respective access patterns. \\
Concurrency Control & Lock Management & InnoDB's lock manager exposes record, gap, next-key (record plus gap, the REPEATABLE READ default), insert-intention (a gap variant allowing concurrent inserts in different positions of the same gap), predicate (for R-Tree spatial indexes), and auto-increment lock modes. \\
Concurrency Control & -- & Metadata locking guards DDL/DML interaction at the SQL layer with a richer mode lattice including SHARED, EXCLUSIVE, and INTENT\_EXCLUSIVE so schema changes do not race with running statements that depend on the same objects. \\
Concurrency Control & Synchronization Primitives & Internal critical sections are protected by mutexes for short exclusive sections and read-write latches offering S-Latch and X-Latch modes for buffer-pool pages and other shared structures. \\
Concurrency Control & Deadlock Handling & InnoDB builds a wait-for graph among transactions to detect cycles and aborts a victim, and as a coarse fallback the innodb\_lock\_wait\_timeout setting bounds how long any lock acquisition may block before erroring out. \\
Concurrency Control & Transaction Management & Transactions support nested SAVEPOINTs for partial rollback and XA two-phase-commit transactions that allow MySQL to act as a resource manager in distributed transactions coordinated by an external TM. \\
Concurrency Control & Isolation Levels (InnoDB) & InnoDB offers all four SQL standard levels: READ UNCOMMITTED, READ COMMITTED, REPEATABLE READ (which uses next-key locks to prevent phantoms while still using MVCC for reads), and SERIALIZABLE which promotes plain SELECTs to locking reads. \\
Recovery \& Logging & InnoDB Redo Log & The redo log uses group commit to batch many transactions into a single fsync for higher throughput, and supports log encryption so on-disk redo records are protected at rest. \\
Recovery \& Logging & InnoDB Undo Log & Undo records are managed in rollback segments living in dedicated undo tablespaces, providing both transaction rollback and the version chain that MVCC reads follow. \\
Recovery \& Logging & -- & InnoDB performs fuzzy checkpoints that advance the redo low-water mark without freezing writes, replays redo from the last checkpoint during crash recovery, and uses the doublewrite buffer to repair partial-page writes detected at recovery. \\
Recovery \& Logging & Binary Log (Binlog) / Binlog Format & The binary log records changes for replication and point-in-time recovery in three formats: statement-based logs the SQL text, row-based logs before/after row images, and mixed switches per-statement based on safety. \\
Replication & Binary Log Replication & Replicas tail the primary's binlog asynchronously by default for maximum throughput, while the semi-synchronous plugin makes the primary wait for at least one replica's acknowledgement before returning success to the client. \\
Replication & -- & GTIDs assign a globally unique identifier to each transaction so failover can resume from the correct position without binlog-coordinate bookkeeping. Replication channels let one replica consume from multiple sources, and the Clone plugin produces a physical snapshot to bootstrap new nodes quickly. \\
Replication & Group Replication & Group Replication runs a Paxos-based atomic broadcast (XCom) and resolves write conflicts through certification of write-sets, applies flow control to keep replicas from falling behind, and operates in either single-primary mode for a fixed writer or multi-primary mode where any node accepts writes. \\
Memory Management & InnoDB Buffer Pool & The buffer pool implements a two-segment young/old LRU to resist scan pollution, can be sharded into multiple instances to reduce mutex contention, performs linear and random read-ahead prefetching, and supports dump/restore so a restarted server can repopulate hot pages quickly. \\
Memory Management & -- & MyISAM owns a separate key cache for index blocks, the TempTable engine provides modern in-memory storage for query temporaries, the sort buffer feeds Filesort, and a dedicated join buffer backs Hash Join and Batch Key Access. \\
Catalog \& Metadata & -- & The 8.0+ atomic data dictionary stores catalog metadata in transactional InnoDB tables instead of legacy .frm files, enabling DDL that is fully atomic across catalog and storage. A dictionary cache layer speeds repeated lookups, INFORMATION\_SCHEMA exposes the dictionary as views, and Performance Schema offers deep instrumentation through hundreds of virtual tables. \\
Process Architecture & Threading Model & The default model assigns one OS thread per client connection, a single-threaded mode exists for debugging, and the Enterprise edition ships a thread-pool plugin that multiplexes connections onto a bounded worker set for high connection counts. \\
Process Architecture & Connection Transports & Clients connect over TCP sockets and Unix domain sockets on Unix-like systems, plus named pipes and shared-memory transports on Windows for local low-latency access. \\
Process Architecture & Background Threads (InnoDB) & InnoDB runs a master thread coordinating periodic maintenance, multiple I/O threads for asynchronous reads and writes, purge threads that retire undo records, page-cleaner threads that flush dirty buffer-pool pages, and a dedicated log writer thread driving redo durability. \\
Networking \& Protocol & Protocol & The classic MySQL wire protocol carries SQL statements and result sets for legacy and standard clients, while the X Protocol underpins MySQL Document Store with CRUD operations over JSON documents and a more modern message framing. \\
Networking \& Protocol & Authentication & Authentication plugins span legacy mysql\_native\_password, the 8.0+ default caching\_sha2\_password, and enterprise integrations with LDAP, PAM, and Kerberos, with multi-factor authentication chaining several methods per account. \\
Networking \& Protocol & -- & Client/server traffic can be wrapped in SSL/TLS for transport encryption and certificate-based mutual authentication. \\
Networking \& Protocol & Authorization & Authorization in 8.0+ adds true role-based access control with grantable roles, dynamic privileges that finely scope administrative actions, and partial revokes that deny access to specific schemas even when a broader privilege is granted. \\
Extensibility & -- & MySQL is heavily plugin-driven: storage engines, authentication, audit, and replication ship as plugins; the 8.0+ component system layers a service-oriented replacement for the older plugin API; user-defined functions extend SQL; the rewriter plugin transforms incoming queries; the audit plugin emits compliance logs; the keyring plugin manages keys for transparent data encryption; and full-text parser plugins such as ngram add tokenizers. \\
\end{longtable}
}

\subsection{MariaDB}
\label{sec:appendix:per-db-tables:mariadb}
{\footnotesize
\renewcommand{\arraystretch}{1.1}
\begin{longtable}{@{}p{0.16\textwidth} p{0.32\textwidth} p{0.46\textwidth}@{}}
\caption{MariaDB per-database extraction (rows merged by (module, path); only implemented leaves shown). \textbf{Module-Tree Path} is the chain of bolded ancestors in MariaDB's module tree; \textbf{Technical Description} summarises how MariaDB realises that path.}\\
\label{tab:appendix:mariadb}\\
\toprule
\textbf{Module} & \textbf{Module-Tree Path} & \textbf{Technical Description} \\
\midrule
\endfirsthead
\multicolumn{3}{c}{\footnotesize\itshape continued from previous page}\\
\toprule
\textbf{Module} & \textbf{Module-Tree Path} & \textbf{Technical Description} \\
\midrule
\endhead
\midrule
\multicolumn{3}{r}{\footnotesize\itshape continued on next page}\\
\endfoot
\bottomrule
\endlastfoot
Parser & DDL Syntax & Parses CREATE/DROP TABLE, CREATE VIEW, CREATE INDEX, CREATE TRIGGER, CREATE FUNCTION/PROCEDURE plus ALTER TABLE with InnoDB Online DDL. MariaDB-specific extensions add CREATE EVENT for the scheduled event scheduler, CREATE SEQUENCE in PostgreSQL style, and CREATE OR REPLACE for views, triggers, procedures and events. \\
Parser & DML Syntax & Implements the standard SELECT, INSERT, UPDATE and DELETE statements together with INSERT ... ON DUPLICATE KEY UPDATE upsert semantics and MySQL-compatible REPLACE. Since 10.5 MariaDB also supports INSERT ... RETURNING, a MariaDB-only extension that returns generated rows directly to the client. \\
Parser & Query Syntax Features & Supports subqueries, recursive CTEs and window functions since 10.2, compound SELECT (UNION/INTERSECT/EXCEPT), VIRTUAL and PERSISTENT generated columns, EXPLAIN ANALYZE, JSON path expressions, optimizer hints and LATERAL derived tables. GROUPING SETS/CUBE/ROLLUP are partially supported from 10.5, and sql\_mode=ORACLE enables a subset of Oracle PL/SQL syntax. \\
Parser & Temporal / Versioning Syntax & Implements WITH SYSTEM VERSIONING to automatically retain prior row versions, application-time period tables for user-defined valid-time ranges, and bitemporal tables that combine both. Historical rows are queried via FOR SYSTEM\_TIME AS OF and BETWEEN clauses. \\
Parser & Constraint Syntax & Provides foreign keys (enforced by InnoDB), CHECK constraints supported well before MySQL 8.0.16, and UNIQUE plus PRIMARY KEY declarations validated by the storage engine. \\
Parser & Built-in Functions & Ships rich function libraries covering JSON manipulation, date/time arithmetic, math, strings, aggregates such as COUNT, SUM, AVG and GROUP\_CONCAT, OGC spatial primitives like ST\_Contains and ST\_Distance, PCRE regular expressions (where MySQL uses ICU), MATCH ... AGAINST full-text search, and NEXT/PREVIOUS VALUE FOR sequence functions unique to MariaDB. \\
Parser & Partitioning Syntax & Offers RANGE, LIST, HASH and KEY partitioning with the LINEAR variant for hash distribution, plus the MariaDB-only SYSTEM\_TIME partitioning that automatically routes historical versioned rows into time-bucketed partitions. \\
Parser & Custom Data Types & Adds first-class scalar types beyond MySQL: native UUID since 10.7, INET4 and INET6 IP-address types since 10.10, and a vector/embedding type for similarity workloads since 11.6. \\
Storage Engine & Storage Engine Selection / InnoDB / Storage Model / Row Formats & InnoDB rows can be stored as REDUNDANT (the original legacy layout), COMPACT (denser header), DYNAMIC (variable-length columns spilled off-page), or COMPRESSED (page-level zlib compression for both data and index pages). \\
Storage Engine & Storage Engine Selection / InnoDB / InnoDB Page Management & The doublewrite buffer guards against torn writes by staging pages in a contiguous area before flushing to their final location, and the change buffer (insert buffer) defers and merges modifications to non-unique secondary indexes so random I/O is amortised. \\
Storage Engine & Storage Engine Selection / InnoDB / InnoDB Tablespace & Supports the shared system tablespace ibdata1, file-per-table layout with one .ibd per table, user-defined general tablespaces grouping multiple tables, dedicated undo tablespaces for rollback segments, and a separate temporary tablespace for intermediate results. \\
Storage Engine & Storage Engine Selection / Aria & Aria is a row-store engine used by default for internal temporary tables (replacing MySQL's TempTable), supports a TRANSACTIONAL flag that enables page-level crash recovery without full ACID, and offers PAGE (default crash-safe), FIXED and DYNAMIC row formats. \\
Indexing & Index Types & Provides B+Tree as the default for InnoDB and MyISAM, hash indexes in the Memory engine and InnoDB's automatically built Adaptive Hash Index, full-text indexes across InnoDB, MyISAM and Mroonga, R-Tree spatial indexes in InnoDB and MyISAM, and LSM-tree indexing through MyRocks. \\
Indexing & Index Organization (InnoDB) & The clustered index is the table's physical layout keyed by primary key, secondary indexes hold primary-key bookmarks at their leaves so non-covering lookups must traverse back to the cluster, and the Adaptive Hash Index opportunistically maps frequently accessed B+Tree pages to a hash structure. \\
Indexing & Index Features & Supports covering indexes that satisfy queries from secondary structures alone, prefix indexes that key only the leading N characters, descending index ordering, invisible indexes hidden from the optimizer for safe rollouts, and functional indexes computed from expressions. \\
Query Processing & -- & Semantic analysis resolves names against the catalogue and performs type checking and coercion before the optimizer plans the statement. \\
Query Processing & Optimizer / MariaDB Unique Optimizations & MariaDB-only transforms include table elimination that drops unreferenced JOIN inputs, group-by cardinality optimization, virtual-column substitution that rewrites expressions to use indexed virtual columns, an enhanced Index Condition Pushdown, and condition pushdown into IN-subqueries. \\
Query Processing & Execution Engine / Join Operators & Supports nested-loop joins, hash join, Batched Key Access for amortising index probes, and Block Nested Loop for joining without an applicable index. \\
Query Processing & Execution Engine / Scan Operators & Provides full table scans, ordered index scans, index range scans, multi-index merge scans, and ref-style unique key lookups for equality probes on primary or unique indexes. \\
Concurrency Control & Isolation Mechanism & InnoDB implements MVCC by chaining old row versions through the undo log (rather than keeping prior versions in the heap as PostgreSQL does), MyISAM and Aria fall back to table-level locking, Memory and Archive run lock-free or with engine-internal serialisation, and MyRocks offers both pessimistic and optimistic transactional modes. \\
Concurrency Control & Lock Management & InnoDB acquires record locks on individual rows, gap locks on the empty ranges between index entries to block phantoms, next-key locks combining record and gap (the REPEATABLE READ default), insert-intention locks as a lightweight gap-lock variant, predicate locks for R-Tree spatial queries, and table-level auto-increment locks for AUTO\_INCREMENT columns. \\
Concurrency Control & -- & Metadata locking arbitrates between DDL and concurrent DML using SHARED, EXCLUSIVE, INTENT\_EXCLUSIVE and other modes so that schema changes wait for in-flight statements without corruption. \\
Concurrency Control & Synchronization Primitives & Internal synchronisation uses mutexes for short critical sections and read/write latches with shared (S) and exclusive (X) modes for buffer-pool pages and other shared structures. \\
Concurrency Control & Deadlock Handling & InnoDB maintains a wait-for graph and aborts the cheapest victim when a cycle appears, while a configurable lock-wait timeout (innodb\_lock\_wait\_timeout) bounds how long any waiter blocks before being rolled back. \\
Concurrency Control & Transaction Management & Transactions support nested SAVEPOINT markers for partial rollback and XA two-phase commit so external transaction managers can coordinate atomic commits across resources. \\
Concurrency Control & Isolation Levels (InnoDB) & InnoDB implements all four standard isolation levels: READ UNCOMMITTED, READ COMMITTED, REPEATABLE READ which uses next-key locking to prevent phantoms, and SERIALIZABLE which promotes plain SELECTs to shared locks. \\
Recovery \& Logging & InnoDB Redo Log & The redo log uses group commit to coalesce many transactions into a single fsync, amortising flush latency, and supports redo log encryption to protect on-disk log contents. \\
Recovery \& Logging & InnoDB Undo Log & Undo records are managed in rollback segments that drive both transaction rollback and MVCC version-chain reconstruction, with a configurable number of segments to reduce contention. \\
Recovery \& Logging & -- & Aria has its own write-ahead log so it can perform crash recovery (unlike MyISAM which has none), fuzzy checkpoints flush dirty pages incrementally, crash recovery replays the redo log to restore committed but unflushed transactions, and the doublewrite buffer protects against partial page writes. \\
Recovery \& Logging & Binary Log (Binlog) / Binlog Format & The binlog can be written in statement-based, row-based, or mixed format; MariaDB defaults to mixed and switches to row-based whenever a statement is unsafe to replay deterministically (whereas MySQL defaults to row-based). \\
Replication & Binary Log Replication & Replication is asynchronous by default with a semi-synchronous mode that delays commit acknowledgement until at least one replica has received the events, trading throughput for stronger durability guarantees. \\
Replication & -- & MariaDB GTID identifies events by a domain-server-sequence triple distinct from MySQL's source\_id:transaction\_id, multi-source replication lets a replica consume binlogs from several primaries simultaneously, and mariabackup (a Percona XtraBackup fork) performs hot physical backups of running instances. \\
Replication & Galera Cluster & Galera provides certification-based replication that detects write-set conflicts using a Paxos-style protocol so all nodes accept writes, flow control throttles fast nodes when slow members lag, and node bootstrap uses incremental state transfer when possible or full state snapshot transfer (via rsync, mariabackup or mysqldump) for fresh joiners. \\
Replication & Parallel Replication & Replicas can apply events in parallel under a conservative scheduler that respects original commit order, an optimistic mode unique to MariaDB that speculatively applies non-conflicting transactions out of order, and an aggressive mode that maximises parallelism. \\
Memory Management & InnoDB Buffer Pool & The buffer pool uses a young/old LRU split to protect hot pages from full-table-scan eviction, supports multiple instances to reduce mutex contention, performs linear and random read-ahead, and can dump and restore its page contents for fast warm-up after restart. \\
Memory Management & -- & Aria has a dedicated page cache that supersedes the legacy MyISAM key cache for index blocks, and per-statement memory comes from sort buffers for filesort and join buffers backing hash join, BKA and BNL; the built-in thread pool greatly reduces per-connection memory compared with thread-per-connection. \\
Catalog \& Metadata & -- & Table definitions are still kept in .frm files combined with InnoDB's internal data dictionary (rather than the transactional data dictionary in MySQL 8.0), while INFORMATION\_SCHEMA exposes standard catalog views, PERFORMANCE\_SCHEMA provides deep instrumentation, and the mysql system database stores users, privileges, time zones and help tables. \\
Process Architecture & Threading Model & MariaDB supports the traditional thread-per-connection model and ships a built-in thread pool in the community edition (MySQL restricts this to Enterprise) that organizes workers into thread groups, auto-tunes the number of active threads, and scales much better under high connection counts. \\
Process Architecture & Connection Transports & Clients can connect over TCP sockets, Unix domain sockets on POSIX hosts, and on Windows additionally via named pipes and shared memory. \\
Process Architecture & Background Threads (InnoDB) & InnoDB runs dedicated background threads: I/O threads driving asynchronous reads and writes, purge threads recycling old undo records, page cleaner threads flushing dirty pages from the buffer pool, and a log writer thread emitting redo records. \\
Networking \& Protocol & Protocol & Speaks the classic MySQL wire protocol so existing MySQL client libraries and connectors interoperate without modification. \\
Networking \& Protocol & Authentication & Supports mysql\_native\_password and SHA-256 password schemes, Unix-socket peer authentication based on the OS user, and several plugins shipped in the community build only by MariaDB: Ed25519 elliptic-curve signatures, GSSAPI/Kerberos, and PAM. \\
Networking \& Protocol & -- & SSL/TLS encrypts client/server traffic and supports certificate verification, cipher selection and modern protocol versions for in-transit confidentiality. \\
Networking \& Protocol & Authorization & Role-based access control has been available since 10.0 (predating MySQL 8.0), built on top of the traditional grant tables that store privileges per user, database, table, column and routine. \\
Encryption & Data-at-Rest Encryption / Encryption Key Management Plugin & Master keys for at-rest encryption can be served by pluggable key managers including a file-based provider, AWS KMS, HashiCorp Vault, and any KMIP-compliant key server. \\
Encryption & Data-at-Rest Encryption / Encrypted Components & Encryption can be applied to InnoDB tablespaces at the tablespace level, the InnoDB redo log and undo log, Aria tablespaces (a MariaDB-only capability), the binary log, and on-disk temporary files used by sorts and large joins. \\
NoSQL Interface & -- & The MariaDB-only HandlerSocket plugin exposes the storage-engine handler interface directly over a network protocol, bypassing SQL parsing and optimization to deliver very high-throughput key-value style index lookups. \\
Extensibility & -- & A pluggable architecture covers storage engines, authentication, audit, encryption and more (with a broader plugin taxonomy than MySQL), users can register UDFs in C/C++, the bundled MariaDB Audit Plugin is community-edition (versus MySQL Enterprise-only), the MariaDB-exclusive Query Response Time plugin captures latency distributions, and full-text parser plugins let users supply custom tokenisers. \\
\end{longtable}
}

\subsection{SQLite}
\label{sec:appendix:per-db-tables:sqlite}
{\footnotesize
\renewcommand{\arraystretch}{1.1}
\begin{longtable}{@{}p{0.16\textwidth} p{0.32\textwidth} p{0.46\textwidth}@{}}
\caption{SQLite per-database extraction (rows merged by (module, path); only implemented leaves shown). \textbf{Module-Tree Path} is the chain of bolded ancestors in SQLite's module tree; \textbf{Technical Description} summarises how SQLite realises that path.}\\
\label{tab:appendix:sqlite}\\
\toprule
\textbf{Module} & \textbf{Module-Tree Path} & \textbf{Technical Description} \\
\midrule
\endfirsthead
\multicolumn{3}{c}{\footnotesize\itshape continued from previous page}\\
\toprule
\textbf{Module} & \textbf{Module-Tree Path} & \textbf{Technical Description} \\
\midrule
\endhead
\midrule
\multicolumn{3}{r}{\footnotesize\itshape continued on next page}\\
\endfoot
\bottomrule
\endlastfoot
Storage Engine & Storage Model / Row Store / Row Organization & Each B-Tree leaf node holds an entire row (rowid plus all column values). Column values are serialised compactly in declaration order using the SQLite Record Format. \\
Storage Engine & B-Tree / B-Tree Page Management / Autovacuum / Autovacuum Mode & Full Autovacuum compacts and reclaims free pages automatically at the end of every commit. Incremental Autovacuum only marks pages reclaimable, deferring the actual shrink until the application invokes PRAGMA incremental\_vacuum(N). \\
Storage Engine & Pager / Journal Mode & Rollback Journal is the default and copies original pages to a side journal before mutation, with DELETE, TRUNCATE, PERSIST, and MEMORY sub-modes. WAL mode writes changes to a log first, allowing readers and a writer to proceed concurrently, and can be omitted at compile time via SQLITE\_OMIT\_WAL. \\
Storage Engine & -- & The Pager fronts a pluggable Page Cache whose default implementation uses LRU replacement. The Database File Format is a single cross-platform big-endian file with a backward-compatible layout. VACUUM rebuilds the file to reclaim space (omittable via SQLITE\_OMIT\_VACUUM), In-Memory Databases are addressable via :memory:, and Incremental BLOB I/O streams large values without loading them whole. \\
Storage Engine & VFS -- Virtual File System / VFS Implementations & Built-in VFS backends ship for Unix (POSIX file syscalls), Windows (Win32 file APIs), and an optional KV-store VFS gated by SQLITE\_OS\_KV\_OPTIONAL for embedded key-value targets. \\
Virtual Machine & -- & The VDBE Core is a bytecode interpreter executing roughly 190 opcodes that SQL is compiled into. The VDBE API surface exposes sqlite3\_step and sqlite3\_finalize to drivers, auxiliary helpers manage Mem value cells, and a dedicated VDBE Sorter implements ORDER BY with in-memory plus external merge passes. \\
Parser & DDL Syntax & The parser handles CREATE TABLE and DROP TABLE through the schema builder, plus CREATE VIEW, CREATE INDEX, CREATE TRIGGER, ALTER TABLE, and ATTACH DATABASE. Most are gated by SQLITE\_OMIT\_* compile flags so embedded builds can drop unused DDL. \\
Parser & DML Syntax & DML covers SELECT, INSERT, UPDATE, and DELETE through dedicated compilation paths, plus UPSERT via INSERT ... ON CONFLICT. UPSERT can be excluded with SQLITE\_OMIT\_UPSERT. \\
Parser & Query Syntax Features & Query features include Subqueries, CTE, Window Functions, Compound SELECT (UNION/INTERSECT/EXCEPT), Generated Columns, EXPLAIN, and CAST, each with its own SQLITE\_OMIT\_* trim switch. PRAGMA is SQLite's signature configuration mechanism for runtime tuning and introspection. \\
Parser & Constraint Syntax & Foreign Keys are parsed unconditionally but only enforced when PRAGMA foreign\_keys is ON, and can be compiled out via SQLITE\_OMIT\_FOREIGN\_KEY. CHECK constraints are similarly trimmable, while UNIQUE and PRIMARY KEY are always available and backed by index B-Trees. \\
Parser & Built-in Functions & Built-ins include the JSON1 function set for JSON manipulation, Date/Time functions for ISO and Julian-day conversions, and the standard Aggregate Functions SUM, AVG, COUNT, MIN, and MAX. JSON and date families are individually trimmable via SQLITE\_OMIT\_* flags. \\
Query Processing & -- & Semantic Analysis and Code Generation are split across the schema builder for DDL, the SELECT/INSERT/UPDATE/DELETE compilers for DML, an expression compiler for scalar evaluation, and a name resolver that binds column and table references before VDBE bytecode is emitted. \\
Query Processing & Query Optimizer / Optimization Strategies & The optimizer chooses among Index Scan, Full Table Scan, Covering Index, and Automatic Index (transient indexes built on the fly). It also applies OR, BETWEEN, and LIKE rewrites to convert predicates into index-friendly range scans, each toggleable via SQLITE\_OMIT\_*\_OPTIMIZATION. \\
Query Processing & Executor Operators / Join & SQLite implements only Nested Loop Join: the planner picks an outer table and probes inner tables (often via an Index Scan) for each outer row, with no hash or sort-merge alternatives. \\
Concurrency Control & Threading Model & Single-Thread mode disables all mutexes for the smallest and fastest build but forbids cross-thread use. Multi-Thread mode protects each connection with its own mutex so different connections can run on different threads. Serialized mode (default) wraps everything in a global mutex so any thread may use any connection. \\
Concurrency Control & Mutex Implementation & The mutex layer is pluggable: a No-op Mutex for single-threaded builds, a pthread-based Unix Mutex, and a Win32 critical-section Windows Mutex. \\
Concurrency Control & File Locking / Locking Protocol & The Unix VFS supports POSIX fcntl byte-range locks (default), whole-file flock for filesystems lacking byte-range support, and a dotfile lock-file scheme for NFS and other environments where kernel locks are unreliable. \\
Concurrency Control & -- & Shared Cache Mode lets multiple connections in the same process share one page cache, adding table-level read/write locks for finer-grained intra-process concurrency. The feature can be removed via SQLITE\_OMIT\_SHARED\_CACHE. \\
Concurrency Control & Transaction & Savepoints provide nested transaction semantics: each SAVEPOINT establishes a named rollback point that can be released or rolled back independently of the enclosing transaction. \\
Concurrency Control & Isolation Level & Outside WAL mode, transactions are Serializable because the writer holds an exclusive lock for the duration of the write. In WAL mode, read transactions see a stable Snapshot Isolation view of the database as of their start while a single writer proceeds in parallel. \\
Memory Management & Memory Allocator & The allocator is selectable at build time: the default System malloc wrapper (mem1), mem2 with guard bytes and leak detection, mem3 carving from a pre-allocated buffer with no system malloc, and mem5 which uses a power-of-two Buddy system over a fixed arena. \\
Memory Management & Lookaside Allocator & Each connection owns a Two-Size Lookaside arena that satisfies small allocations (such as parse trees and statement structures) from a pre-allocated slab to bypass the general allocator. SQLITE\_OMIT\_TWOSIZE\_LOOKASIDE disables it. \\
Memory Management & -- & The Memory Journal stores rollback data for temporary tables and transient databases entirely in RAM to avoid disk I/O. The Soft Heap Limit, set via sqlite3\_soft\_heap\_limit64, asks the engine to release cached pages when usage crosses a threshold, and SQLITE\_ENABLE\_MEMORY\_MANAGEMENT enables proactive memory release. \\
Extensions & R-Tree & The R-Tree extension supports spatial indexing of bounding boxes, and the GeoPoly add-on extends it with arbitrary 2D polygon containment and intersection queries on top of the same R-Tree index. \\
Extensions & -- & Dynamic Extension Loading allows runtime loading of shared libraries (.so or .dll) when SQLITE\_ENABLE\_LOAD\_EXTENSION is set. The Virtual Table mechanism lets external data sources appear as ordinary SQL tables, and FTS5, R-Tree, and dbstat are all built on this framework; SQLITE\_OMIT\_VIRTUALTABLE removes it. \\
API \& Callbacks & -- & The Core API exposes sqlite3\_open/close, sqlite3\_prepare\_v2, and sqlite3\_step/sqlite3\_column\_* for connection, compilation, and result retrieval. The Authorization Callback lets applications intercept and forbid specific SQL operations, the Backup API performs online hot copies, and Trace/Profile and Progress callbacks let applications observe engine behaviour, each individually trimmable via SQLITE\_OMIT\_* flags. \\
Build \& Deployment & Build Mode & The recommended Amalgamation build concatenates every source file into a single sqlite3.c of 200k+ lines, which compiles faster and runs about 5\% faster thanks to whole-program optimization. Separate Compilation builds each translation unit individually and is mainly used by SQLite developers. \\
Build \& Deployment & Platform Targets & SQLite ships as a Native C Library and is also packaged as a WASM build for browsers, a JNI binding for Java and Android, and a long-standing Tcl binding (the language in which SQLite's test suite is written). \\
Build \& Deployment & -- & The Shell is the sqlite3 command-line tool that provides interactive SQL execution, dot-commands for import, export, and schema inspection, and is itself an example application built on the public C API. \\
\end{longtable}
}

\subsection{DuckDB}
\label{sec:appendix:per-db-tables:duckdb}
{\footnotesize
\renewcommand{\arraystretch}{1.1}
\begin{longtable}{@{}p{0.16\textwidth} p{0.32\textwidth} p{0.46\textwidth}@{}}
\caption{DuckDB per-database extraction (rows merged by (module, path); only implemented leaves shown). \textbf{Module-Tree Path} is the chain of bolded ancestors in DuckDB's module tree; \textbf{Technical Description} summarises how DuckDB realises that path.}\\
\label{tab:appendix:duckdb}\\
\toprule
\textbf{Module} & \textbf{Module-Tree Path} & \textbf{Technical Description} \\
\midrule
\endfirsthead
\multicolumn{3}{c}{\footnotesize\itshape continued from previous page}\\
\toprule
\textbf{Module} & \textbf{Module-Tree Path} & \textbf{Technical Description} \\
\midrule
\endhead
\midrule
\multicolumn{3}{r}{\footnotesize\itshape continued on next page}\\
\endfoot
\bottomrule
\endlastfoot
Parser & DDL Syntax & Parses CREATE/DROP TABLE, CREATE VIEW, and CREATE INDEX statements that build ART indexes. Also supports CREATE MACRO for scalar and table macros, CREATE SEQUENCE, CREATE TYPE for user-defined types, CREATE SCHEMA, ALTER TABLE for column add/drop/rename, and ATTACH/DETACH DATABASE for mounting multiple database files. \\
Parser & DML Syntax & Accepts SELECT, INSERT, UPDATE, DELETE, and INSERT OR REPLACE / INSERT OR IGNORE upsert variants. The COPY TO and COPY FROM statements provide bulk import and export against CSV, Parquet, and JSON files. \\
Parser & Query Syntax Features & Supports subqueries, recursive CTEs, window functions, and compound SELECT via UNION/INTERSECT/EXCEPT, plus INNER, LEFT, RIGHT, FULL, CROSS, SEMI, ANTI, ASOF, and POSITIONAL joins. DuckDB-specific extensions include QUALIFY for window filtering, SAMPLE, EXPLAIN/EXPLAIN ANALYZE, PIVOT/UNPIVOT, GROUP BY ALL, SELECT * EXCLUDE/REPLACE, LATERAL JOIN, VALUES clauses, and RETURNING. \\
Parser & Constraint Syntax & Recognises PRIMARY KEY, UNIQUE, NOT NULL, and CHECK constraints with full enforcement. FOREIGN KEY clauses are accepted declaratively but are not actively enforced at runtime. \\
Parser & Built-in Functions / Scalar Functions & Provides a broad scalar library covering string, math, date/time, list/array, struct, map, and type-conversion functions, plus pattern matching via LIKE, GLOB, and regular expressions. JSON functions are exposed through the bundled JSON extension. \\
Parser & Built-in Functions / Aggregate Functions & Ships standard aggregates (COUNT, SUM, AVG, MIN, MAX), statistical aggregates (stddev, variance, corr), approximate aggregates such as approx\_count\_distinct and reservoir\_quantile, and list/string aggregates including list\_agg and string\_agg. \\
Parser & Built-in Functions / Table Functions & Exposes table-producing functions including read\_csv and read\_csv\_auto with automatic type detection, read\_parquet, read\_json and read\_json\_auto, glob for file enumeration, and range and generate\_series for synthetic row sources. \\
Storage Engine & Storage Model & Stores user data in a pure column-store layout where each column lives independently on disk. The columnar layout naturally enables vectorized execution and high-ratio compression. \\
Storage Engine & Data Organization & Tables are partitioned into Row Groups of roughly 122880 rows that act as the unit of parallelism and I/O. Within a Row Group each column is held as an independently compressed and encoded Column Segment, and each Column Segment carries min, max, count, and null\_count statistics used for segment-level pruning. \\
Storage Engine & Compression Algorithms & Selects per-segment among RLE, Dictionary plus FSST, Bit-packing, ZSTD, Chimp for time-series floats, Roaring Bitmaps, ALP and ALPRD for floating-point data, PATAS, Constant for uniform columns, and Uncompressed as a fallback. The choice is made by a sampling-based analyzer that picks the smallest representation per Column Segment. \\
Storage Engine & Index / ART (Adaptive Radix Tree) / ART Node Types & Implements an Adaptive Radix Tree whose internal nodes adapt their fan-out using Node4, Node16, Node48, and Node256 layouts depending on the number of children. The ART backs PRIMARY KEY and UNIQUE constraint enforcement and accelerates point and range lookups. \\
Storage Engine & Buffer Manager & Performs all disk traffic in fixed-size blocks, treating block-level I/O as the unit of read, write, and eviction between memory and storage. \\
Storage Engine & -- & Persists the entire database in a single file similar to SQLite, with explicit forward and backward compatibility management across file format versions. A checkpoint mechanism flushes in-memory state and merges WAL changes into the main database file. \\
Query Processing & -- & A dedicated Binder spread across many files performs name resolution, type inference, and expression binding to turn the AST into bound logical expressions. The Planner then converts the bound AST into a logical plan tree composed of Scan, Filter, Projection, Join, Aggregate, Sort, and Limit operators. \\
Query Processing & Optimizer / Optimization Strategies & Applies filter pushdown and pullup, projection pullup, column pruning, and join order optimization driven by a cost model, cardinality estimator, and plan enumerator. Further passes include join elimination, CTE inlining and CTE filter pushing, common subexpression elimination, late materialization, Top-N rewrites for ORDER BY plus LIMIT, Row Group pruning via column statistics, compressed materialization of intermediates, build/probe side selection for Hash Join, and outer join simplification. \\
Query Processing & Execution Engine & Executes plans as vectorized column batches rather than tuple-at-a-time iterators, in the spirit of ClickHouse and unlike the SQLite VDBE or PostgreSQL Volcano model. Queries are decomposed into Pipelines connected at materialization points, and a Task Scheduler dispatches Pipeline tasks across multiple worker threads for intra-Pipeline parallelism. \\
Query Processing & Join Algorithms & Defaults to Hash Join and additionally provides Nested Loop Join, Sort-Merge Join, Cross Product, ASOF Join for time-series nearest matches, Range Join, IEJoin for inequalities, Blockwise Nested Loop, Positional Join, Delim Join for delimiter-driven correlation, and Perfect Hash Join when key domains permit. \\
Query Processing & Aggregation & Builds aggregations on a radix-partitioned hash table that splits keys by hash radix into independent partitions, enabling parallel aggregation across worker threads with low contention. \\
Type System & Scalar Types & Offers TINYINT, SMALLINT, INTEGER, BIGINT, and 128-bit HUGEINT integers, FLOAT and DOUBLE floats, arbitrary-precision DECIMAL, and BOOLEAN. \\
Type System & String Types & Provides VARCHAR for variable-length text and BLOB for arbitrary binary payloads. \\
Type System & Temporal Types & Supports DATE, TIME, TIMESTAMP with TIMESTAMP\_S, TIMESTAMP\_MS, and TIMESTAMP\_NS precision variants, and INTERVAL for durations. \\
Type System & Complex Types & Implements nested STRUCT records, variable-length LIST, fixed-length ARRAY, key-value MAP, and tagged UNION types, allowing arbitrary composition for semi-structured workloads. \\
Type System & Special Types & Adds VARIANT for JSON-like semi-structured data, geospatial types delivered through the spatial extension, and custom user-defined types registered through the Catalog. \\
Concurrency \& Transaction & MVCC & Uses optimistic concurrency control where writers do not acquire row locks eagerly but check for conflicts at commit time. An Undo Buffer retains prior row versions so concurrent readers see a consistent snapshot without blocking writers. \\
Concurrency \& Transaction & Writer Model & Adopts a single-writer multiple-reader model in which only one write transaction is active at a time while any number of read transactions proceed concurrently against MVCC snapshots. \\
Concurrency \& Transaction & Transaction States & Transactions move through Commit and Rollback terminal states, followed by a Cleanup phase that releases the Undo Buffer entries once no snapshot still needs them. \\
Concurrency \& Transaction & -- & A Write-Ahead Log records changes before the corresponding data pages are mutated, providing the foundation for crash recovery. Checkpoint operations periodically or manually fold WAL contents into the main database file and truncate the log. \\
Catalog & Catalog Manager / Catalog Entry Types & Maintains catalog entries for Schema, Table, Index, View, Function, Sequence, Macro, and Type objects. Each entry kind is versioned alongside the transaction system so DDL respects MVCC visibility. \\
Catalog & -- & A Dependency Manager tracks references between catalog objects, for example views depending on tables, to prevent unsafe drops. The Attached Databases mechanism allows mounting multiple database files at once and issuing cross-database queries. \\
Data Formats & Built-in Formats & Reads and writes Parquet with support for geometry and variant types, CSV with an automatic sniffer for delimiters and types, and JSON in both directions. These formats are first-class table sources usable directly in SELECT statements. \\
Data Formats & Extension-based Formats & Adds Arrow, Delta Lake, and Excel support through dedicated extensions that register additional table functions and writers, keeping the core engine slim. \\
Extension System & -- & An Extension Manager dynamically loads and unloads compiled extensions and can auto-install them from a remote extension repository. \\
Extension System & Built-in Extensions & Bundles core\_functions, parquet, json, icu for ICU collations, autocomplete for SQL completion, jemalloc for the Jemalloc allocator integration, and tpch and tpcds for benchmark data generation. \\
Extension System & External Extensions & Offers httpfs for S3 and HTTP file access, spatial for GIS functions, postgres\_scanner, mysql\_scanner, and sqlite\_scanner for federated reads, arrow for Apache Arrow interoperability, excel for spreadsheet I/O, fts for full-text search, and aws and azure for cloud credentials and Azure Blob Storage access, plus delta for Delta Lake reads. \\
Memory Management & -- & A Buffer Manager governs block-level data exchange between memory and disk and can spill blocks to a temporary directory when memory is exhausted. An Arena Allocator handles bulk allocation and release during query execution, all data is managed in fixed-size blocks, and the optional jemalloc extension supplies a more efficient system allocator. \\
Process Architecture & -- & Runs as an embedded library linked directly into the host application, executing in-process with no server and no client-server protocol, mirroring the SQLite model. Inside that process a Task Scheduler drives multi-threaded Pipeline execution for intra-query parallelism, and although there is no native network protocol, community wrappers expose HTTP and PostgreSQL-compatible interfaces. \\
\end{longtable}
}

\subsection{ClickHouse}
\label{sec:appendix:per-db-tables:clickhouse}
{\footnotesize
\renewcommand{\arraystretch}{1.1}
\begin{longtable}{@{}p{0.16\textwidth} p{0.32\textwidth} p{0.46\textwidth}@{}}
\caption{ClickHouse per-database extraction (rows merged by (module, path); only implemented leaves shown). \textbf{Module-Tree Path} is the chain of bolded ancestors in ClickHouse's module tree; \textbf{Technical Description} summarises how ClickHouse realises that path.}\\
\label{tab:appendix:clickhouse}\\
\toprule
\textbf{Module} & \textbf{Module-Tree Path} & \textbf{Technical Description} \\
\midrule
\endfirsthead
\multicolumn{3}{c}{\footnotesize\itshape continued from previous page}\\
\toprule
\textbf{Module} & \textbf{Module-Tree Path} & \textbf{Technical Description} \\
\midrule
\endhead
\midrule
\multicolumn{3}{r}{\footnotesize\itshape continued on next page}\\
\endfoot
\bottomrule
\endlastfoot
Parser & DDL Syntax & DDL covers CREATE and DROP TABLE with mandatory engine selection, CREATE VIEW and MaterializedView with auto-refresh, external dictionaries, ALTER TABLE column and index changes, CREATE DATABASE across multiple database engines like Atomic, Lazy, MySQL and PostgreSQL, ATTACH and DETACH for table mounting, and access control DDL for users, roles, quotas and policies. \\
Parser & DML Syntax & DML supports SELECT, INSERT, INSERT from SELECT, and TRUNCATE; UPDATE and DELETE are expressed as ALTER TABLE mutations that asynchronously rewrite affected data parts rather than performing in-place row modification. \\
Parser & Query Syntax Features & Queries support subqueries, CTEs, window functions, UNION ALL composition, ARRAY JOIN to unnest array columns, the full join taxonomy including ANTI, SEMI and ASOF, the PREWHERE pre-filter for column pruning, SAMPLE clauses on MergeTree, EXPLAIN variants for AST and pipeline, the FORMAT clause spanning many output formats, per-query SETTINGS overrides, LIMIT BY for per-group caps, and grouping extensions WITH TOTALS, ROLLUP and CUBE. \\
Parser & Built-in Functions / Regular Functions & Hundreds of scalar functions span math, string with URL and regex helpers, date and time, JSON, array with higher-order operators like map and filter, map, tuple, type conversion, conditional forms including multi-branch and CASE, hash families such as cityHash, sipHash and xxHash, geospatial via H3, S2 and GeoHash, Roaring bitmap operations, IP address and UUID helpers, encoding utilities, and machine learning evaluators such as catboost. \\
Parser & Built-in Functions / Aggregate Functions & Aggregates include the standard COUNT, SUM, AVG, MIN and MAX, statistical functions like variance, stddev and correlation, approximate distinct and quantile estimators, parametric sequence and funnel analyzers, group array collectors, regression fitters, and the unique combinator system that suffixes any aggregate with If, Array, Map, State or Merge to compose behaviour. \\
Storage Engine & Storage Engine Selection / MergeTree Family / Storage Model / Data Part Format & Each MergeTree data part is laid out either in Wide format with one file per column, the default for large parts, or in Compact format with all columns packed into a single file for small parts. \\
Storage Engine & Storage Engine Selection / MergeTree Family / MergeTree Variants & The MergeTree family extends the base columnar engine with merge-time semantics: ReplacingMergeTree deduplicates by version, SummingMergeTree sums numeric columns, AggregatingMergeTree retains pre-aggregated states paired with AggregateFunction types, CollapsingMergeTree folds rows via a sign column, VersionedCollapsingMergeTree adds a version, and GraphiteMergeTree downsamples Graphite time series. \\
Storage Engine & Storage Engine Selection / MergeTree Family / Replicated Variants & Each MergeTree variant has a Replicated counterpart such as ReplicatedMergeTree, ReplicatedReplacingMergeTree and ReplicatedSummingMergeTree that adds multi-replica coordination while preserving the merge semantics of its base. \\
Storage Engine & Storage Engine Selection / MergeTree Family / MergeTree Data Organization / Skip Indexes & Optional skip indexes attached to granules accelerate filters: minmax stores per-granule extremes, set stores the value set, bloom filters in plain, tokenised and n-gram variants give probabilistic membership, and annoy and usearch provide approximate nearest neighbour search. \\
Storage Engine & Storage Engine Selection / Non-MergeTree Engines / Log Family & The Log family offers minimal columnar storage without concurrent writes: Log uses a basic per-column layout, TinyLog is the simplest with one file per column, and StripeLog packs all columns into a single file. \\
Storage Engine & Storage Engine Selection / Non-MergeTree Engines / In-Memory & In-memory engines include Memory which loses data on restart, Buffer which caches rows in RAM and periodically flushes into a target table, Set which stores deduplicated values for use in IN predicates, and Join which preloads join inputs into memory. \\
Storage Engine & Storage Engine Selection / Non-MergeTree Engines / Special Purpose & Special engines cover Null which discards inserts, Distributed which fans out queries across shards, MaterializedView for incremental view maintenance, Dictionary for exposing external dictionaries as tables, Merge for unioning structurally identical tables virtually, and View together with LiveView for query reuse. \\
Storage Engine & Storage Engine Selection / Non-MergeTree Engines / External Integrations & External engines stream from or query remote systems including Kafka topics, RabbitMQ, NATS, S3, HTTP URLs, local files, HDFS, remote MySQL and PostgreSQL tables, MongoDB, and SQLite databases. \\
Column Types \& Encoding & Column In-Memory Representations & Columnar in-memory layouts span fixed-width vector columns for integer and float types, variable string and fixed string columns, nested array, tuple and map columns, a nullable wrapper that adds a null-mask column, low-cardinality dictionary encoding for repeated strings, sparse representation for mostly-default data, in-memory compressed columns, and semi-structured forms covering JSON object columns plus dynamic and variant union-like columns. \\
Column Types \& Encoding & Compression Codecs & Per-column codecs include LZ4 as the speed-first default and ZSTD for higher ratio, time-series specialists Delta, DoubleDelta and Gorilla, integer encodings T64, GCD and Bitpacking, floating-point compressors FPC and ALP, an Encrypted codec for at-rest encryption, and an explicit None option. \\
Query Processing & -- & Query processing is split into an Analyzer for name resolution, type inference and expression analysis, a Planner that builds a logical query plan, and a large family of Interpreters that translate AST and plans into executable operations with one interpreter class per statement type. \\
Query Processing & Processor-Based Execution Engine / Execution Mode & The processor graph supports both pull-based execution where downstream stages request chunks and push-based execution where producers actively forward chunks, allowing each pipeline edge to choose the most efficient mode. \\
Query Processing & Processor-Based Execution Engine / Source Processors & Source processors feed the pipeline from MergeTree parts, remote nodes, Kafka topics, and local files or S3 objects. \\
Query Processing & Processor-Based Execution Engine / Transform Processors & Transform processors implement WHERE and PREWHERE filtering, expression evaluation, sorting and merge sorting, aggregation, joins, window functions, LIMIT truncation, HAVING with TOTALS, and ROLLUP or CUBE expansion. \\
Query Processing & Processor-Based Execution Engine / Sink Processors & Sink processors persist pipeline output to MergeTree parts, forward chunks to remote nodes, or discard rows via a null sink. \\
Query Processing & Join Algorithms & Join execution chooses among hash join as the default, parallel hash join, sort-merge join and full sorting merge join, direct lookup joins through a Dictionary or Join engine, grace hash join for spill-to-disk operation on large inputs, cross join, and ASOF join for nearest-time matching unique to ClickHouse. \\
Data Formats & Input/Output Formats & Pluggable input and output formats include text forms TabSeparated, CSV, several JSON dialects, Values, Pretty and Vertical for terminals, and Markdown plus XML, alongside binary forms Parquet, ORC, Avro, Arrow, the Native ClickHouse encoding, RowBinary, Protobuf, MsgPack and Cap'n Proto. \\
Distributed Architecture & Sharding / Sharding Strategy & Sharding strategies include hash-based routing on a column, random distribution, and arbitrary custom expressions evaluated per row. \\
Distributed Architecture & Replication / Coordination / ClickHouse Keeper & ClickHouse Keeper is a built-in Raft-based replacement for ZooKeeper that speaks the same client protocol so existing replicated tables can switch coordinators without code changes. \\
Distributed Architecture & -- & Remote query execution ships intermediate columnar chunks between nodes over TCP using the Native binary format for low-overhead transfer. \\
Memory Management & -- & Per-query memory is bounded by tracking allocations and aborting queries that exceed the configured maximum, columnar buffers are sourced from arena allocators and POD-style array structures so memory is allocated and released per column, and a global overcommit mechanism arbitrates memory across concurrent queries. \\
Memory Management & Caches & Multiple caches accelerate reads: a mark cache holds granule offset metadata, an uncompressed block cache retains decoded blocks, a compiled expression cache retains JIT-generated code, a query result cache memoises full results, and the engine otherwise relies on the operating system page cache instead of maintaining its own buffer pool. \\
Concurrency \& Consistency & -- & Concurrency is part-centric: each INSERT writes one immutable data part atomically with no row-level transactions, ALTER TABLE mutations are asynchronous rewrites with eventual consistency and no undo log, readers and writers never block each other thanks to immutable parts swapped via atomic pointers, DDL takes coarse table locks with no row or gap locks, and inserted blocks are deduplicated by content hash so retries are idempotent and ReplicatedMergeTree drops duplicate replicated blocks automatically. \\
Recovery \& Data Integrity & -- & Each data part carries a checksum manifest to detect corruption, small inserts can be protected by a write-ahead log to survive in-memory buffer loss, ReplicatedMergeTree recovers missing parts by syncing from healthy replicas through ZooKeeper or Keeper, immutable parts and merge-to-new-part semantics avoid partial writes without needing a double-write buffer, and BACKUP and RESTORE statements introduced in 22.x support incremental backups. \\
Materialized Views & Materialized View Types & Materialized views come in two flavours: incremental views are triggered on INSERT so new rows flow continuously into a backing MergeTree-family table for real-time pre-aggregation, while refreshable views perform full periodic recomputation under a REFRESH schedule. \\
Materialized Views & -- & Materialized views may be chained so the output of one view feeds another, building multi-stage pre-aggregation pipelines. \\
Process Architecture & Threading Model / Background Thread Pools & Dedicated background pools handle merging of data parts, executing mutations, fetching parts from replicas, and moving parts between storage volumes. \\
Process Architecture & Connection Handling & Clients connect over the high-performance native TCP protocol, a RESTful HTTP interface, gRPC, and wire-protocol shims that emulate MySQL and PostgreSQL clients. \\
Access Control \& Security & Authentication & Authentication accepts password hashes such as SHA256, double SHA1 and bcrypt, integrates with LDAP and Kerberos, and supports SSL client certificates. \\
Access Control \& Security & Authorization & Authorization layers role-based access control, per-row policies for row-level security, quotas that cap query frequency and resource usage, and reusable settings profiles. \\
Access Control \& Security & Encryption & Encryption covers SSL and TLS for transport and an Encrypted column codec for at-rest protection of selected columns. \\
External Integrations & Streaming Sources & Streaming integrations consume from Kafka, RabbitMQ and NATS through dedicated table engines that materialise messages into ClickHouse tables. \\
External Integrations & Object Storage & Object storage integrations read and write S3 with a queue-based variant for continuous ingestion, Azure Blob Storage, and HDFS. \\
External Integrations & Database Engines & Database-level engines mount entire remote schemas, mapping a remote MySQL, PostgreSQL or SQLite database as a local ClickHouse database. \\
External Integrations & Table Functions & Ad-hoc table functions query remote ClickHouse clusters via remote and remoteSecure, hit external systems through mysql, postgresql, jdbc and odbc, and read files directly from S3, URLs or the local filesystem. \\
\end{longtable}
}

\subsection{TiDB}
\label{sec:appendix:per-db-tables:tidb}
{\footnotesize
\renewcommand{\arraystretch}{1.1}
\begin{longtable}{@{}p{0.16\textwidth} p{0.32\textwidth} p{0.46\textwidth}@{}}
\caption{TiDB per-database extraction (rows merged by (module, path); only implemented leaves shown). \textbf{Module-Tree Path} is the chain of bolded ancestors in TiDB's module tree; \textbf{Technical Description} summarises how TiDB realises that path.}\\
\label{tab:appendix:tidb}\\
\toprule
\textbf{Module} & \textbf{Module-Tree Path} & \textbf{Technical Description} \\
\midrule
\endfirsthead
\multicolumn{3}{c}{\footnotesize\itshape continued from previous page}\\
\toprule
\textbf{Module} & \textbf{Module-Tree Path} & \textbf{Technical Description} \\
\midrule
\endhead
\midrule
\multicolumn{3}{r}{\footnotesize\itshape continued on next page}\\
\endfoot
\bottomrule
\endlastfoot
Parser & DDL Syntax & TiDB parses MySQL-compatible DDL covering CREATE/DROP TABLE, CREATE VIEW, CREATE INDEX, CREATE SEQUENCE and CREATE/DROP DATABASE, plus TiDB extensions: ALTER TABLE driven through Online DDL with distributed backfill, FLASHBACK for table and database recovery, table-level TTL clauses for automatic data expiration, and PLACEMENT POLICY for data-placement rules. \\
Parser & DML Syntax & The parser accepts SELECT, INSERT, UPDATE and DELETE alongside MySQL-style INSERT ... ON DUPLICATE KEY UPDATE upsert and REPLACE; LOAD DATA performs bulk client-side ingest while IMPORT INTO triggers a distributed parallel import path for large volumes. \\
Parser & Query Syntax Features & Query syntax covers subqueries, CTEs including recursive WITH, window functions via OVER, compound SELECT with UNION/INTERSECT/EXCEPT, VIRTUAL and STORED generated columns, EXPLAIN and EXPLAIN ANALYZE, a rich optimizer Hint vocabulary embedded in /*+ ... */ comments, and partition definitions of RANGE, LIST, HASH and KEY flavours. \\
Parser & Constraint Syntax & Table-level constraints recognized include foreign keys, CHECK predicates, and UNIQUE plus PRIMARY KEY declarations, with PRIMARY KEY potentially backed either by a clustered or non-clustered organisation. \\
Parser & Built-in Functions & The built-in catalog spans JSON manipulation, date/time arithmetic, math, string handling, aggregates such as COUNT, SUM, AVG and GROUP\_CONCAT, encryption helpers, and control-flow primitives like IF, CASE and COALESCE, all aiming for MySQL semantic parity. \\
Parser & Administrative Syntax & Administrative grammar includes TiDB-specific ADMIN statements, SET and SHOW for variables, ANALYZE TABLE to refresh statistics, SPLIT REGION for manual Region pre-splitting, RESOURCE GROUP management, and CALIBRATE RESOURCE for capacity calibration. \\
Query Processing & -- & Semantic analysis resolves names against the InfoSchema, infers types, and pre-processes expressions before logical planning, ensuring identifiers, columns and function signatures are bound prior to optimization. \\
Query Processing & Optimizer / Logical Optimization & Logical rewrites include column pruning, predicate push-down, aggregation push-down, join reorder, subquery decorrelation, partition pruning, outer-join elimination, and projection elimination, each applied as a transformation on the logical plan tree. \\
Query Processing & Optimizer / Physical Optimization & Physical optimization estimates cost from collected statistics, picks Root, Coprocessor or MPP task placement, prunes index candidates by Skyline comparison across multiple dimensions, and finally chooses an access path balancing scan cost and index utility. \\
Query Processing & Optimizer / Statistics & Statistics combine equi-depth histograms for range estimation, Count-Min Sketch for point-query frequency, FM-Sketch for distinct value counts (NDV), and Top-N tracking of high-frequency values, with a background Auto Analyze loop refreshing them as data shifts. \\
Query Processing & Execution Engine / Join Operators & Joins are realized as Hash Join as the default in-memory algorithm, Index Join (index nested loop) when an inner index can be probed per outer row, and Merge Join when both inputs arrive sorted on the join key. \\
Query Processing & Execution Engine / Scan Operators & Scans include full table scan, secondary index scan, index range scan bounded by computed ranges, and a Point Get fast path that bypasses the iterator stack for single-row lookups by primary key or unique index. \\
Query Processing & Execution Engine / Aggregate & Aggregation is performed either by Hash Aggregate, which builds a hash table keyed on the grouping columns, or Stream Aggregate, which exploits sorted input to emit groups incrementally with constant memory. \\
Query Processing & Distributed Execution / MPP Execution & For TiFlash-backed analytical workloads TiDB runs an MPP engine using shuffle exchanges to repartition data, broadcast joins for small build sides, and shuffle hash joins when both sides need hash repartitioning across compute nodes. \\
Storage Interface & KV Interface & TiDB talks to TiKV through a transactional KV API exposing Get, BatchGet and Scan for reads and Set and Delete for mutations, with all higher-level table and index operations expressed as compositions of these primitives. \\
Storage Interface & -- & A key-encoding layer maps tables and indexes onto KV pairs: row data is encoded as a tablePrefix/tableID/recordPrefixSep/rowID key bound to the row payload, and each secondary index becomes its own tablePrefix/tableID/indexPrefixSep/indexID/indexValues key pointing back to the rowID. A Region cache on the SQL layer remembers which TiKV node currently serves each Region so requests can be routed without consulting PD on every call. \\
Indexing & Index Types & Indexes include the primary key (clustered or non-clustered), secondary indexes encoded as separate KV pairs in TiKV, unique indexes that constrain the indexed values, expression indexes over computed expressions, invisible indexes hidden from the optimizer for safe rollout, and multi-valued indexes that index elements of JSON arrays. \\
Indexing & Index Organization & A clustered index, default since 5.0, stores row data directly under the primary key so the row is co-located with its PK; a non-clustered table instead allocates an internal \_tidb\_rowid as the heap address and the primary key becomes an ordinary unique index pointing to it. \\
Indexing & Index Selection \& Range Analysis & Index selection uses Skyline Pruning to discard candidate indexes dominated on every dimension (covering, equal-prefix length, scan order), and a range builder converts WHERE predicates into the disjoint key intervals that the chosen index must scan. \\
Transaction Coordination & Transaction Model & Distributed transactions follow the Google Percolator design: a Prewrite phase places primary and secondary locks on participating keys and a Commit phase marks them committed once the primary succeeds, with the protocol implemented client-side via the TiKV client library. \\
Transaction Coordination & Locking Mode / Pessimistic Locking & Pessimistic transactions take row locks during execution like MySQL InnoDB; Fair Locking arranges waiters in a queue so the longest-waiting request is granted next, reducing starvation under contention. \\
Transaction Coordination & Isolation Levels & Read Committed reads the latest committed value per statement, Repeatable Read takes a single MVCC snapshot at transaction start which behaves as Snapshot Isolation, and Serializable is accepted as a syntax-compatible alias rather than a separate enforcement mechanism. \\
Transaction Coordination & -- & Every transaction obtains start and commit timestamps from the Timestamp Oracle hosted by PD, which hands out globally monotonic TSOs that order all reads and writes across the cluster. \\
Transaction Coordination & Transaction Features & Optional features include Savepoint for partial rollback, Lazy Uniqueness Check that batches uniqueness verification to commit, Async Commit which returns success once Prewrite completes to cut latency, and 1PC which collapses Prewrite and Commit when all keys live in a single Region. \\
DDL & Online DDL Framework & Online DDL implements the F1 multi-state schema-change protocol, advancing the cluster through compatible intermediate schema versions; pending changes go through a job queue scheduled by a single elected DDL Owner so that exactly one node drives state transitions at a time. \\
DDL & DDL Operations & Supported operations include adding and dropping columns, building indexes via distributed parallel backfill, modifying column types, renaming and truncating tables, and DDL over partitioned tables such as adding, dropping and reorganising partitions. \\
DDL & Fast DDL & Fast DDL accelerates index creation by ingesting pre-sorted SST files directly into TiKV instead of issuing per-row writes, and by running the backfill phase using a Lightning-style parallel worker pool. \\
DDL & -- & Multi-Schema Change lets a single ALTER TABLE statement bundle several column or index modifications, executing them under one schema version transition rather than as separate DDL jobs. \\
Session \& Connection & -- & The frontend speaks the MySQL wire protocol so unmodified MySQL clients connect directly, with optional SSL/TLS encryption negotiated during the handshake. \\
Session \& Connection & Session Management & Sessions expose MySQL-compatible system variables augmented with TiDB-specific ones, support Prepared Statements with parameter binding and plan caching, and accept multi-statement requests in a single round-trip. \\
Session \& Connection & Authentication & Authentication plugins include the legacy mysql\_native\_password challenge-response, caching\_sha2\_password, an auth\_token mechanism for token-based logins, and LDAP-backed verification against an external directory. \\
Session \& Connection & Authorization & Authorization is role-based: privileges are granted to roles that users assume, and a set of dynamic privileges names cluster-level capabilities (such as backup or system management) that can be delegated independently of the static MySQL grant table. \\
Catalog \& Metadata & InfoSchema & Two InfoSchema implementations coexist: V1 loads the entire schema snapshot into memory at startup and on each version bump, while V2 fetches and caches schema objects on demand to bound memory use on clusters with very many tables. \\
Catalog \& Metadata & -- & Table, column and index metadata is encoded as KV pairs and stored inside TiKV itself rather than in a separate catalog store, with read/write helpers managing the encoding; INFORMATION\_SCHEMA exposes MySQL-compatible introspection views and the mysql.* system tables hold users, privileges and persisted statistics. \\
Resource Management & Resource Control / Resource Group & Resource Groups meter consumption in a unified Request Unit (RU) that combines CPU, IO and network costs, prioritise traffic as HIGH, MEDIUM or LOW, allow burstable groups to exceed quota when capacity is idle, and apply Runaway Query Management to tag, throttle or kill queries that overrun their budget. \\
Resource Management & Memory Management & A per-query Memory Tracker accumulates allocations across operators; when the limit is exceeded a configurable OOM Action either cancels the query or triggers Disk Spill, allowing Sort and Hash Join operators to spill intermediate state to local files and continue executing. \\
Resource Management & -- & A temporary file manager owns the on-disk scratch space used by sort, aggregation and other spillable operators, handling allocation, cleanup on query end and bounding aggregate disk usage across concurrent queries. \\
TTL & TTL Framework & TTL is implemented as a background framework with a job scheduler that periodically launches expiry jobs per TTL-enabled table, scan-and-delete workers that locate expired rows and remove them in batches, and integration with partitioned tables so whole expired partitions can be dropped instead of row-by-row deletion. \\
CDC \& Data Replication & -- & TiDB itself only exposes the change-event interface; the actual change-data-capture pipeline runs as the separately deployed TiCDC component, while legacy MySQL Binlog compatibility was provided by the Pump/Drainer pair, which is now largely superseded by TiCDC. \\
Extensibility & Plugin System & A plugin interface allows loadable Audit plugins that observe statements and connections, and Authentication plugins that introduce custom credential validation alongside the built-in handshake methods. \\
Extensibility & -- & A lightweight Extension Framework lets operators register custom scalar functions and session hooks without forking the server, and a Coprocessor Cache memoises Coprocessor request results so repeated pushed-down sub-queries can be served from cache rather than re-executed in TiKV. \\
\end{longtable}
}

\subsection{CockroachDB}
\label{sec:appendix:per-db-tables:cockroachdb}
{\footnotesize
\renewcommand{\arraystretch}{1.1}
\begin{longtable}{@{}p{0.16\textwidth} p{0.32\textwidth} p{0.46\textwidth}@{}}
\caption{CockroachDB per-database extraction (rows merged by (module, path); only implemented leaves shown). \textbf{Module-Tree Path} is the chain of bolded ancestors in CockroachDB's module tree; \textbf{Technical Description} summarises how CockroachDB realises that path.}\\
\label{tab:appendix:cockroachdb}\\
\toprule
\textbf{Module} & \textbf{Module-Tree Path} & \textbf{Technical Description} \\
\midrule
\endfirsthead
\multicolumn{3}{c}{\footnotesize\itshape continued from previous page}\\
\toprule
\textbf{Module} & \textbf{Module-Tree Path} & \textbf{Technical Description} \\
\midrule
\endhead
\midrule
\multicolumn{3}{r}{\footnotesize\itshape continued on next page}\\
\endfoot
\bottomrule
\endlastfoot
Parser & DDL Syntax & Parses CREATE/DROP TABLE, CREATE VIEW for view definitions, CREATE INDEX with distributed Online Schema Change semantics, CREATE TRIGGER, ALTER TABLE for distributed-safe online schema changes, CREATE FUNCTION/PROCEDURE for user routines, distributed CREATE SEQUENCE, CREATE TYPE for enums and composites, CONFIGURE ZONE for distribution-specific zone configs, and SET LOCALITY plus ALTER DATABASE SET PRIMARY REGION for multi-region placement. \\
Parser & DML Syntax & Recognises SELECT, INSERT, UPDATE, DELETE, and UPSERT including INSERT ... ON CONFLICT, producing a typed AST that downstream optimization and distributed execution layers consume. \\
Parser & Query Syntax Features & Supports subqueries, recursive CTEs, window functions, compound SELECT with UNION/INTERSECT/EXCEPT, generated columns, JSONB path and function operators, EXPLAIN and EXPLAIN ANALYZE that surface distributed plan shapes, LATERAL joins, GROUPING SETS/CUBE/ROLLUP, and AS OF SYSTEM TIME for MVCC historical reads. \\
Parser & Constraint Syntax & Accepts foreign keys enforced across distributed Ranges, CHECK constraints, UNIQUE and PRIMARY KEY declarations, and NOT NULL, with constraint validation woven into the distributed schema-change state machine. \\
Parser & Built-in Functions & Exposes date/time, math, and string libraries, aggregates such as SUM, AVG, COUNT, the ST\_* geo-spatial family backed by S2, and set-returning generators like generate\_series, all dispatched through the typed function registry. \\
Storage Engine & Storage Model / Row Store / Row Organization & Encodes each SQL row as one or more KV pairs with the primary key composed into the lexicographic Key and column values packed into the Value, stored in sorted order inside the Pebble LSM tree so range scans become contiguous KV iteration. \\
Storage Engine & Pebble / LSM Tree (Log-Structured Merge Tree) / Compaction & Runs Leveled Compaction so SSTables migrate from L0 down through size-tiered levels, bounding read amplification while merging overlapping key ranges and discarding superseded MVCC versions. \\
Storage Engine & Pebble / LSM Tree (Log-Structured Merge Tree) / Compression & Compresses SSTable data blocks with a configurable codec, offering Snappy as the default low-CPU option, Zstd for stronger ratios, and LZ4 as an alternative fast codec selectable per store. \\
Storage Engine & MVCC & Implements versioned Get, Scan, Put, and Delete keyed by HLC timestamp suffixes; an MVCC GC process periodically removes versions older than the configured TTL, and Intent Resolution converts uncommitted write markers into committed versions or discards them based on the transaction record outcome. \\
Indexing & Index Types / Secondary Index & Adds Covering Indexes via the STORING clause to avoid base-row lookups, Partial Indexes filtered by a WHERE predicate, Expression Indexes over functions of columns, Hash-Sharded Indexes that prepend a hash bucket to spread sequential write hotspots, and Multicolumn Indexes whose key tuples follow declared column order. \\
Query Processing & -- & The analyzer performs semantic analysis, name resolution, and type checking over the parsed AST and emits a fully typed AST that the optimizer consumes. \\
Query Processing & Optimizer & A Cascades-style optimizer represents the search space in a compact Memo, applies Normalization Rules to canonicalise expressions before exploration, uses Exploration Rules to enumerate equivalent logical plans, performs Physical Planning that picks implementations and derives required properties, scores alternatives with a statistics-driven Cost Model, and feeds on automatically collected Statistics including histograms. \\
Query Processing & Execution Engines / Row-Oriented Engine / Join Operators & Provides Nested Loop Join, Hash Join, Merge Join over sorted inputs, Lookup Join that probes a remote index across distributed Ranges, and Inverted Join for JSONB, Array, and geo predicates backed by inverted indexes. \\
Query Processing & Execution Engines / Row-Oriented Engine / Scan Operators & Offers full Table Scan, Index Scan over secondary indexes, and Reverse Scan that walks the underlying Pebble iterator backwards to satisfy ORDER BY DESC without an explicit sort. \\
Query Processing & Execution Engines / Row-Oriented Engine / Other Operators & Implements Sort with disk spill, Aggregate in both Hash and Ordered variants, Window functions with framing, Limit pushdown, and Distinct deduplication, all expressible as row-by-row processors in the row engine. \\
Query Processing & DistSQL & Flow Scheduling fragments the physical plan into processor flows distributed across nodes, Row-based Distribution assigns scan ranges to the nodes that hold the corresponding Range leases, and Result Aggregation streams partial outputs back through gateway processors that merge them. \\
KV Layer & -- & Exposes a transactional KV API with Get, Put, ConditionalPut, Scan, ReverseScan, and Delete primitives plus Batch operations that group multiple requests into a single round trip subject to atomic commit. \\
KV Layer & Range Management / Allocator & The Allocator performs Load-based Rebalancing by moving replicas toward nodes with spare CPU and QPS headroom, and Locality-aware Placement constrains replicas to honour zone, region, and diversity preferences. \\
KV Layer & Lease Management & Epoch-based Leases bind validity to a node liveness epoch so leases stay cheap to renew while the holder is alive, Expiration-based Leases use a wall-clock expiration for the meta and system Ranges, and Lease Transfers hand the leaseholder role to peers when the allocator needs to balance load. \\
Consensus \& Replication & Raft Consensus & Each Range runs a Raft group with Leader Election, Log Replication that durably commits writes once a majority acknowledges, Snapshot transfer to bootstrap new or far-behind followers, Joint Consensus for safe membership changes, and Pre-Vote to suppress disruptive elections during transient partitions. \\
Consensus \& Replication & Replica Management & Maintains a default Replication Factor of three voting replicas per Range, uses Learner Replicas as non-voting catch-up targets during rebalancing, and supports Non-Voting Replicas placed in remote regions to serve follower reads without contributing to quorum. \\
Distributed Transactions & Transaction Protocol & Writes deposit Intent markers visible alongside committed MVCC versions, a Transaction Record stored on the coordinator's Range tracks pending/committed/aborted state, Parallel Commits acknowledge the client once intents and the staging record reach consensus, and Transaction Pipelining lets later writes proceed without waiting for prior writes to fully replicate. \\
Distributed Transactions & Timestamp Management & An HLC combines physical wall-clock time with a logical counter to produce monotonic causally consistent timestamps, the Timestamp Cache records the highest read time per key span to detect write-after-read conflicts, and Closed Timestamps publish a horizon below which no new writes will appear so follower reads are safe. \\
Distributed Transactions & Isolation Levels & Defaults to Serializable isolation enforced through timestamp ordering and conflict detection, with Read Committed added in later versions for workloads that prefer fewer retries at the cost of weaker guarantees. \\
Concurrency Control & -- & A Concurrency Manager coordinates incoming requests, the Latch Manager grants short-lived span latches that serialise overlapping requests during execution, distributed Deadlock Detection inspects waits-for graphs across nodes, and a Transaction Wait Queue parks conflicting transactions until the blocking transaction commits or aborts. \\
Concurrency Control & Lock Table & The lock table tracks Exclusive Locks acquired by writers, Shared Locks for SELECT FOR SHARE, and Intent Locks materialized from MVCC intents so distributed lock state is reconstructed from the same versioned data the storage engine already holds. \\
Recovery \& WAL & -- & Durability comes primarily from the Raft Log acting as a replicated write-ahead log so a write is recoverable once a quorum has persisted it, supplemented by Pebble's local WAL that protects MemTable contents between flushes, and Snapshot Recovery brings far-behind or fresh replicas up to date by streaming a Raft snapshot of the Range. \\
Backup \& Restore & BACKUP / RESTORE & The BACKUP statement produces a Full Backup of all targeted spans and Incremental Backups derived from MVCC timestamp ranges since the prior backup, writing artefacts into Cloud Storage destinations such as S3, GCS, or Azure Blob. \\
Backup \& Restore & -- & Scheduled Backups attach CRON expressions to BACKUP jobs so full and incremental snapshots run automatically on the configured cadence and retention. \\
Change Data Capture & Changefeeds / Sink Types & Changefeeds can deliver events to Kafka topics with Avro or JSON encoding, generic Webhook endpoints, Cloud Storage sinks emitting Parquet or CSV files, and Google PubSub topics, each with its own batching and resolved-timestamp semantics. \\
Multi-Region & -- & Span Configuration v2 expresses zone settings declaratively over key spans for data placement, Follower Reads serve historical queries from non-leaseholder replicas to cut cross-region latency, and Locality-Optimized Search prefers locally resident copies before falling back to remote regions. \\
Multi-Region & Table Locality & REGIONAL BY TABLE pins an entire table to one home region, REGIONAL BY ROW lets each row carry its own region column that drives placement, and GLOBAL tables replicate fully and use closed timestamps so reads from any region stay strongly consistent without crossing the WAN. \\
Security & Authentication & Clients can authenticate with Password credentials, mutual TLS Certificates issued from the cluster CA, GSSAPI/Kerberos tickets, or OIDC tokens validated against an external identity provider. \\
Security & Authorization & Privileges are managed with GRANT and REVOKE over databases, schemas, tables, and other objects, layered with role membership for hierarchical permission management. \\
Security & Encryption & Encryption at Rest applies AES to on-disk store files with rotatable data keys protected by a store key, while Encryption in Transit secures inter-node and SQL client links using TLS with certificate-based peer verification. \\
Security & -- & Multi-Tenant support isolates SQL tenants on a shared KV layer so each tenant sees its own catalogues and key space, and Row-Level Security policies filter rows visible to a session based on declared access predicates. \\
SQL Schema Management & -- & Online Schema Changes drive DDL through a multi-stage state machine that propagates intermediate index and column states across the cluster so concurrent traffic stays correct, and the Declarative Schema Changer is a newer framework that plans transitions from a desired target state rather than imperative steps. \\
Observability & -- & SQL Activity surfaces statement statistics and slow-query traces, the DB Console offers a built-in web UI for cluster monitoring, Metrics export in a Prometheus-compatible format, and Tracing emits distributed query spans with OpenTelemetry integration for end-to-end visibility. \\
\end{longtable}
}

\subsection{YugabyteDB}
\label{sec:appendix:per-db-tables:yugabytedb}
{\footnotesize
\renewcommand{\arraystretch}{1.1}
\begin{longtable}{@{}p{0.16\textwidth} p{0.32\textwidth} p{0.46\textwidth}@{}}
\caption{YugabyteDB per-database extraction (rows merged by (module, path); only implemented leaves shown). \textbf{Module-Tree Path} is the chain of bolded ancestors in YugabyteDB's module tree; \textbf{Technical Description} summarises how YugabyteDB realises that path.}\\
\label{tab:appendix:yugabytedb}\\
\toprule
\textbf{Module} & \textbf{Module-Tree Path} & \textbf{Technical Description} \\
\midrule
\endfirsthead
\multicolumn{3}{c}{\footnotesize\itshape continued from previous page}\\
\toprule
\textbf{Module} & \textbf{Module-Tree Path} & \textbf{Technical Description} \\
\midrule
\endhead
\midrule
\multicolumn{3}{r}{\footnotesize\itshape continued on next page}\\
\endfoot
\bottomrule
\endlastfoot
Parser & Query API Selection / YSQL -- PostgreSQL 15 Fork / DDL Syntax & YSQL inherits PostgreSQL 15 DDL grammar, accepting CREATE TABLE and DROP TABLE with HASH or RANGE partitioning clauses, CREATE INDEX that materialises as an LSM-backed DocDB index, and CREATE VIEW, CREATE TRIGGER, CREATE FUNCTION, CREATE PROCEDURE in PL/pgSQL, plus ALTER TABLE for online schema evolution. \\
Parser & Query API Selection / YSQL -- PostgreSQL 15 Fork / DML Syntax & The PostgreSQL-derived parser accepts SELECT, INSERT, UPDATE, and DELETE statements as well as UPSERT expressed via INSERT ... ON CONFLICT, with the resulting plans dispatched through pggate to DocDB. \\
Parser & Query API Selection / YSQL -- PostgreSQL 15 Fork / Query Syntax Features & YSQL preserves PostgreSQL query features including subqueries, common table expressions with recursion, OVER-clause window functions, compound SELECT via UNION, INTERSECT, and EXCEPT, EXPLAIN and EXPLAIN ANALYZE diagnostics, and JSON or JSONB path expressions. \\
Parser & Query API Selection / YSQL -- PostgreSQL 15 Fork / Constraint Syntax & The constraint surface covers PRIMARY KEY declarations whose HASH or RANGE choice directly drives Tablet sharding, UNIQUE constraints, FOREIGN KEY references, and CHECK predicates evaluated within the YSQL layer. \\
Parser & Query API Selection / YSQL -- PostgreSQL 15 Fork / Built-in Functions & The fork exposes PostgreSQL built-in libraries for JSON and JSONB manipulation, date and time arithmetic, mathematical operations, string handling, and aggregate functions, all reusing the upstream implementations. \\
Storage Engine & DocDB & DocDB encodes SQL and CQL rows as document key-value pairs supporting composite keys, sub-documents, and collection types, with each Tablet maintaining a separate Intent store for provisional writes during distributed transactions and a Regular store for committed data. \\
Storage Engine & RocksDB / LSM-Tree Architecture / Compaction & Both Leveled compaction, which keeps each level size-bounded and non-overlapping, and Universal compaction, which favours larger merges for write-heavy workloads, are available as configurable strategies per column family. \\
Storage Engine & RocksDB / Compression & SSTable blocks may be compressed with Snappy, LZ4, or Zlib, letting operators trade CPU cost for on-disk footprint based on workload characteristics. \\
Storage Engine & Storage Model & Row data is materialized as a Document Store where each Tablet maps to its own RocksDB instance and rows are encoded as document key-value pairs keyed by the primary key. \\
Sharding \& Partitioning & Tablet-based Sharding / Partitioning Strategy & Tables can be partitioned by Hash, which spreads rows uniformly across Tablets via primary-key hashing, by Range for contiguous key distribution, or by a combined Hash plus Range scheme that hashes a leading prefix and ranges within each shard. \\
Indexing & Index Types & DocDB supports LSM-based indexes as the default form, Covering indexes via the INCLUDE clause to avoid base-row lookups, Partial indexes filtered by WHERE, Expression indexes over functions, GIN indexes for full-text and JSONB search, and Unique indexes enforced across the distributed cluster. \\
Indexing & -- & A vector index implementation provides HNSW approximate nearest neighbour search for AI and ML similarity queries, while index storage itself is realized as a separate DocDB table sharing the Tablet and Raft machinery of base data, and YCQL requires explicit index declarations. \\
Query Processing & YSQL Query Processing / Optimizer / Scan Operators & The optimizer can select Sequential Scan over the base relation, Index Scan that fetches matching keys then base rows, Index Only Scan when the index covers all referenced columns, and Bitmap Scan that combines multiple index lookups. \\
Query Processing & YSQL Query Processing / Optimizer / Join Operators & Join physical operators include Nested Loop for small or indexed inputs, Hash Join for equi-joins on larger inputs, and Merge Join when both sides are already ordered on the join keys. \\
Query Processing & -- & YCQL ships an independent query processor with its own semantic analysis and execution logic, intentionally restricted to single-table operations without JOIN support and without a sophisticated cost-based optimizer. \\
Query Processing & Pushdown & DocDB accepts Predicate Pushdown so WHERE filters run inside the Tablet, Aggregate Pushdown so reductions execute at the TServer to shrink result sets, and Expression Pushdown so projections evaluate close to storage. \\
Consensus \& Replication & Raft Consensus & Each Tablet replica group runs Raft to perform Leader Election, replicate Log entries to followers, hold a Leader Lease that allows safe local reads, persist a Raft Log that doubles as the Tablet write-ahead log, and serve bounded-staleness Follower Reads to reduce client latency. \\
Consensus \& Replication & xCluster Replication & Asynchronous xCluster replication supports both Unidirectional pipelines from a primary cluster to a standby and Bidirectional topologies where two clusters exchange writes for active-active deployments. \\
Consensus \& Replication & -- & Change Data Capture streams committed row mutations out of Tablets to external systems such as Kafka sinks, allowing downstream consumers to react to database events in near real time. \\
Distributed Transactions & Transaction Management & Distributed writes follow an intent-based protocol that first stages provisional records as intents in each Tablet's dedicated Intent RocksDB instance, while a Transaction Status Tablet authoritatively tracks each transaction as pending, committed, or aborted. \\
Distributed Transactions & MVCC & Multi-version concurrency uses HybridTime, a globally ordered hybrid logical clock timestamp combining physical and logical components, together with a Safe Time barrier that fixes a consistent read point across replicas. \\
Distributed Transactions & Isolation Levels & YSQL offers Snapshot Isolation as the default, Serializable isolation enforced through conflict detection on read and write sets, and Read Committed semantics matching PostgreSQL behaviour. \\
Distributed Transactions & -- & A distributed deadlock detector builds the wait-for graph across TServers to break cycles among contending transactions, and Savepoint statements are supported so YSQL clients can roll back to intermediate transaction points. \\
Cluster Architecture & Master Server & The Master process hosts the Catalog Manager for table, index, and namespace metadata, a Placement Policy engine for multi-region and multi-AZ replica rules, a Tablet Split Manager that triggers automatic splits, a Load Balancer that migrates Tablet replicas across TServers, and Master HA via its own Raft group. \\
Cluster Architecture & TServer & Each TServer serves local Tablet read and write requests, embeds a PostgreSQL backend process for YSQL clients, exposes a Backup Service for snapshot operations, and performs Remote Bootstrap to copy Tablet data when new replicas join. \\
Memory Management & -- & Memory is partitioned among the RocksDB Block Cache for SSTable data with LRU eviction, the RocksDB MemTable as a flushable write buffer, PostgreSQL Shared Buffers playing a limited role under YSQL since most caching lives in DocDB, and Per-Tablet memory accounting that bounds resource usage per shard. \\
Recovery \& Durability & -- & Durability layers a per-Raft-group WAL that guarantees committed entries survive crashes, the underlying RocksDB WAL for storage-level recovery, Crash Recovery that replays Raft logs and reopens RocksDB to restore Tablet consistency, Snapshot and Restore at cluster or table granularity with Point-in-Time Recovery, and a Backup pipeline for distributed backup and restore. \\
Networking \& RPC & -- & Inter-node and Master-TServer communication runs over a custom Protobuf-based RPC framework with asynchronous I/O, and SSL or TLS can be enabled for both peer traffic and client connections. \\
Networking \& RPC & Connection Pooler & Odyssey is bundled as a built-in connection pooler in front of the YSQL endpoint to multiplex many short-lived client sessions onto a smaller backend pool. \\
Networking \& RPC & Client Protocols & Clients connect via the PostgreSQL wire protocol for YSQL, the Cassandra CQL binary protocol for YCQL, and the Redis protocol for the legacy key-value API, allowing reuse of existing ecosystem drivers. \\
Networking \& RPC & Authentication & Authentication methods include Password credentials, LDAP integration with directory services, Kerberos for enterprise SSO, and Certificate-based mutual TLS for service-to-service trust. \\
Networking \& RPC & Authorization & Authorization is enforced through YSQL RBAC, which inherits the PostgreSQL role and privilege model, and YCQL RBAC, which implements Cassandra-style permission grants on keyspaces and tables. \\
Security & -- & Security features include Encryption at Rest using AES-256 over data files, Encryption in Transit via TLS for both node-to-node and client-to-node channels, and Row-Level Security in YSQL inherited from PostgreSQL row policies. \\
Observability \& Administration & -- & Operators interact through Admin UIs hosted on the Master and TServer web ports, scrape Prometheus-formatted Metrics for monitoring, and inspect System Catalog Tables surfaced as PostgreSQL system catalogs under YSQL and Cassandra-style System tables under YCQL. \\
\end{longtable}
}

\end{document}